\documentclass{optica-article}

\journal{opticajournal}
\articletype{Research Article}

\usepackage{siunitx}
\usepackage{wrapfig}
\usepackage{lineno}
\begin{document}
\title{Controlling light propagation in multimode fibers for imaging, spectroscopy and beyond}

\author{Hui Cao,\authormark{1,*} Tom\'{a}\v{s} \v{C}i\v{z}m\'{a}r,\authormark{2,3,4,*} Sergey Turtaev,\authormark{2} Tom\'{a}\v{s} Tyc,\authormark{5,3} and Stefan Rotter\authormark{6,*}}

\address{\authormark{1}Department of Applied Physics, Yale University, New Haven, CT 06511, USA\\
\authormark{2}Leibniz Institute of Photonic Technology, Albert-Einstein-Straße 9, Jena, 07745, Germany\\
\authormark{3}Institute of Scientific Instruments of CAS, Královopolská 147, 612 64, Brno, Czech Republic\\
\authormark{4}Institute of Applied Optics, Friedrich Schiller University Jena, Fr\"{o}belstieg 1, 07743 Jena, Germany\\
\authormark{5}Department of Theoretical Physics and Astrophysics, Faculty of Science, Masaryk University, Kotl\'a\v rsk\'{a} 2, 61137 Brno, Czech Republic\\
\authormark{6}Institute for Theoretical Physics, Vienna University of Technology (TU Wien), A–1040 Vienna, Austria}

\email{\authormark{*}Corresponding authors: hui.cao@yale.edu, tomas.cizmar@leibniz-ipht.de, stefan.rotter@tuwien.ac.at} 

\begin{abstract}	
Light transport in a highly multimode fiber exhibits complex behavior in space, time, frequency and polarization, especially in the presence of mode coupling.
The newly developed techniques of spatial wavefront shaping turn out to be highly suitable to harness such enormous complexity:
a spatial light modulator enables precise characterization of  field propagation through a multimode fiber, and by adjusting the incident wavefront it can accurately tailor the transmitted spatial pattern, temporal profile and polarization state. 
This unprecedented control leads to multimode fiber applications in imaging, endoscopy, optical trapping and microfabrication. Furthermore, the output speckle pattern from a multimode fiber encodes spatial, temporal, spectral and polarization properties of the input light, allowing such information to be retrieved from spatial measurements only. This article provides an overview of recent advances and breakthroughs in controlling light propagation in multimode fibers, and discusses newly emerging applications. 
\end{abstract}


	

\tableofcontents
\pagebreak
	\section{Introduction}
	\label{sec:intro}
 Multimode fibers (MMFs) are waveguides of microscopic dimensions with close to perfect cylindrical symmetry, which lend themselves to large magnitudes of bending and twisting. These attributes drove their dominant exploitation in the past, particularly towards the low-attenuation delivery of optical signals over long distances and for conveniently bridging remote sites of optical systems, which are to a large extent free to move with respect to one another. 
Sending information through MMFs by modulating the transported light intensity in time is nowadays well established and frequently exploited in short-distance communication \cite{richardson2013space}.  In some cases, the data transmission speed is enhanced by parallelizing this process on separable intervals of wavelengths (wavelength division multiplexing~\cite{9052826}). Yet this is far from utilizing the complete information capacity such systems can offer. MMFs can simultaneously support tens to hundreds of thousands of information channels, encoded in the spatial properties of the light signals. 

\begin{figure}[htbp]\centering
\includegraphics[width=.8\textwidth]{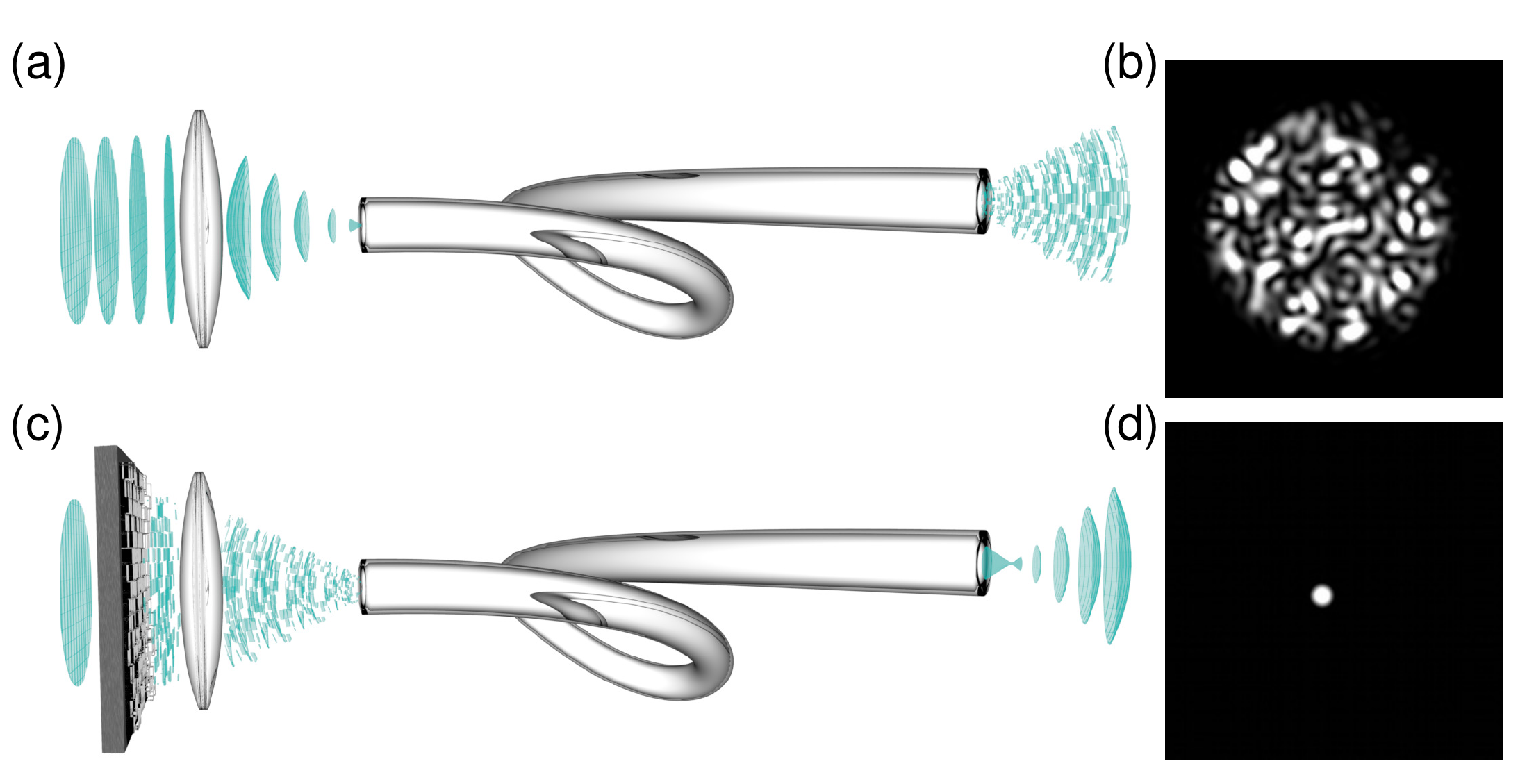}
\caption{Randomization of light propagating through a MMF. (a) Coherent light coupled into a MMF results in randomized outputs.  (b) Typical MMF output signal forming a randomly distributed speckle.
(c) Controlled light propagation through a MMF using a spatial modulator of light. (d) Desired output taking the shape of a diffraction limited focus.}
\label{fig:intro}
\end{figure}

Coupling monochromatic light to a MMF, for example by projecting a tight focus to a specific location at the input facet, results in a unique, randomly distributed speckle pattern at the output facet (see Fig. \ref{fig:intro}). Coupling the light at the same frequency to a different location of the fiber core produces a different speckle pattern, which is completely uncorrelated with the initial one if the displacement of the input is sufficiently large. The seemingly random yet deterministic transmission process therefore indicates the possibility to distinguish between multiple different input signals sharing the same frequency from the spatial distribution of the received output. Furthermore, if the input wavefront is fixed but the frequency shifts, the output speckle pattern also changes. Hence, the output speckle pattern encodes both spatial and spectral information of input light. 

The technology with which we may harness the full richness of such complex light transport became available at the turn of the century. Especially a whole spectrum of spatial light modulators (SLMs)~~\cite{https://doi.org/10.1002/andp.201500111,https://doi.org/10.1002/lpor.200900047} introduced a relatively low-cost mechanism for sculpturing light and for rapidly reconfiguring from a computer interface, all with fidelity not available ever before. 
Being able to shape optical fields in space  brought a direct route towards handling highly complex optical systems~\cite{gigan2022roadmap}. Light transport through MMFs at low power is linear and deterministic. Regardless of the basis of light modes we choose, the input-output mapping can be mathematically expressed as a linear operator, nowadays commonly known as the transmission matrix (TM)~\cite{popoff2010measuring}. Knowing the exact TM of a given optical system containing a MMF would enable the synthesis of any desired vector optical field at the output within the space and spatial frequency constraints of the MMF. The technology of spatial light modulators can be used to accurately acquire all TM components of a MMF supporting even tens of thousands of optical modes. The realization, that the same device can use the measured TM in order to synthesize any desired optical outputs, moreover in a manner which is immune to alignment imperfections and optical aberrations in other parts of the optical system~\cite{Cizmar2011}, brought a revolution to the field. 
As a result, the utility of MMFs has been reevaluated in the context of many new applications, especially those, where the minuscule dimensions of MMFs make a significant difference, and where the role of the transmitted light goes beyond that of a pure information carrier. Consider here especially the cases where the delivered signals actively participate in information acquisition or interact with microscopic samples in a desired manner.    

\textcolor{black}{ It is well known, that optical fibres have played a major role in imaging applications. Especially the emergence of fibre bundles in the 50s\cite{VanHeel_1954, HOPKINS:1954kq} turned this prospect into a global industry\cite{hecht2004city}. The potential of MMFs for imaging applications have been identified already in the 60s and 70s\cite{spitz1967transmission, yariv1976transmission, Gover:76}, yet the technology necessary for handling the complexity of light transport was missing. This is very different now:} exploiting a low-cost commercially available MMF and the above methodology, one can create a hair-thin endoscope, which can be employed through even the most sensitive tissues of living organisms (brain), sending back high-resolution images without causing significant damage to the overlying structures~\cite{bianchi2012multi,cizmar2012exploiting,choi2012scanner,papadopoulos2013high,loterie2015digital,ploschner2015LightSheet,caravaca2017single,Ohayon2018,Turtaev2018,EmptageLSI19}. 
MMFs with very high numerical aperture can be produced, opening up the prospect of other bio-photonics methods such as 3-D optical tweezers to be introduced in previously unthinkable studies~\cite{Cizmar2011,bianchi2012multi,Leite2018}.

A whole class of further applications emerges, when broad-band light signals are considered, since MMFs conversely act as a mixer between the spatial and the spectral/temporal domain \cite{mosk2012controlling, cao2022shaping}. Although beam-shaping will not affect the overall power carried by individual wavelengths, it can manipulate their phase and polarization relations. By appropriate manipulation of fields in spatial channels, each carrying a spectrum of frequencies, one can achieve focusing in both, space and time. In this way it becomes possible to produce for example spatially and temporally focused pulses carrying high energies~\cite{morales2015delivery}, which can be exploited in multi-photon and non-linear imaging modalities~\cite{morales2015two,tragaardh2019label} and in microfabrication~\cite{morales2017three}.  
Further in relevance to sensing, this mixing of domains allows for information of an input signal in one domain to be recovered from measurements of the output light in a different domain. As an example, the spatial intensity distribution of transmitted light encodes the information of the spectrum and of the temporal shape of an incident wave. Such information of the input signal may be retrieved from spatial measurements of the output light thereby turning the MMF into a high-resolution, broad-band spectrometer~\cite{redding2012using, redding2013all, redding2014high, liew2016broadband, bruce2019overcoming, meng2019multimode, bruce2020femtometer, wang2020study} or profiler of ultrashort pulses~\cite{xiong2020multimode, lee2020single, xiong2020deep, ziv2020deep}. It is even possible to simultaneously recover the input information in multiple domains. Reconstructing both spatial and spectral distributions of a signal from one speckle pattern even enables snapshot hyperspectral imaging~\cite{french2018snapshot, kurum2019deep}.

Further, the new experimental possibilities keep re-shaping our understanding of MMF transmission in general and drive a vibrant discussion of the relevant fields. 
The availability of the theoretical framework for predicting details of light propagation in perfect, straight or even bent MMFs predates our current investigations by many decades~\cite{Snitzer:1961vy,SnyderLove1983} and, on its own, it infers immensely complex behavior. 
However, real-life MMFs suffer from manufacturing imperfections and external perturbations manifesting themselves as unpredictable disorder that leads to random spatial- and polarization-mode coupling.  Thus light propagation through a MMF bears similarities to coherent transport of electron wave in a narrow metallic waveguide, which has been widely studied in mesoscopic physics \cite{RevModPhys.69.731,akkermans2007mesoscopic,rotter2017light}. Physical concepts and theoretical models, previously developed for light transport in complex systems such as random scattering media or chaotic optic cavities, may be applicable to MMFs. However, caution must be exercised as there are notable differences between those systems.  First, the transmission through a fiber is extremely high, even in the presence of strong mixing between spatial- and polarization modes, thus information on the input state of the light is only scrambled but not lost. This is in contrast to strong-scattering (diffusive) samples where most of the incident light is reflected. Second, unlike wide slabs with open (side) boundaries, fibers allow to fully control the coupling of the input light to all the guided modes thanks to the finite numerical aperture, and  to collect all output fields. Thanks to negligible reflection, a sharply bounded number of spatial modes, weak loss, flexibility, and reconfigurability \cite{resisi2020wavefront}, MMFs provide a powerful platform for fundamental studies of mesoscopic physics \cite{cao2022shaping}.

This review aims to introduce the theoretical background of light transport in ideal, perturbed and disordered fibers; to assess the available technology for manipulating and harnessing the delivery of light through MMFs in  experimental settings; to discuss the broad spectrum of applicability of MMFs; and finally, to outline the current research trends and open questions of this exciting field. This review is limited to classical light and to linear optical processes in MMFs, thereby excluding topics such as multimode nonlinear optical processes \cite{wright2022physics} and quantum optics with MMFs \cite{lib2022quantum}, as well as MMF-based lasers and amplifiers \cite{zervas2014high}. Further, we neither cover  single- or few-mode fibers, nor fiber bundles. Lastly, our selection of applications excludes the use of MMFs for telecommunications, especially spatial division multiplexing \cite{richardson2013space}, as this material is already covered by numerous other sources \cite{bottacchi_multi-gigabit_2006,senior_optical_2009,agrawal_fiber-optic_2012,lecoy_fiber-optic_2013}; moreover, many of the considerations explained here would not be scalable to the required dimensions (in terms of fiber lengths and environment stability). Multimode fiber sensors based on mode- and spatial-division have been reviewed recently \cite{caucheteur2022mode}, here we will not consider the MMF application for sensing environmental changes, instead we will focus on recovering the information of an input light from the output signals.

	\section{Principles}
	\label{sec:principle}
 
\newcommand{\vecc}[1]{\mathbf{#1}}
\newcommand{\e}{\mathrm{e}}
\newcommand{\ii}{\mathrm{i}}
\def\dd{{\rm d}}
\def\fii{\varphi}

In this section, we revisit the fundamental principles of light propagation in optical fibers. The various theoretical tools are selected here to prepare the readers for the discussions on advanced  experimental methods and emerging trends. Some important phenomena related to light transport through multimode optical fibers, including derivation of numerical aperture (NA), polarization transport and modal dispersion, can be derived even from simple principles based on ray optics, which next to its simplicity offers good level of intuitive understanding. 

Much more accurate wave-optics analysis is, however, required in order to model the exact distribution of the optical field as it propagates through a given MMF and describe the influences of imperfections of the fiber refractive index profile, subtle polarization effects, or the effects of fiber bending. 
We proceed to the wave description of light in terms of Maxwell's equations that can be applied on different levels of accuracy. The most rough description, the scalar approximation, separates the spatial and polarization degrees of freedom; within its domain of applicability, the polarization state is preserved upon propagation in the fiber, which is true only for very short fibers. However, the scalar approximation gives a good estimate of the number of modes,  their propagation constants and other properties of light in the fiber. The next step is the weak guidance approximation that takes into account the interaction of the spatial degrees of freedom with polarization; it provides a very accurate description of fibers with low numerical aperture. 

It is also important to take into account deviation from the ideal case of straight cylindrical waveguide. In realistic situations, fibers are often deformed, the most common deformation being fiber bending. Also, the refractive index profile often departs from the ideal or desired one due to imperfections in the manufacturing process. Studying these effects assists greatly when describing light transport through MMFs in real experimental conditions. Further, fiber imperfections have direct impact on the performance of imaging or other desirable applications. Knowing the influence of the imperfection on fiber performance enables the design of fibers that are more resilient to bending, or suitable for other purposes. 

\subsection{Ideal multimode fiber} 
\label{sec:light}

Here we consider optical fibers as perfectly predictable waveguides, free from scattering and invariant along their length. They are first studied as ideal cylindrically symmetric wave-guides and we introduce several common models of light transport applicable in diverse cases. Further, we will introduce how to extend these models for bent fibers, featuring axially independent perturbations in their refractive index distributions.  
Optical fibers are usually designed to have a cylindrically symmetric refractive index profile, $n(r)$, in which the index depends only on the coordinate $r$ of the cylindrical coordinate system $(r,\varphi,z)$. The index usually has a maximum value on the axis, $r=0$, and decreases with growing $r$. It is the refractive index profile that keeps light within the fiber and prevents it from escaping. In step-index (SI) fibers, $n$ has a constant value of $n_{\mathrm{co}}$ for $r$ smaller than a certain radius $R$ (this region is called the {\em core}), and another constant value $n_{\mathrm{cl}}<n_0$ for $n\ge R$ (the {\em cladding}). In graded-index (GRIN) fibers the refractive index varies smoothly with $r$. The most commonly used GRIN fibers have a parabolic index profile, 
\begin{equation}
n^2(r)=n^2_{\mathrm{co}}(1-r^2/b^2),
\label{nparabolic}
\end{equation}
where the parameter $b$ has a dimension of length and determines the index profile steepness. 


\subsubsection{Geometrical optics approach} 
\label{sec:GO}

In fibers with a core radius that is much larger than the wavelength, it is possible to think in terms of geometrical optics and to consider light {\em rays} propagating within the fiber.

\begin{figure}[htbp]\centering
\includegraphics[width=.7\textwidth]{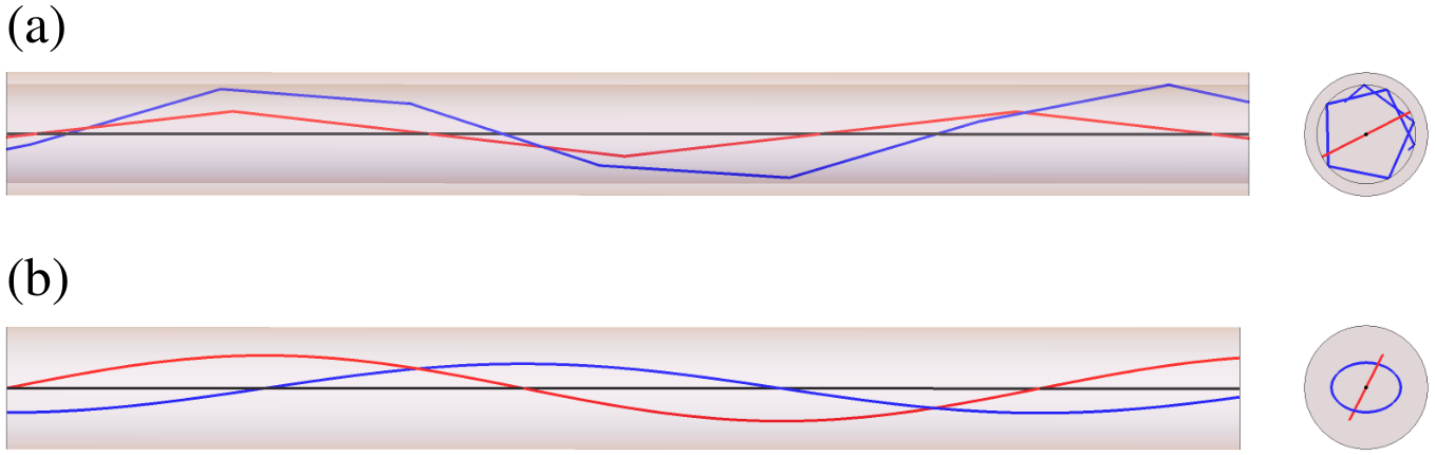}
\caption{(a) Rays in a SI fiber shown in sideview and frontview. The black ray propagates along the fiber axis. The red ray propagates in a single plane, repeatedly intersects the $z$ axis and carries \textcolor{black}{zero} orbital angular momentum. The blue ray forms a spiraling curve and carries  nonzero orbital angular momentum.
(b) The same for a parabolic GRIN fiber, where the rays form smooth trajectories.}
\label{fig:rays}
\end{figure}

In SI fibers, light rays undergo total internal reflection at the interface between the core and the cladding and orbit the fiber axis along a broken line resembling a helix composed of straight line segments, see Fig.~\ref{fig:rays}~(a). There is a maximum angle $\alpha_{\mathrm{max}}$ that the ray can make with the fiber axis to undergo the total internal reflection. A simple calculation shows that the  relation  $n_{\mathrm{co}}\sin\alpha_{\mathrm{max}}=\sqrt{n_{\mathrm{co}}^2-n_{\mathrm{cl}}^2}\equiv N\!A$ holds for this case. This value is called {\em numerical aperture} ($N\!A$) of the fiber and describes the maximum angle $\alpha'_{\mathrm{max}}$ of a ray entering the fiber from vacuum that will be guided by the fiber as $N\!A=\sin\alpha'_{\mathrm{max}}$.

Rays making larger angles with the fiber axis penetrate into the cladding where they usually get quickly attenuated since the reflection on the outer cladding boundary is not perfect. Obviously, rays making larger angles $\alpha$ with the fiber axis have a longer path and hence travel through a given fiber section longer than those with smaller $\alpha$. If a short optical pulse is launched into a SI fiber, it excites optical rays with different angles of propagation, and hence different portions of the pulse take different times to get through the fiber. As a result, the pulse gets elongated when leaving the fiber, which is a manifestation of {\em modal dispersion}, expressed in terms of geometrical optics.

Instead of total internal reflection as in SI fibers, rays in GRIN fibers are subject to bending and orbit the fiber axis along a smooth spatial curve, see Fig.~\ref{fig:rays}~(b). The radial extent of the ray motion is not the same for all rays as in SI fibers but depends on the entrance angle. This way, some rays explore large portions of the index profile while others stay near the fiber axis. The length of the rays in a given fiber segment thus varies just as in SI fibers. However, the rays with a larger length travel on average in lower refractive index regions (farther from the axis), which partially compensates the time difference between different rays. It can be shown that in fibers with parabolic profiles this compensation is almost perfect \cite{SnyderLove1983,ghatak_introduction_1998} and the time of travel through a given fiber segment is essentially the same for all rays. This way, modal dispersion is almost eliminated in such fibers and a short pulse launched into the fiber remains short at the output as well; this designates parabolic GRIN fibers for communication applications if large information flows are relevant.

\subsubsection{Wave description} 
\label{sec:wave}

In many practical cases it is desirable to model the distribution of the light field propagating through the MMF. Here, the geometrical optics can no longer be used and wave optics has to be employed instead.
For this purpose, it is appropriate to write Maxwell's equations in the cylindrically symmetric profile of relative permittivity $\varepsilon_{\mathrm r}(r)=n^2(r)$ and unit relative permeability $\mu_{\mathrm r}=1$. One then obtains six coupled equations for the components of the electric and magnetic fields. Due to the cylindrical symmetry of the index profile, it is natural to look for solutions of Maxwell's equations that separate in cylindrical coordinates. The evolution of such a wave along the fiber is given simply by acquiring a phase without changing the spatial distribution of the fields; therefore such solutions are called propagation invariant modes (PIMs) or simply ``fiber modes''. The rate at which the phase of the mode shifts along the $z$ axis is called {\em propagation constant} and denoted by $\beta$. It turns out \cite{SnyderLove1983} that for each mode, all the field components can be expressed in terms of the transverse part (i.e., the part perpendicular to the $z$ axis) $\vecc E_{\mathrm t}$  of the electric field $\vecc E$. This way, the mode is completely described by $\vecc E_{\mathrm t}$ whose spatial and temporal evolution is
\begin{equation}
 \vecc E_{\rm t}(r,\varphi,z,t)=\e^{\ii(\beta z-\omega t)} \vecc E_{\rm t}(r,\varphi)\,,
\label{}
\end{equation}
where $\omega$ is the angular frequency and $\vecc E_{\rm t}(r,\varphi)$ is the transverse field distribution at $t=0$ in the plane $z=0$. The field $\vecc E_{\mathrm t}(r,\varphi)$ is governed by the transverse equation that follows from Maxwell's equations \cite{SnyderLove1983}:
\begin{equation}
   (\nabla_{\rm t}^2+k^2n^2-\beta^2)\vecc E_{\rm t}
         =-\nabla_{\rm t}(\vecc E_{\rm t}\nabla_{\rm t}\ln n^2)\,.
\label{vectorEq}
\end{equation}
If the refractive index difference between the core and cladding is small with respect to unity, $n_{\mathrm{co}}-n_{\mathrm{cl}}\ll1$, we talk about a {\em weakly guiding fiber}.

\subsubsection{Scalar approximation} 
\label{sec:scalarapprox}

\begin{figure}[htbp]
\centering
\includegraphics[width=.8\textwidth]{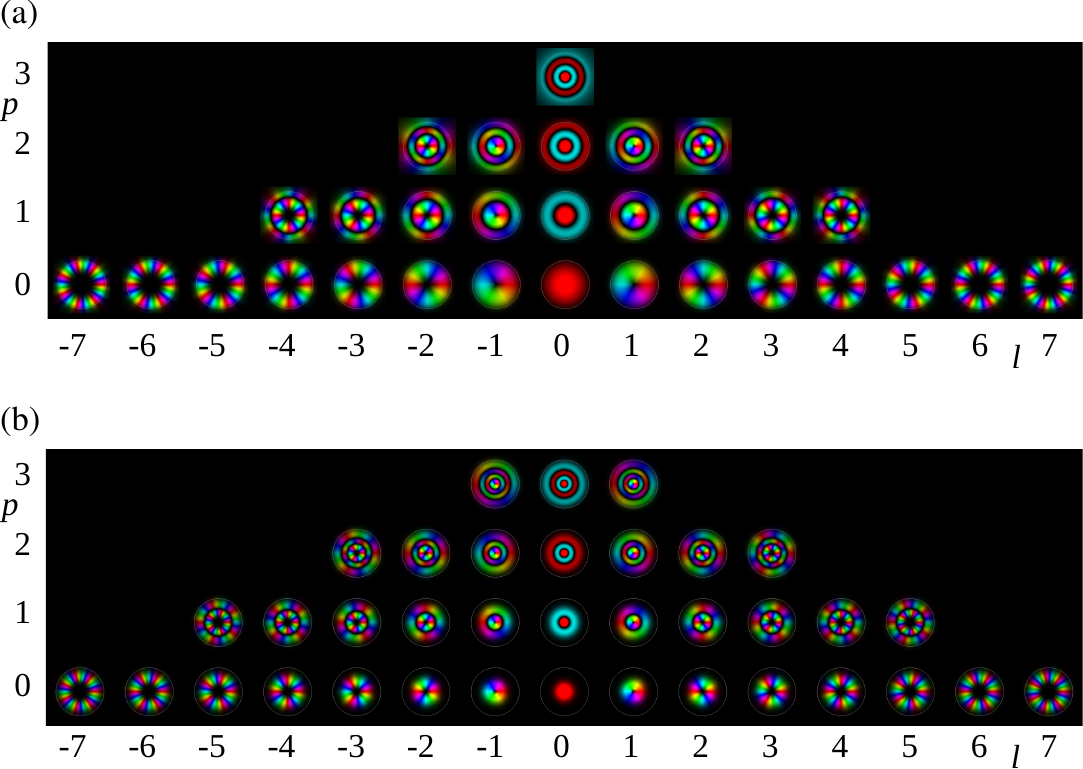}
\caption{(a) Scalar modes in a SI fiber with parameters  $N\!A=0.15$, $R=6\,\mu$m, $\lambda=532$ nm. (b) Laguerre-Gaussian scalar modes for parabolic GRIN fiber with parameters $N\!A=0.25$, $R=6\,\mu$m, $\lambda=532$ nm. The brightness encodes the wave amplitude and hue its phase. The modes are arranged according to index $l$ (horizontal axis) and $p$ (vertical axis). Note the different radial extents of the GRIN fiber modes as opposed to those for the SI fiber.}
\label{fig:modes}
\end{figure}

For a weakly guiding fiber, one could, in the roughest approximation, completely neglect the RHS of Eq.~(\ref{vectorEq}). Then  the $x$ and $y$ components of $\vecc E_{\mathrm t}$ completely decouple and each of them is governed by the scalar Helmholtz equation,
\begin{equation}
 (\nabla_{\rm t}^2+k^2n^2-\beta^2)\psi=0\,.
\label{scalarEq}
\end{equation}
Here, $\psi$ denotes the $x$ or $y$ component of the transverse field so that \begin{equation}
 \vecc E_{\mathrm t}=\hat{\vecc e}\psi\,,
\label{Efrompsi}
\end{equation} 
where $\hat{\vecc e}$ is the polarization vector (either $\hat{\vecc x}$ or $\hat{\vecc y}$). This way, the modes are linearly polarized and the propagation constant is not influenced by polarization. The scalar Helmholtz equation can further be separated in polar coordinates and yields the solutions
\begin{equation}
 \psi(r,\varphi)=\e^{\ii l\varphi}f_{lp}(r)\,.
\label{scalarmodes}
\end{equation}
Here $l=0,\pm1,\pm2,\dots$ is the {\em orbital angular momentum index} determining the phase change of the wave when encircling the $z$ axis, and $p=0,1,2,\dots$ is the {\em radial index} of the mode, respectively. The radial functions $f_{lp}(r)$ describe the radial distribution of the modes and have $p$ radial nodes for $r>0$. In SI fibers, they are given by Bessel functions and have an oscillatory character in the core while in the cladding they die out exponentially with growing $r$ (evanescent waves), see Fig.~\ref{fig:modes}~(a).  In GRIN fibers, the radius $R_{lp}$ at which the radial function changes from oscillatory to evanescent form is different for different modes, in analogy to what has been said about rays in GRIN fibers---$R_{lp}$ then describes the radial turning point of the corresponding rays. In parabolic fibers, the spatial field distribution $\e^{\ii l\varphi}f_{lp}(r)$ corresponds to Laguerre-Gaussian modes, see Fig.~\ref{fig:modes}~(b).

The fiber modes have to be normalized such that the energy flux in each of them is the same. For low-$N\!A$ fibers, this is equivalent to normalizing the mode function $\psi(r,\varphi)$ over the fiber cross-section, for larger $N\!A$ the flux normalization is essential~\cite{rotter2017light}. 

There is a degeneracy with respect to the propagation constant $\beta$ in the scalar approximation: the modes with different polarizations as well as with opposite values of $l$ have the same values of $\beta$. This way, most modes are four-fold degenerate and those with $l=0$ are two-fold degenerate. Due to this degeneracy, the selection of the modes is not unique. For example, instead of the linear polarization basis $\hat{\vecc x}, \hat{\vecc y}$ one can use in Eq.~(\ref{Efrompsi}) the circular polarization basis 
$\hat{\sigma}_\pm=(\hat{\vecc x}\pm\ii\hat{\vecc y})/\sqrt2\,$. It turns out that this basis is more suitable than the linear one when going beyond the scalar approximation, as will be explained below. 

\begin{figure}[htbp]\centering
\includegraphics[width=0.85\textwidth]{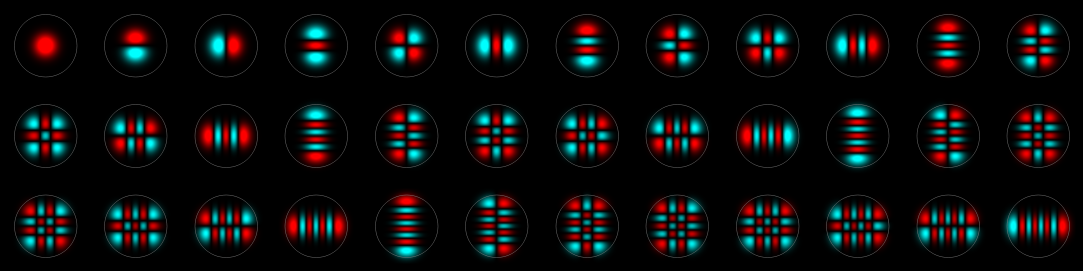}
\caption{The Hermite-Gaussian scalar modes for the GRIN fiber of Fig.~\ref{fig:modes}~(b). Due to the strong degeneracy of scalar PIMs in an ideal parabolic fiber, their choice is not unique, and alternative modes can be defined via superpositions in the subspace of the original modes corresponding to the same propagation constant.}
\label{fig:hermitemodes}
\end{figure}

In addition to this degeneracy, for certain index profiles $n(r)$ there might also be an additional degeneracy. In particular, in parabolic GRIN fibers such a degeneracy is very strong and $\beta_{lp}$ depends only on the linear combination $|l|+2p$. Due to this degeneracy, the choice of the modes is not unique, and alternative modes can be defined via superpositions in the subspace of the original modes corresponding to the same propagation constant. Figure~\ref{fig:hermitemodes} shows the Hermite-Gaussian modes, which form an alternative choice to the LG modes.

Interestingly, this situation is very similar to the energy level degeneracy in the quantum mechanical isotropic 2D harmonic oscillator; there is a close analogy between such an oscillator and the parabolic fiber, including the similarity of the governing equations (in case of the oscillator, it is the stationary Schr\"odinger equation). As is well known, the stationary Schr\"odinger equation for the oscillator can be separated both in polar coordinates, leading to Laguerre-Gaussian states, and in Cartesian coordinates, leading to Hermite-Gaussian states. This is in complete analogy to the situation in parabolic fibers.

The scalar approximation \textcolor{black}{provides a sufficient description for very short fibers. However, for fibers over a few millimeters long, polarization effects become important and} the term on the RHS of Eq.~(\ref{vectorEq}) must be taken into account. 

\subsubsection{Weak guidance approximation} 
\label{sec:WGA}

A more accurate description than the scalar approximation is provided by the weak guidance approximation (WGA)~\cite{SnyderLove1983}. It takes into account the mutual influence of light polarization and the spatial distribution of the fields, the so-called spin-orbit (SO) interaction~\cite{bliokh2015}. This term is used also in quantum mechanics where it describes interaction of the electron spin with its spatial wavefunction; in optics the situation is analogous, but its description is different. The origin of the optical SO interaction lies in the fundamental vector character of Maxwell's equations~\cite{bliokh2015}, and it is closely related to the concept of the geometric phase~\cite{bhandari1997}.

The approach is based on the expansion of an unknown modal field $\vecc E_{\rm t}(r,\varphi)$ into the modes given by Eqs.~(\ref{Efrompsi}) and~(\ref{scalarmodes}) with unknown coefficients and then using the  perturbation \textcolor{black}{calculation} to find these coefficients and the corresponding propagation constant. The perturbation terms describe the SO interaction in the fiber. By this procedure one finds the approximate vector modes and their propagation constants.
 
The perturbation partially removes the degeneracy present in the scalar approximation, which reflects a typical situation in perturbation theory. Most transverse modal fields are now approximately described by Eq.~(\ref{Efrompsi}) with $\hat{\vecc e}\in\{\hat{\sigma}_-,\hat{\sigma}_+\}$. We can denote these modes by $|l,\sigma\rangle_p$ according to their index $l$ and circular polarization index $\sigma=\pm1$, where $\sigma=+1$ and $\sigma=-1$ corresponds to polarization state $\hat\sigma_+$ and $\hat\sigma_-$, respectively, as defined above. These modes have circular polarization, and modes with the same value of $l$ and opposite circular polarizations have slightly different propagation constants---a manifestation of the SO interaction. Changing the sign of $l$ and $\sigma$ simultaneously does not change the propagation constant---a manifestation of the mirror symmetry of the fiber. 

Even though most of the circularly polarized states $|l,\sigma\rangle_p$ are PIMs of the fiber, there is an exception: the states $|1,-1\rangle_p$ and  $|-1,1\rangle_p$ are not propagation invariant; the PIMs are actually their equal superpositions:
\begin{equation}
  |\mathrm H\rangle_p=\frac{|1,-1\rangle_p+|-1,1\rangle_p}{\sqrt2},\quad
  |\mathrm B\rangle_p=\frac{|1,-1\rangle_p-|-1,1\rangle_p}{\sqrt2}\,.
\label{HB}
\end{equation}
The state $|\mathrm H\rangle_p$ is sometimes called ``hedgehog state'' because the vectors $\vecc E_{\rm t}(r,\varphi)$ are directed radially, resembling pins of a hedgehog. The state $|\mathrm B\rangle_p$ is sometimes called ``bagel state'' and the vectors $\vecc E_{\rm t}(r,\varphi)$ are directed in the angular direction. The bagel modes are specific by having zero electric field component $E_z$ along the $z$ axis, so they are transversely electric modes. Similarly, hedgehog modes are transversely magnetic; for all other modes the components $E_z$  and  $B_z$ are both nonzero (but still small in weakly guiding fibers). 

In general, the modes $|\mathrm H\rangle_p$ and $|\mathrm B\rangle_p$ are the only nondegenerate ones. All other modes form pairs $|l,1\rangle_p$ and $|-l,-1\rangle_p$ with the same propagation constant. Therefore, here again the choice of the modes is not unique, and often superpositions $(|l,1\rangle_p\pm|-l,-1\rangle_p)/\sqrt2$ are preferred in the literature~\cite{SnyderLove1983}; these modes are called even and odd HE and EH modes. 

Propagation constants in the WGA are very close to the scalar propagation constants to which there are small corrections, depending on the mode. These corrections follow from the perturbation theory described above and can be expressed as integrals over the plane $(r,\varphi)$ of the derivative $\mathrm{d}n/\mathrm{d}r$ with the corresponding modal fields~\cite{SnyderLove1983}.

The SO interaction has an interesting consequence: consider an $x$-linearly polarized wave with orbital angular momentum index $l>1$ and some radial index $p$ injected into the fiber at $z=0$. Such a wave corresponds to the superposition $(|l,-1\rangle_p+|l,1\rangle_p)/{\sqrt2}$ of the fiber modes. As it propagates along the fiber, the two modes pick up slightly different phases due to the difference in their propagation constants, so the state stays linearly polarized, but the polarization vector slowly rotates in the direction of the angular momentum. This way, the orbital angular momentum ``drags'' the polarization direction. \textcolor{black}{In a typical weakly-guiding SI fiber with $N\!A=0.22$ and $R=25\,\mu$m, the full $2\pi$ polarization rotation occurs on distances starting at about 5 cm (depending on the mode indexes $l$ and $p$), so the SO interaction must be taken into account for fibers longer than a few millimeters.}

\begin{figure}[htbp]\centering
\includegraphics[width=0.9\textwidth]{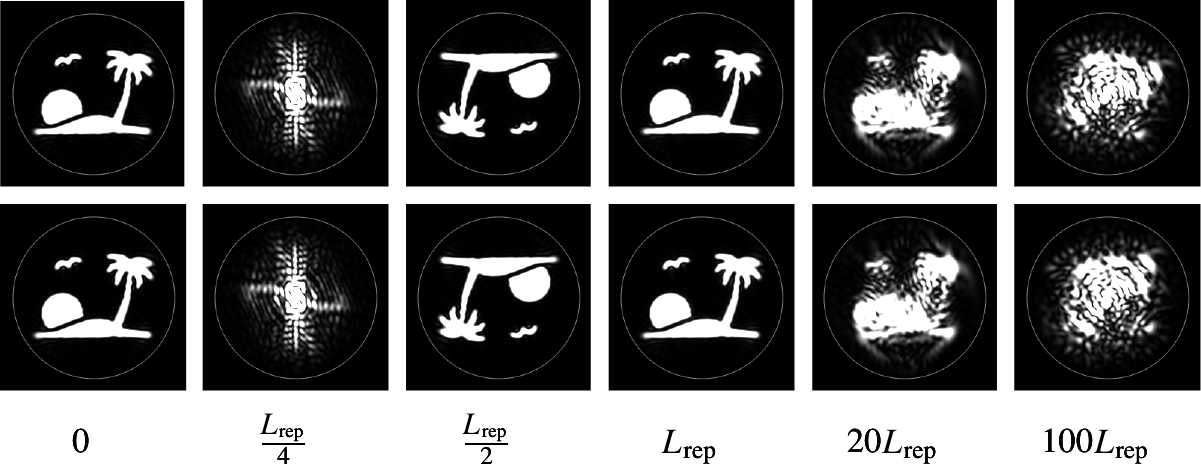}
\caption{\textcolor{black}{Simulation} of the evolution of the left (first row) and right (second row) circularly polarized state along the parabolic GRIN fiber with parameters $N\!A=0.3$ and $R=25\,\mu$m. The distances $z$ along the fiber are shown at the bottom. The input state at $z=0$ is a plane wave modulated by a binary mask in the shape of a tropical island scenery (courtesy: Une Butaite and David Phillips). At the distance $z=L_{\mathrm{rep}}/4$ one gets essentially the Fourier transform of the input image. At $z=L_{\mathrm{rep}}/2$ one gets a $\pi$ rotated input image, and at  $z=L_{\mathrm{rep}}$ the input image is reproduced. This process repeats many times along the fiber, but the image gradually gets degraded, as is clear for $z=20L_{\mathrm{rep}}$ and $100L_{\mathrm{rep}}$. Comparing the two polarizations at $z=100L_{\mathrm{rep}}$ , one sees that they are clearly rotated with respect to each other due to the spin-orbit (SO) interaction. The revival distance $L_{\mathrm{rep}}$ is 777.5 $\mu$m and the full rotation distance $L_{2\pi}$ is 3.48 meters.}
\label{fig:grin-revivals}
\end{figure}

Conversely, one can instead launch into the fiber the superposition $(|l,1\rangle_p+|-l,1\rangle_p)/{\sqrt2}$, which is 
a standing-wave pattern with right circular polarization. This time, it is the spatial pattern that slowly rotates, being dragged by the circular polarization. The distance $L_{2\pi}$ at which a full $2\pi$ rotation occurs depends in general on both indexes $l$ and $p$. Remarkably, in parabolic GRIN fibers this dependence vanishes \cite{flaes2018}, which results in the fact that changing the polarization of the input field (at $z=0$), e.g., from left to right results in a rotation of the whole output field (at $z=L$) by an angle $4\pi L/L_{2\pi}$. This effect can be seen in the last column of Fig.~\ref{fig:grin-revivals}.


\begin{figure}[htbp]\centering
\includegraphics[width = 0.3\textwidth]{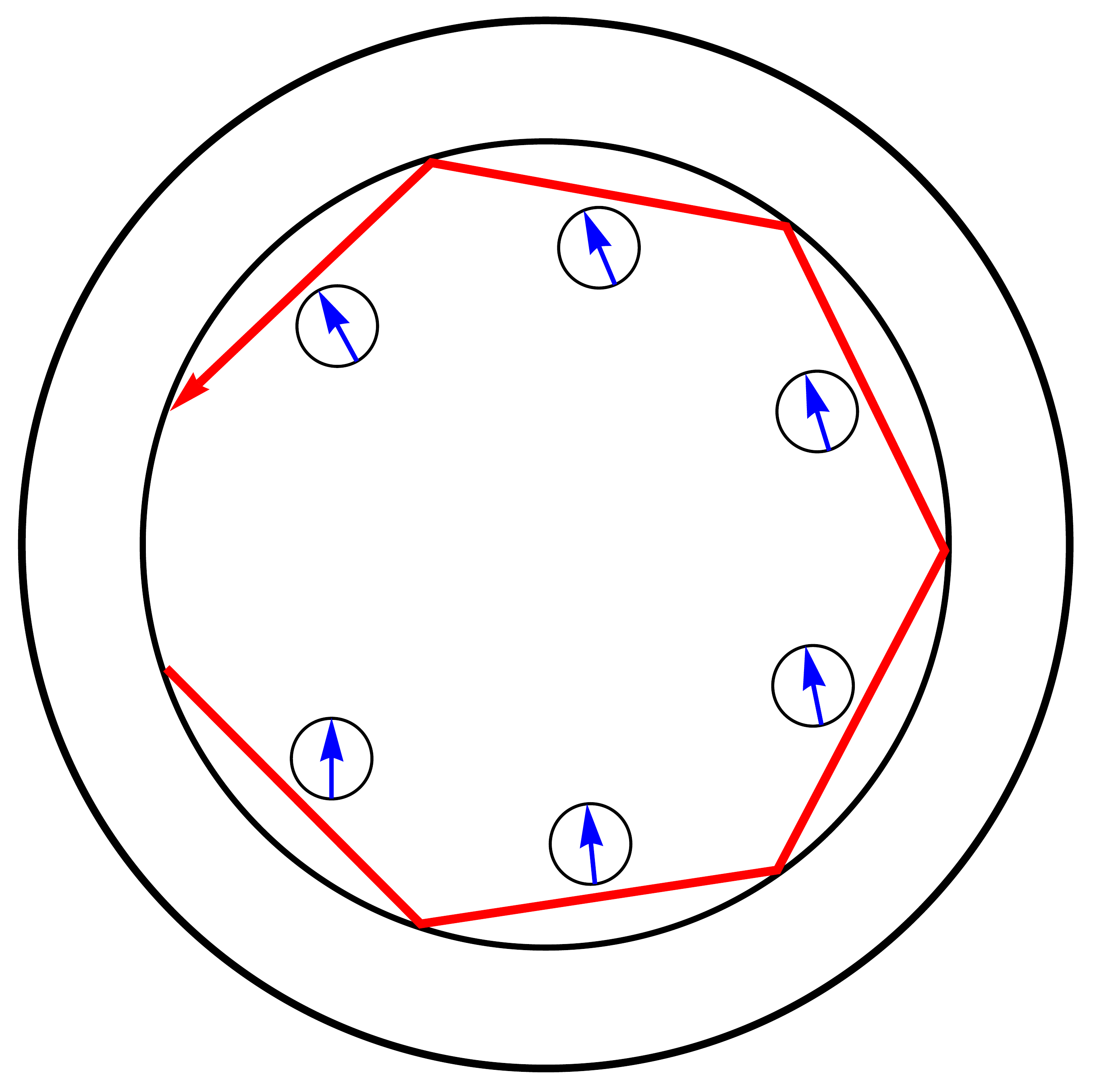}
\caption{\textcolor{black}{Illustration of the drag of the linear polarization by the orbital angular momentum. As the ray (shown in red color in the front view) carrying orbital angular momentum propagates in a weakly guiding SI fiber, its linear polarization vector (shown by the blue arrows) is parallel transported along the ray with a good accuracy. This results in a slow polarization rotation. The rotation is greatly exaggerated here to be visible.} }
\label{fig:polarization_rotation}
\end{figure}

Interestingly, the drag of the linear polarization by the orbital angular momentum can be described simply within geometrical optics. As the ray moves through the fiber, its tangent vector changes and describes a kind of a pyramid (in an SI fiber) or a cone (in a GRIN fiber). The linear polarization direction is then parallel-transported along the ray, which results in its rotation with respect to a fixed direction, \textcolor{black}{as is illustrated in Fig. \ref{fig:polarization_rotation}}; the angle of rotation in a given fiber segment is numerically equal to the total solid angle enclosed by the direction vector during motion in that segment. The rotation angle calculated in this way agrees quite well with the more accurate calculation using WGA. Alternatively, the polarization rotation can be interpreted in terms of the geometric phase~\cite{Berryphase1984,bliokh2015}. This phase along a given ray is different for left and right circularly polarized states, and it is proportional to the solid angle enclosed by the ray direction vector. The result is again the polarization rotation.

\subsubsection{Self-imaging in parabolic fibers} 
\label{sec:ref-parabolic}

An interesting and useful effect~\cite{grin_lens2008,grin_lens2010} occurs in parabolic fibers with a refractive index as in Eq.~(\ref{nparabolic}). The scalar propagation constants in this case are given by~\cite{flaes2018}
\begin{eqnarray}
\beta_{lp}&=&\sqrt{k^2n_{\mathrm{co}}^2-\frac{2kn_{\mathrm{co}}(|l|+2p+1)}{b}} \nonumber\\
            &=& kn_{\mathrm{co}}-\frac{|l|+2p+1}b-\frac{(|l|+2p+1)^2}{2kn_{\mathrm{co}}b^2}-\dots\,,
\label{eq:betasgrin}
\end{eqnarray} 
where the square root was expanded by binomial expansion. If the numerical aperture is not large, the first two terms of the expansion provide a good approximation to propagation constants. Consider for this case the phase change of the modes on the distance $\Delta z=L_{0}\equiv2\pi b$. The first term $kn_{\mathrm{co}}$ in Eq.~(\ref{eq:betasgrin}) then gives the same phase for all modes, which is just a global phase; the phase $2\pi(|l|+2p+1)$ coming from the second term is an integer multiple of $2\pi$, so it cancels. This way, the wave pattern repeats in the parabolic fiber after the distance $L_0$. However, due to dephasing caused by the third and higher-order terms in Eq.~(\ref{eq:betasgrin}), the pattern revivals gradually degrade, as illustrated in Fig.~\ref{fig:grin-revivals}. It can be shown that the effect of the third term in Eq.~(\ref{eq:betasgrin}) is partially eliminated if the distance $L_0$ is replaced by $L_{\mathrm{rep}}=2\pi b-\pi R N\!A/(2kn_{\mathrm{co}})$. This can be regarded as a more accurate expression for the revival distance.

A closer inspection of the modes' parity reveals that at the distance $L_{\mathrm{rep}}/2$ the pattern is repeated too, just rotated by $\pi$, and at the distance $L_{\mathrm{rep}}/4$ one gets the Fourier transform of the input image, so the segment of the fiber of this length works as a lens.
 
In addition to all this, the SO interaction causes rotation of the repeated pattern in the negative or positive direction for left or right circular polarizations, respectively, as explained above and illustrated in Fig.~\ref{fig:grin-revivals}.

\subsection{Imperfection and perturbation}
\label{sec:imperfection}

In reality, all fibers have inherent imperfections such as refractive index variations and distortions of the fiber cross-section along the fiber. Also external perturbations due to fiber bending or twisting and ambient temperature inhomogeneity, strains, etc. affect light transport in an MMF. We will discuss the influence of these effects here below.
 
\subsubsection{Fiber bending} 
\label{sec:bending}

Optical fibers are rarely completely straight, and fiber bending can influence light propagation significantly. An important effect often discussed in the literature is energy loss because the light from some of the guided modes can tunnel to the cladding on the outer side of the bend. In addition to the mode-dependent loss (MDL), fiber bending causes mode mixing and phase shifting in bent fibers. Understanding in detail how the bending influences the transmission matrix is important in applications where bending occurs and where it changes in time, especially because it is not always possible to measure the TM repeatedly on short time scales.

For instance, consider a SI fiber where a specific speckle pattern is sent onto the proximal end of a straight fiber so that a focused diffraction-limited spot is obtained on the distal end; this is a typical situation in fluorescent imaging through the fiber. Now suppose that the fiber gets bent. Due to the changes in the TM, this results in a complete degradation of the focused spot at the distal end. In order to recover it, one would need to take into account these changes and modify the input pattern accordingly.  
The modification can be found theoretically, based on the known fiber shape and the knowledge of light propagation in the bent fiber~\cite{ploschner2015}. However, the most advantageous way is to use a bending resilient fiber where the transmission matrix almost does not change with fiber deformation. This is possible with near-perfect parabolic fibers but not with SI fibers. Therefore, the knowledge of bending effects on the TM is very important. 

Let us discuss first light propagation in bent fibers within the scalar theory, and then mention how the more accurate description within WGA works. Let the fiber be bent in the positive $x$ direction and have curvature $\rho$. This means that the center of curvature lies in the positive $x$ direction at the distance $1/\rho$ from the fiber axis. In a straight fiber, the phase in PIMs changes uniformly along the fiber, and the wavefronts are perpendicular to the fiber axis (when not considering the helicity of wavefronts caused by the orbital angular momentum). In the PIMs of bent fibers, it is natural to require the same property; however, due to bending the separation of the wavefronts in the $z$ direction now changes with $x$ as $d(x)=(1-\rho x)d_0$, where $d_0=2\pi/\beta$ is the wavefront separation on the fiber axis, as can be seen from a simple geometric consideration. This way, the ``local propagation constant'' now depends on $x$ as $\beta(x)=2\pi/d(x)=\beta/(1-\rho x)$.

To put this into the mathematical form, let the PIM of the bent fiber with a given polarization state be described by the wavefunction $\psi(x,y)$ according to Eq.~(\ref{Efrompsi}). To account for the local propagation constant dependence on $x$, we replace $\beta$ in Eq.~(\ref{scalarEq}) by $\beta(x)$, which yields an equation for $\psi$ 
\begin{equation}
   \left[\nabla_{\rm t}^2+k^2n^2-\frac{\beta'^2}{(1-\rho x)^2}\right]\psi=0\,.
\label{scalarbent}
\end{equation}
Here we are using $\beta'$ for the propagation constant on the bent fiber axis to distinguish it from the unprimed propagation constants for a straight fiber.

The most straightforward way of solving Eq.~(\ref{scalarbent}) is to multiply it with the expression $(1-\rho x)^2$ and plug into it the unknown bent fiber mode $\psi$ expanded into the straight fiber PIMs $\psi_j$ as $\psi=\sum_{j=1}^n c_j\psi_j$. The subsequent calculation is relatively straightforward~\cite{Stopka2017} and enables us to reduce the problem of finding the modes of the bent fiber and their propagation constants to the eigenvalue problem for the matrix 
\begin{equation}
   B'=(1-\rho X)B\,,
\label{Bprime}
\end{equation}
where $B$ is the diagonal matrix of propagation constants of the straight fiber, i.e., the matrix with entries $B_{ij}=\delta_{ij}\beta_i$, and the matrix $X$ is defined by
\begin{equation}
  X_{ij}\equiv\int_{\mathbb R^2}x
  \psi^*_i(x,y)\psi_j(x,y)\,\mathrm{d}x\,\mathrm{d}y \,.
\label{Xmatrix}
\end{equation}
The eigenvectors of $B'$ then express vectors of the superposition coefficients $c_i$ corresponding to the PIMs of the bent fiber, and the propagation constants are given by the corresponding eigenvalues of $B'$. 

This way, the matrix $B'$ represents the modes of the bent fiber just as the matrix $B$ represents the modes of the straight fiber. Moreover, the state evolution in the bent fiber takes a very simple form in terms of the matrix $B'$: in the fiber segment of length $L$ and a constant curvature $\rho$, the transmission matrix in the basis $\psi_i$ is simply $T(\rho,L)=\exp(\ii B'L)=\exp[\ii(1-\rho X) BL]$. 


It can be useful to be able to estimate how fiber bending  influences the form of the modes and their propagation constants  based on some simple insights. This will enable us to estimate what index profiles will give more bending-resilient propagation etc. Insights that will be relevant in this regard proceed along the following direction. For small curvatures, the term $\rho x$ is very small compared to unity, which allows us to perform in Eq.~(\ref{scalarbent}) the approximations $\beta'^2/(1-\rho x)^2\approx\beta'^2(1+2\rho x)\approx \beta'^2+2k^2n_{\mathrm{co}}^2\rho x$, where in the second approximation we replaced $\beta'$ by $kn_{\mathrm{co}}$; in weakly guiding fibers, propagation constants of all modes are very close to this value. In this way, Eq.~(\ref{scalarbent}) can be rewritten as follows
\begin{equation}
 \left[\nabla_{\rm t}^2+k^2
    (n^2-2n_{\mathrm{co}}^2\rho x)-\beta'^2\right]\psi=0 \,.
\label{scalarbent2}
\end{equation}
This equation is formally equivalent to the equation for straight fibers with a modified refractive index $n'^2=n^2-2n_{\mathrm{co}}^2\rho x$. The index modification, $\Delta n^2\equiv n'^2-n^2=-2n_{\mathrm{co}}^2\rho x$, has the character of a uniform slope in the $x$ direction, increasing the index in the $x<0$ region (the outer side of the bend) and  decreasing it in the $x>0$ region. As larger index values generally tend to attract the light, at least for lower modes, we see that the light will be ``pulled'' to the outer side of the bend. This could be interpreted as an effect of the ``centrifugal force'' acting on the light in the bend. \textcolor{black}{Equivalent results can be obtained by the method of conformal transformation of the refractive index profile \cite{fiber_conformal}.}

\begin{figure}[htbp]\centering
\includegraphics[width=0.6\textwidth]{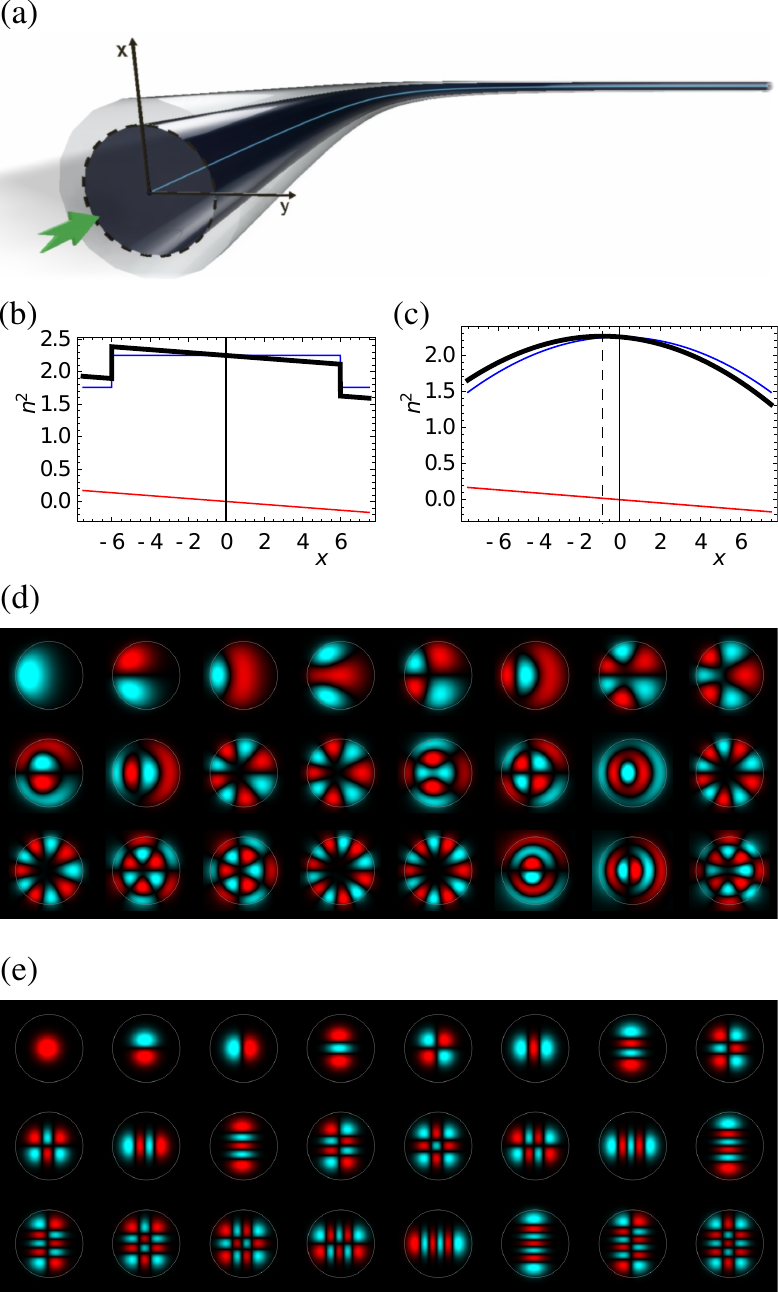}
\caption{Bending of the fiber (a) is approximately equivalent to adding a constant slope to the refractive index. The plots (b-c) show the original square index $n^2$ (blue), the added slope $\Delta n^2$ (red) and the resulting effective index $n^2+\Delta n^2$ (thick black) for (b) SI fiber and (c) parabolic GRIN fiber. In (c), the position of the effective index maximum is shown by the dashed line. The parameters are chosen rather unrealistic so that the effects can be seen in the plots: for both fibers, $\rho=(200\,\mu\mathrm m)^{-1}$, $N\!A=0.7$, $R=6\,\mu$m.
(d-e) Scalar modes in bent fibers with curvature $\rho=(5\mbox{ mm})^{-1}$, for (d) bent SI fiber with the same parameters as in Fig.~\ref{fig:modes}~(a), and for (e) bent parabolic GRIN as in Fig.~\ref{fig:modes}~(b). Only the first 24 modes are shown for each fiber as some of the highest modes are no more bound due to the bending. The effect of bending on the SI fiber modes is dramatic while the GRIN fiber modes are just slightly shifted in the negative $x$ direction by $\rho b^2=0.26\,\mu$m. Note that for the GRIN fiber, the shifted Hermite-Gaussian rather than Laguerre-Gaussian modes are obtained, the result of breaking the original rotational symmetry of the index profile by bending.}
\label{fig:bending_index}
\end{figure}

The analogy of fiber bending to modification of the refractive index also helps to understand the dramatic difference between the behavior of modes in SI and GRIN fibers upon fiber bending. Consider first a SI fiber: its index profile, $n^2$, along the $x$ axis is a partwise constant function, see Fig.~\ref{fig:bending_index}~(a), blue curve. Adding a constant slope $\Delta n^2$ (red curve) changes this profile significantly, creating the maximum index region at the edge of the fiber core, see Fig.~\ref{fig:bending_index}~(a), black curve. This is the reason why the lowest modes are localized in that region, as Fig.~\ref{fig:bending_index}~(c) shows.

Next, consider a parabolic GRIN fiber: its index profile, $n^2$, along the $x$ axis is parabolic, see Fig.~\ref{fig:bending_index}~(b), blue curve. Adding a constant slope $\Delta n^2$ (red curve) has a much less significant effect on the index, simply shifting the parabola slightly in the negative $x$ direction by the distance $\rho b^2$ [where $b$ is the parameter of the index profile from Eq.~(\ref{nparabolic})] and simultaneously lifting it slightly up (see Fig.~\ref{fig:bending_index}~(b), black curve). The effect of bending on an ideal parabolic fiber is therefore much less dramatic than for SI fibers: the modes are simply shifted to the outer side of the bend by a constant value and their propagation constants are shifted by a constant value, as is demonstrated in  Fig.~\ref{fig:bending_index}~(d). This makes parabolic fibers  substantially more resilient to bending than SI fibers. 

The effect of bending can, of course, also be described within the weak guidance approximation. The idea is the same as has been described in Sec.~\ref{sec:WGA}: expand the unknown bent fiber mode in the mode basis desribed by Eqs.~(\ref{Efrompsi}) and~(\ref{scalarmodes}) with unknown coefficients $c_j$, and substitute this into Eq.~(\ref{vectorEq}) with $\beta$ replaced by $\beta'/(1-\rho x)$. After some manipulation, one gets a matrix equation for the coefficients $c_j$, so the solution again reduces to an eigenvalue problem.

The influence of fiber bending on the transmission matrix can be expressed in terms of the {\em deformation matrix}. It is defined as the matrix product~\cite{ploschner2015}
\begin{equation}
  D=T_{\mathrm{bent}}T^{-1}_{\mathrm{straight}}\,,
\label{deformoperator}\end{equation}
where $T_{\mathrm{straight}}$ and $T_{\mathrm{bent}}$ is the TM of the same fiber in the  straight and bent layout, respectively. The physical meaning behind the deformation matrix is simple: it expresses the relation between the output state of the bent fiber and of the straight fiber if the same input state is used. For bending resilient fibers, $D$ is close to the unit matrix (up to a possible global phase factor) while, e.g., for SI fibers, it differs from the unity matrix significantly even for slight bending. 

Other definitions of the deformation matrix have also been used in literature when, for instance, one subtracts the identity matrix from the product~(\ref{deformoperator}); this way, one obtains zero deformation matrix rather than unity matrix in case of no deformation \cite{matthes_learning_2021}.


\subsubsection{Adiabaticity of bending} 
\label{sec:adiabatic}

If an optical fiber is bent, usually the curvature is not constant (that case  would correspond to a coiled fiber) but varies along its length. To find the TM theoretically in this case, one can follow a simple strategy: divide the fiber into a large number $N$ of segments, the curvature in each of which can be regarded as constant. Let us assume for simplicity that the bending occurs in the $x$ direction, so the fiber lies in one plane. The TM of the $i$th segment in the basis of straight fiber modes is then $T(L_i,\rho_i)=\exp[\ii(1-\rho_i X) BL_i]$ as explained in Sec.~\ref{sec:bending}, and the total TM is then simply the matrix product $T_{\mathrm{bent}}=T(L_N,\rho_N)T(L_{N-1},\rho_{N-1})\cdots T(L_1,\rho_1)$. The corresponding deformation matrix is $D=T_{\mathrm{bent}}T^{-1}(L,0)$.

\begin{figure}[htbp]\centering
\includegraphics[width=0.75\textwidth]{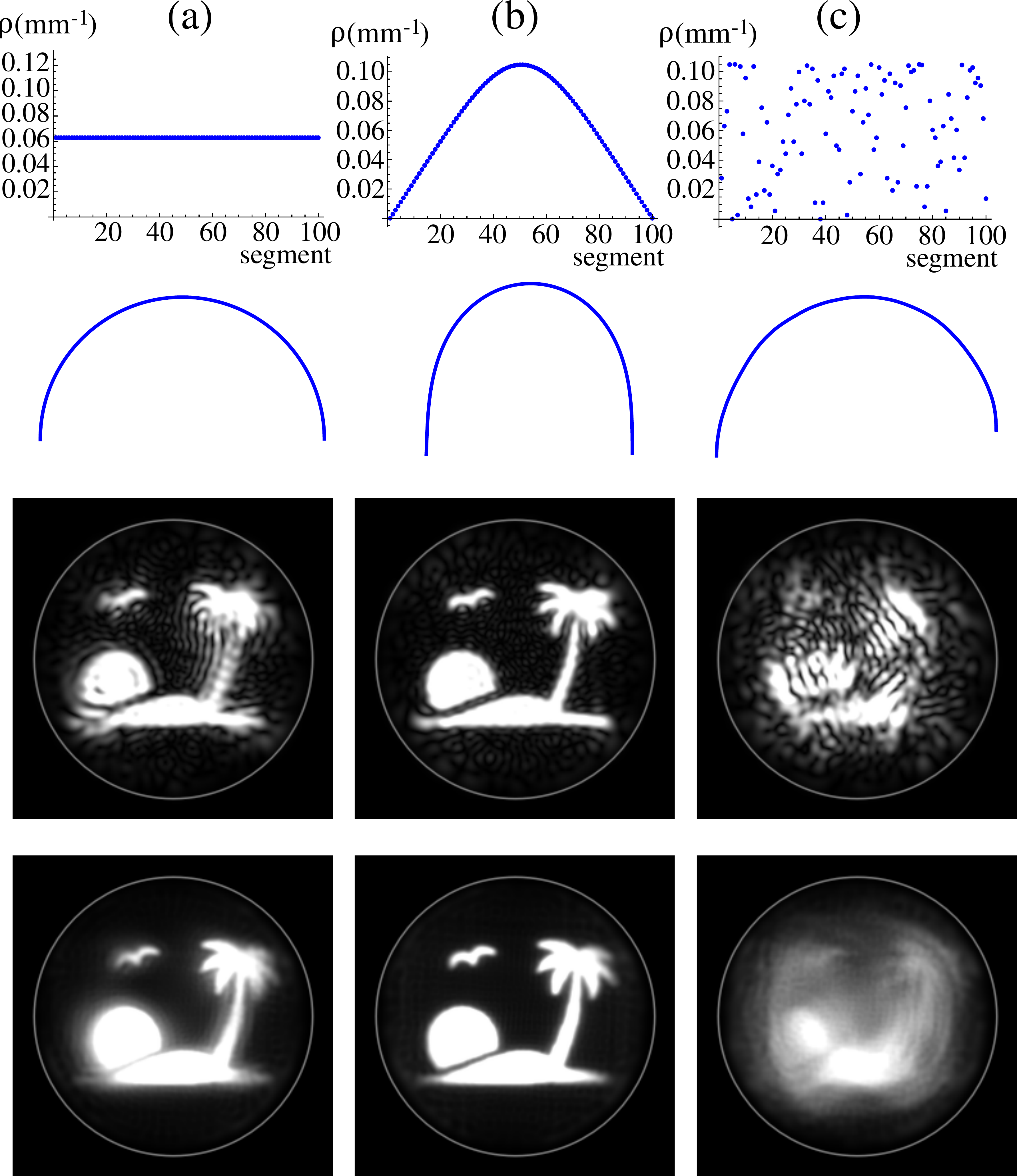}
\caption{Simulation of the influence of the type of fiber bending on the transmitted image. The top row shows curvatures of 100 fiber segments of equal length for (a) bending of constant curvature, (b) adiabatic bending and (c) random bending. The second row shows the corresponding fiber shapes. The third row shows the intensity pattern at the output of a bent fiber, with the input speckle pattern that would yield the desired scenery (with a flat phase) at the output of a straight fiber. The fourth row shows the simulated fluorescence imaging if the straight fiber TM is used for generating focused spots at the output. In (c), the same curvatues of 100 segments as in (b) were used, but with random sequence, so adjacent segments have very different curvatures. The fiber parameters were as in Fig.~\ref{fig:grin-revivals} and the fiber length $L=50$ mm.}
\label{fig:grin-resilience}
\end{figure}

It turns out that it is not only the values of the curvatures in the set $\{\rho_i,i=1,\dots,N\}$ that influence the resulting deformation matrix $D$, but also their order. In particular, if curvatures of adjacent fiber segments differ only slightly, the deformation matrix will be much closer to the unity matrix than if they differ significantly. The reason is that if the curvature changes smoothly along the fiber, light can adapt adiabatically to the new conditions, while abrupt curvature changes cause abrupt mode changes, which will degrade the DM strongly. This is illustrated in Fig.~\ref{fig:grin-resilience} where in each case, the specific input speckle pattern is launched into the fiber that would make the desired scenery (as of Fig.~\ref{fig:grin-revivals}, the first column) at the output facet of the fiber if it were straight. Panel (a) in Fig.~\ref{fig:grin-resilience} corresponds to a fiber bent to a constant curvature; since the modes in this case do not quite match the modes of a straight fiber, the image is not perfect. Panel (b) corresponds to smooth fiber bending, starting and ending with zero curvature. Since the bending is adiabatic, the image has a much better quality. In panel (c), the curvatures were taken from the case of the fiber in (b) divided into 100 segments of equal length, but randomly reordered. The distorsion is now very strong since the bending is far from being adiabatic, even though the curvatures had the same values as in (b).

This way, if one needs a high bending resilience in a specific experiment, it is not only important to choose a bending-resilient fiber, but also to make sure that the curvature changes smoothly. In particular, concentrated forces acting on the fiber as well as torques acting on its ends should be avoided. This is also the case of Fig.~\ref{fig:grin-resilience}(b) where the forces deforming the fiber act only at the fiber ends (no torques are applied there).

\subsubsection{Refractive index perturbations} 
\label{sec:perturbations}

Most optical fibers have a rotationally symmetric refractive index profile. Finding the PIMs and the TM in this case is quite straightforward because the Helmholtz equation~(\ref{scalarEq}) can be separated in polar coordinates, and WGA can subsequently be applied as explained in Sec.~\ref{sec:WGA}. However, often there are imperfections in the refractive index profile that do not have cylindrical symmetry, and it is desirable to be able to describe the PIMs of the fiber that has such imperfections. An example is a transverse index perturbation that does not change along the fiber. As the index profile can be measured by tomographic methods~\cite{yablon2009multi,yablon2013multifocus}, one can then theoretically calculate PIMs and the TM of such a fiber with high accuracy, which can be very useful in interpreting results of experiments. In the following we briefly describe how to find scalar modes of a fiber with index perturbations; a generalization of this method to WGA is relatively straightforward. 

We start with the scalar Helmholtz equation~(\ref{scalarEq}) where we express the squared refractive index as an ideal profile plus perturbation, $ n^2=n^2_{\mathrm{id}}+\Delta n^2$, which yields the equation
\begin{equation}
[\nabla_{\rm t}^2+k^2(n^2_{\mathrm{id}}+\Delta n^2)-\beta'^2]\psi=0\,.
\label{nperturb}
\end{equation} 
Expressing, as usual, a perturbed mode as a superposition of unperturbed modes, $\psi=\sum_j  c_j \psi_j$ with unknown coefficients $c_j$, substituting into Eq.~(\ref{nperturb}), multiplying with $\psi^*_i$ and integrating over the $xy$ plane, we get a matrix equation 
\begin{equation}
 \sum_j (\beta_i^2\delta_{ij}+k^2\Delta N_{ij}) c_j=\beta'^2 c_i\,,
\label{perturbation-matrix}
\end{equation}
where $\Delta N_{ij}=\int_{\mathbb R^2}\Delta n^2(x,y)\psi^*_i(x,y)
 \psi_j(x,y)\,\mathrm{d}x\,\mathrm{d}y$
are the matrix elements of the index perturbation. This way, we again obtain an eigenvalue problem for the matrix 
$B'^2=B^2+k^2\Delta N$, where $B^2$ is the diagonal matrix with squares of unperturbed propagation constants. 

\begin{figure}[htbp]\centering
\includegraphics[width=0.62\textwidth]{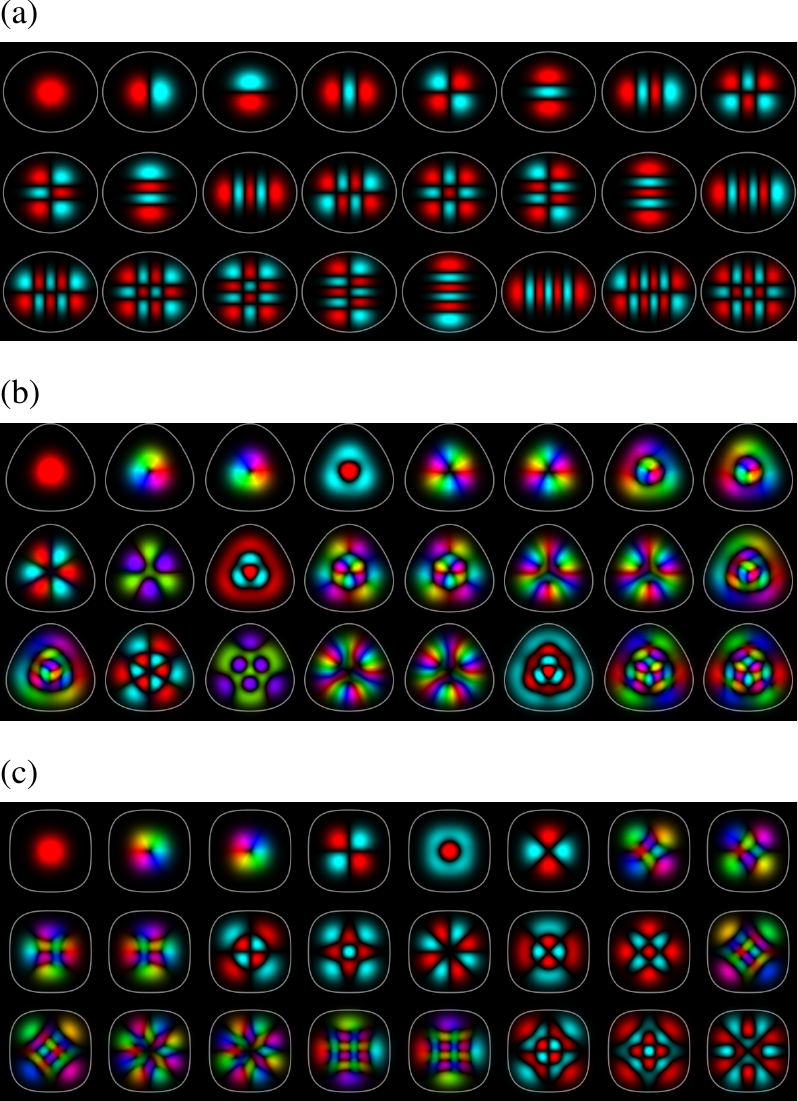}
\caption{The first 24 PIMs in a parabolic GRIN fiber ($N\!A=0.25$, $R=6\,\mu$m, $\lambda=532$ nm) with refractive index perturbations: (a) Elliptic perturbation of squared refractive index, $\Delta n^2(r,\varphi)=-0.01(r/R)^2\cos[2(\varphi+\pi/2)]$, (b) triangular perturbation $\Delta n^2(r,\varphi)=-0.01(r/R)^2\cos[3(\varphi+\pi/2)]$, and (c) four-fold perturbation $\Delta n^2(r,\varphi)=-0.0075(r/R)^2\cos[4(\varphi+\pi/2)]$. The gray curves show the contour of refractive index with the value corresponding to radius $R$ in the unperturbed fiber.}
\label{fig:deformedmodes}
\end{figure}

Fig.~\ref{fig:deformedmodes} shows the PIMs calculated with this method for an elliptic, triangular and square perturbation, respectively. We see that the symmetry of the modes reflects clearly the symmetry of the perturbation. 

Similarly as in the case of bending, one can treat the index perturbations also within WGA (a generalization of the method is straightforward). What is most demanding in numerical calculations is to express the matrix elements of the right-hand side of Eq.~(\ref{vectorEq}).

	\subsection{Mode coupling} 
	\label{sec:coupling}
	
As summarized in the previous section, very elaborate models are available for light transport through straight or bent optical fibers, including a wide spectrum of aberrations. The models already predict very complex behavior including coupling between polarization states and spectral decorrelations.  
Light transport through real optical fibers is, moreover, affected by  non-circularity of the core, roughness at the core-cladding boundary, variations in the core radius or the index profile, which are experimentally very hard or even impossible to determine. For certain applications, perturbations are even intentionally induced e.g. by local mechanical stress, microbends or material selection, to cause the fiber modes (derived from models of unperturbed fibers) to couple each other. Consequently,  energy transfers from one fiber mode to another, as light propagates in the fiber. Often such coupling is neither predictable nor traceable, and thus referred to as {\it random} mode coupling.

Ignoring these perturbations in the previous models therefore makes the predictions diverge from the experimental reality, with the deviation increasing with the length of the fiber.
For each fiber, there is a length limit beyond which the exact prediction of the light transport becomes impossible. Exceeding the limit, the coupling of optical power between modes can only be treated statistically. 

\subsubsection{Spatial- and polarization-mode coupling}

Traditionally, modal dispersion and coupling in MMFs have been described using power-coupling models \cite{kutz1998mode}. Such models are effective in describing the modal power distribution as a function of time and fiber length, and also provide a good understanding of signal distortion, pulse broadening as a function of fiber length, and fiber loss. Such models \textcolor{black}{fail} to consider phase effects, making them generally appropriate only for incoherent light sources. For coherent light sources, field-coupling models are needed instead to describe phase-dependent coupling between complex-valued electric field amplitudes.

On the numerical level, bending of a fiber has been used to induce spatial-mode coupling and birefringence \cite{shemirani2009principal}. Concatenated multiple sections with differently oriented bends cause polarization-mode coupling. Because spatial-mode coupling is phase dependent, and birefringence leads to different phase shifts for different polarizations, this model naturally leads to polarization-dependent spatial-mode coupling.

\subsubsection{Weak vs.~strong mode coupling}

The pairwise coupling strength between two modes depends on the ratio between the coupling coefficient (per unit length) and the difference between two modal propagation constants. Hence, a given perturbation may strongly couple modes with nearly equal propagation constants, but weakly couple modes with highly unequal propagation constants. 

Compared to light scattering in disordered photonic structures, mode coupling can be treated as a scattering process taking place in fiber mode space \cite{xiong2017principal}. The associated scattering mean free path $\ell_s$ gives the average distance that light travels in the fiber before hopping from one spatial mode to another. One also defines the transport mean free path $\ell_t\geq\ell_s$ as the minimum propagation distance beyond which light is spread over all fiber modes, no matter which mode it is initially launched into. Even if the fiber length $L$ is already longer than $\ell_s$, but still shorter than $\ell_t$, the mode coupling is considered weak. Once $L \gg l_t $, the mode coupling is strong enough to initiate a random walk of light in mode space. As light still propagates only forward in the fiber, this optical diffusion process does not result in significant back-reflection or loss. This is in stark contrast to disordered photonic structures that induce strong backscattering \cite{RevModPhys.69.731, akkermans2007mesoscopic, mosk2012controlling, rotter2017light, cao2022shaping}.

In the weak-coupling regime, the group delays (derivative of the change in spectral phase with respect to the angular frequency) are weakly dependent on mode coupling, and the differential group delays (difference in group delays) are linearly proportional to fiber length. By contrast, in the strong-coupling regime, the group delays are strongly dependent on mode coupling. Differential group delays are reduced as compared to the low-coupling regime, and are proportional to the square root of fiber length. In the strong-coupling regime, the statistics of modal dispersion and mode-dependent loss depend only on the number of modes and the variance of accumulated group delay or loss, and can be derived from the eigenvalue distributions of certain Gaussian random matrices \cite{ho2011statistics}. 

In optical telecommunication, strong mode coupling reduces the group delay spread from modal dispersion, minimizing signal processing complexity for spatial-division-multiplexing systems. Likewise, it reduces the variations of loss due to mode-dependent loss MDL, maximizing the channel capacity for long-haul communication \cite{ho2013linear, juarez2012perspectives}. 

\subsubsection{Disorder and complexity}

With no doubt, the presence of disorder in MMFs limits the spectrum of applications greatly, and it remains highly desirable to develop fibers which would be predictable to ever larger distances. 
Further, it is very important to develop methods, with which one can discriminate disorder form other manifestations of complexity. More specifically, it is very desirable to ascertain reliably, how far a specific fiber can be considered predictable. For example, many applications relying on the exact spatial control of light outputs (imaging, optical trapping, micromanufacturing), which shall function through bendable fibers and without direct optical access to the output, may be enabled 
in predictable fibers with the use of sufficiently precise theoretical models. But once disorder overtakes the dominant role, such possibilities would be immensely challenging and extremely demanding technologically.   

Assessing the predictability of an optical fiber, i.e., to what extent it follows a theoretical prediction, is however associated with yet another important problem, related to limitations in our experimental possibilities. Here, one has to consider further uncertainties in the parameters of the given fiber (core diameter, NA) as well as the uncertainties in spatial alignment under which the light signals have been coupled into and collected from the fiber. \textcolor{black}{Unless these uncertainties are identified and eliminated\cite{ploschner2015,matthes_learning_2021}, they} manifest themselves practically as coupling between modes and can be easily confused with disorder. 

Importantly, disorder does not only bring limitations, but also benefits. Specifically, random mode coupling not only suppresses negative effects such as modal dispersion and mode-dependent loss, but it is also  the crucial tool for certain applications. Consider here, e.g., disorder-induced spatial- and polarization-mode coupling, which allows one to use the spatial degrees of freedom in the input light to a MMF to control the polarization degrees of freedom of the output field, which will be described in a later section.

	\subsubsection{Optical memory effects}
	
	One notable difference between optical fibers and disordered photonic structures lies in their optical memory. The angular memory effect, also referred to as intrinsic isoplanatism, has been well studied for random scattering media \cite{feng1988correlations, freund1988memory, Katz2012, Katz2014}. When the spatial wavefront of a coherent beam incident on a disordered slab is tilted by a small angle, the transmitted wavefront is tilted by the same angle. The angular range of the memory effect is inversely proportional to the system thickness, thus a thin scattering layer has a large memory effect range. Correspondingly, the angular memory effect is usually absent in a MMF, because its length typically lets the memory effect range go to zero. 
	
	However, short fibers with weak mode coupling have rotational and quasi-radial memory effects \cite{amitonova2015rotational, rosen2015focusing, Li2021memory}. Rotating the incident wavefront around the fiber axis leads to a rotation of the transmitted intensity pattern without any significant change of the pattern itself. When monochromatic light propagates through a MMF with a specific propagation constant, a quadratic radial phase modulation of the input wavefront will cause an axial shift of the output pattern \cite{cizmar2012exploiting}. More recently, the translational memory effect has been observed in MMFs with square cross-section \cite{caravaca2021optical}. Symmetry properties of the square-core fiber lead to speckle patterns shifting along four directions at the fiber output for any given shift direction at the input. \textcolor{black}{The memory effect has been extended to the spectral domain \cite{devaud2021chromato}. When the input wavefront of a monochromatic light is shaped to focus through a step-index multimode fiber, a frequency change induces an axial shift of the output focus. This broadband chromato-axial memory effect originates from the conservation of the transverse component under spectral detuning in the MMF.}  Nevertheless, none of these memory effects will survive in long MMFs with strong mode coupling.
	
	A phenomenon closely related to optical memory effects is ``coherent backscattering'' (CBS). It is manifested by an enhancement by a factor
	of two of the backscattered light intensity, due to constructive interference of waves which propagate along time-reversed paths \cite{RevModPhys.69.731,akkermans2007mesoscopic}. While CBS is commonly known as weak localization of waves in random scattering systems, it also exists for light reflected from the distal facet of a MMF, which has random mode coupling \cite{bromberg2016control}. By tuning the nonreciprocal phase with the magneto-optical effect, it is possible to control the interference between time-reversed paths in a MMF and realize a continuous transition from enhancement to suppression of coherent backscattering \cite{bromberg2016control}. 
	
	\section{Methods}
	\label{sec:method}

In this section, we will introduce the experimental tool that enables optical wavefront shaping - spatial light modulator (SLM). There are several types of SLMs, operating with different principles. Their specifications will be summarized in section 3.A to facilitate the choice of an appropriate SLM for the specific application. In section 3.B, we describe how the SLM is used to measure the field transmission matrix of a multimode fiber (MMF). The transmission eigenchannels are then introduced. The SLM is used to control the spatial profile of transmitted light through a MMF. In addition to monochromatic light, broadband transmission and short-pulse propagation through MMFs are characterized by multi-spectral and time-gated transmission matrices in section 3.C. They allow for temporal control of MMF transmission, including spatio-temporal focusing and global temporal focusing of an optical pulse through MMFs. In section 3.D, novel states of light, such as principal modes, super- or anti-principal modes, are introduced to manipulate modal dispersion in MMFs. Finally, section 3.E shows full control of output polarization states are realized by shaping input wavefront to a MMF.  	

\subsection{Spatial Light Modulators, Optical Wavefront Shaping and transmission matrix measurements}
\label{sec:Modulators_WFS}

First appearing in the 70s \cite{spitz1967transmission}, the idea of taming a complex and seemingly random light transfer in multimode fibers using holography was based on photographic plates as a static wavefront shaping element. Being able to recreate a single prerecorded image at the distal end of the fiber, this remarkable fundamental concept based on a time-reversal technique was ahead of its time for any practical outputs. \textcolor{black}{An elegant extension of this concept was proposed by Yariv a decade later \cite{yariv1976transmission}, utilising not a single but two identical fiber segments of multimode fiber for image transmission. The spatial distortion, accumulated over propagation through a first piece of the fibre, can be compensated, or as the author referred "healed" \cite{yariv1978phase}, by complex conjugating of the propagated field and sending it down to the second, identical segment. However, the manufacturing of two identical fiber segments of practical length remains a technological challenge even today.}
A new round of development started only in the early 2010s \cite{vellekoop2007focusing}, fueled by novel technology of computer-controlled spatial light modulators and principles of digital holography, allowing for fast, dynamic and on-demand generation of desired optical fields.

\begin{figure}\centering
\includegraphics[width = 0.8\textwidth]{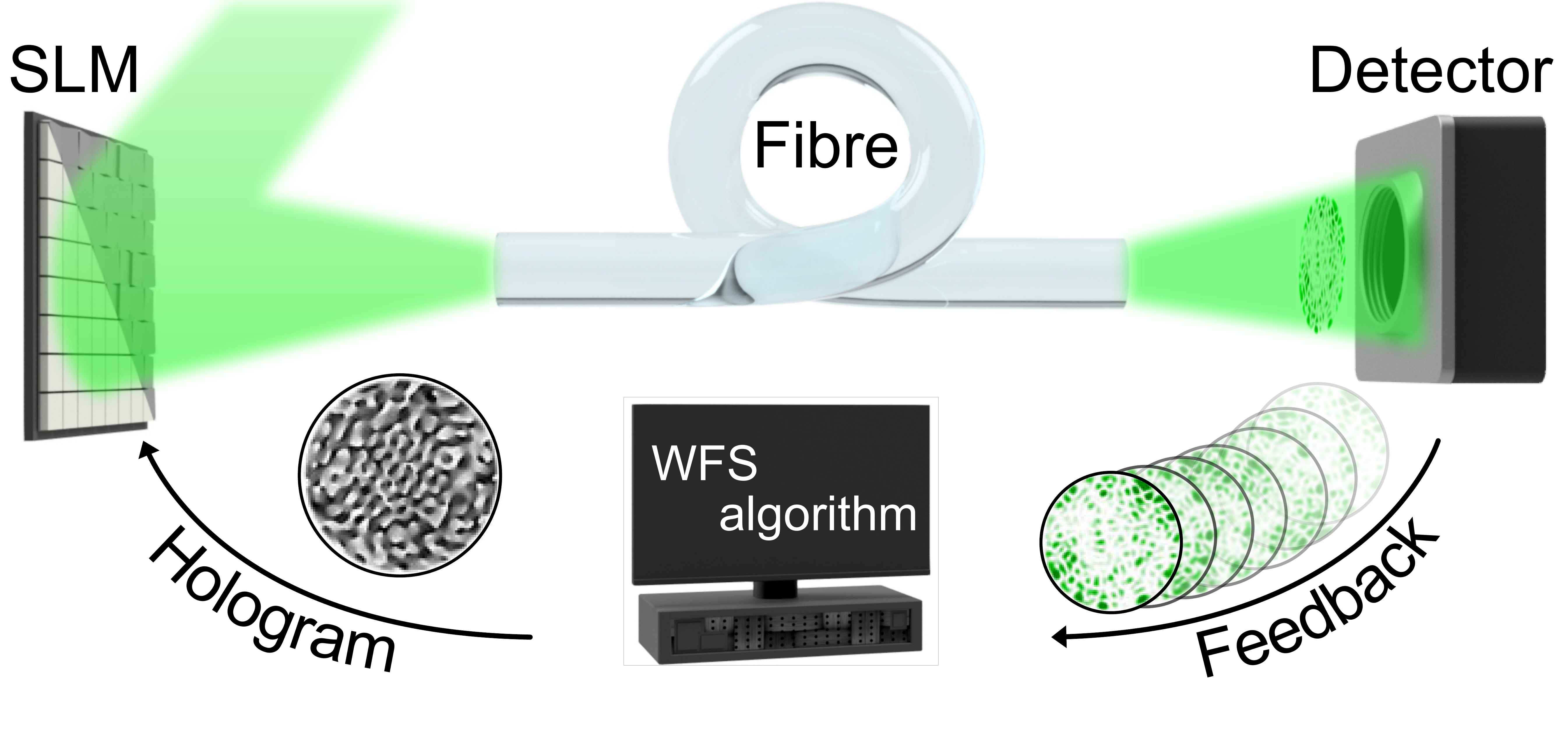}
\caption{Wavefront shaping through a multimode fiber. Coherent light is modulated by a SLM prior coupling to the fiber. The detector captures speckled responses to the applied input wavefronts. Measured feedback, fed to the WFS algorithm, results in a computer-generated hologram. Such pattern is imprinted on the input wavefront by the SLM to generate the desired output field after propagating through the fiber.}
\label{fig:WFS_generalised}
\end{figure}

The general case depicted in Fig.\ref{fig:WFS_generalised} represents a typical wavefront shaping (WFS) scheme. The laser light passes by the spatial light modulator (SLM) and couples to the fiber at the proximal end, allowing a user to gain control over the input wavefront. The detector, which is usually located at the distal end, enables recording the optical response of the fiber to the coupled input wavefronts and serves as feedback for wavefront shaping algorithms. Using these response measurements to characterize light transfer through the fiber or as part of iteration-based approaches, one can tailor the output wavefront of a MMF to the desired light distribution.

The choice of the SLM with appropriate specifications is an essential aspect to address when considering an application. One of the critical specifications is the diffraction efficiency of the device, defining the portion of illumination power redirected to the fiber. Another is the pixel count of SLM, which limits the spatial definition of the input wavefronts or the number of modes to be controlled in the fiber, while the type of modulation and its depth dictates the precision of the WFS. Finally, frame rate and upload latency are critical for both quick fiber response measurements and highly-dynamic wavefront shaping.
A perfect modulator, in this case, would offer high pixel resolution, modulation depth and diffraction efficiency,  while providing a high refresh rate with low upload overheads.
Wavefront shaping through multimode fibers is a highlight of complex photonics applications since the existing multi-megapixel modulators provide enough degrees of freedom to manage the number of available spatial channels or fiber modes for large variety of commercially available fibers. However, when looking at the broader picture involving the other vital parameters, the choice of spatial modulator for a particular application becomes an exercise of balancing the tradeoffs.

\begin{figure}[htbp]\centering
\includegraphics[width = 0.75\textwidth]{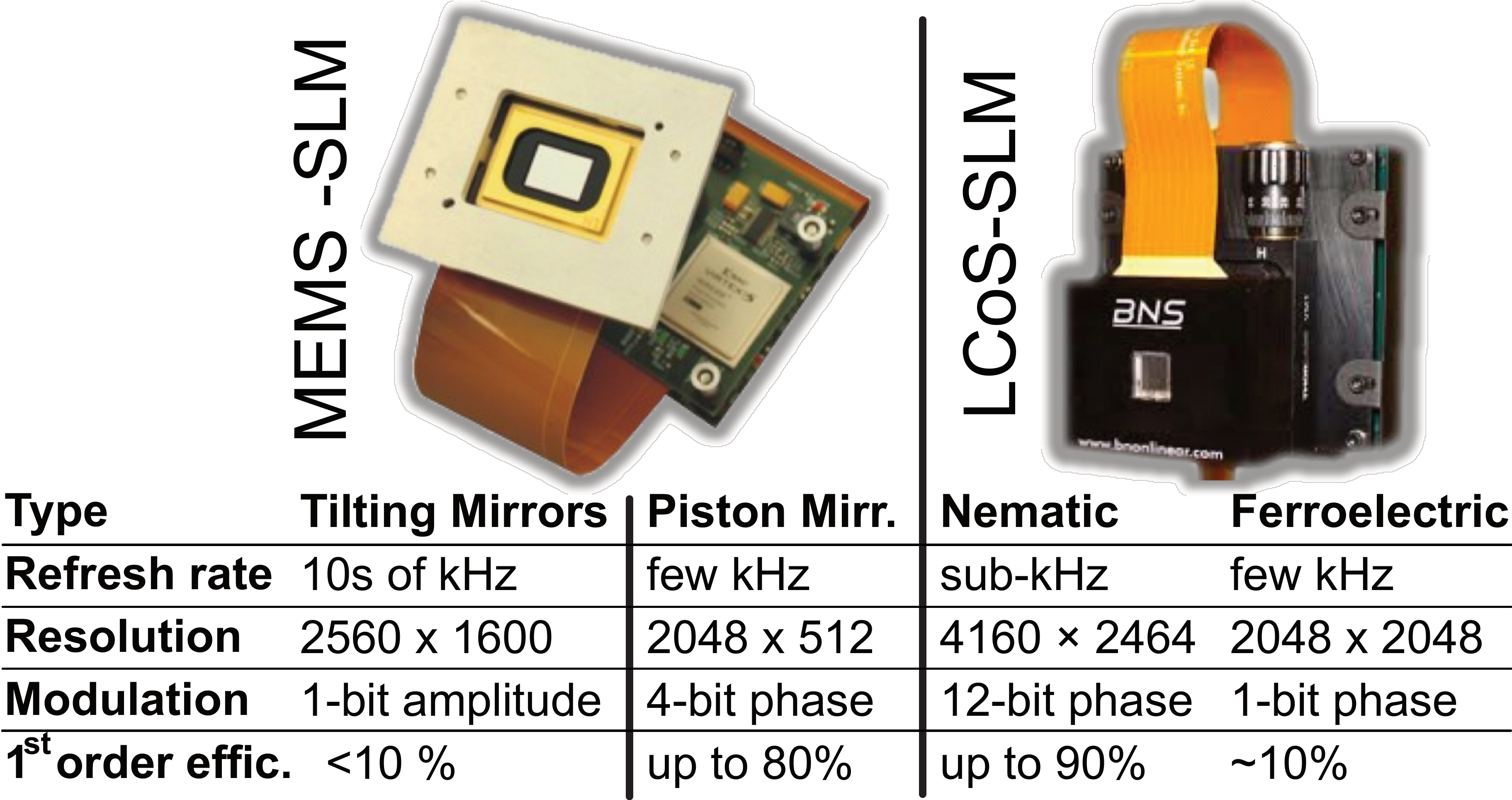}
\caption{A table summarizing specifications of modern spatial light modulators. Values are not associated with particular SLM models, trade-offs between parameters should always be expected.
}
\label{fig:SLMs}
\end{figure}

Nowadays, commercially available SLMs are based on either liquid crystal (LC) technology, or optical microelectromechanical system MEMS, that owe their rapid development and affordable price to mass-market success of display and projection devices (see Fig. \ref{fig:SLMs} for comparison of these two technologies).

\subsubsection{Liquid crystal SLM}
Initially developed for video projectors, LC-SLM promptly found new applications in holographic optical tweezers, advanced microscopy, holographic displays, data storage and optical computing \cite{zhang2014fundamentals}.
In particular, the parallel-aligned liquid crystal on silicon microdisplays became a gold standard in complex photonics applications offering a direct full-phase control with high spatial resolution and modulation depth. This modulator remains the first choice for power-efficient applications due to its high diffraction efficiency (up to 90\%) in the off-axis regime. While the standard liquid-crystal SLM modulates only the phase of light field, a computer-generated phase hologram enables modulation of both amplitude and phase \cite{arrizon2007pixelated}.  For highly dynamic wavefront shaping scenarios, however, liquid crystal technology often becomes a severe obstacle for practical applications due to the relatively limited frame rate of a few 100s Hz. An example of the limitation mentioned is raster-scan imaging via multimode fiber, which takes more than a minute to acquire a single 120x120 pixels image when implemented on LC-SLM \cite{vasquez2018subcellular}. Another modulator in the liquid crystal family - the ferroelectric liquid-crystal-on-silicon display allows for accelerated wavefront shaping with a kilohertz-level refresh rate, which comes at the cost of a limited modulation depth. Less efficient binary phase modulation (0 or $\pi$ states) modality is suitable for applications not demanding power efficiency \cite{maurer2008suppression}, which usually does not exceed 10 per cent for the first diffraction order to this type of device.

\subsubsection{Micro-Electro-Mechanical System based SLM}
In parallel to the liquid crystal technology, Micro-Electro-Mechanical or Micro-OptoElectro-Mechanical Systems (MEMS or MOEMS), began their rapid development in the '80s. MEMS devices have common major advantages when compared to LC. These are broad spectral range, high frame rate, possibility to operate with non-polarised light and long lifetime \cite{song2018review}. Digital Micromirror Devices (DMDs) have emerged as a powerful solution to high modulation speed applications, reaching framerates of more than 20 kHz. Unlike LC-SLMs, which typically modulate the phase of the reflected wavefront directly, DMDs operate as purely binary amplitude modulators, posing a limit to the precision and efficiency with which each degree of freedom can be controlled. Nevertheless, it has already been shown that using a DMD in the off-axis regime \cite{lee1979binary, conkey2012high, mitchell2016high, goorden2014superpixel} makes it possible to perform beam shaping through a MMF with the fidelity of generated fields matching and, for some instances, even outperforming that for LC-SLMs \cite{turtaev2017comparison, caravaca2017single}. 

\begin{figure}[htbp]\centering
\includegraphics[width=82mm]{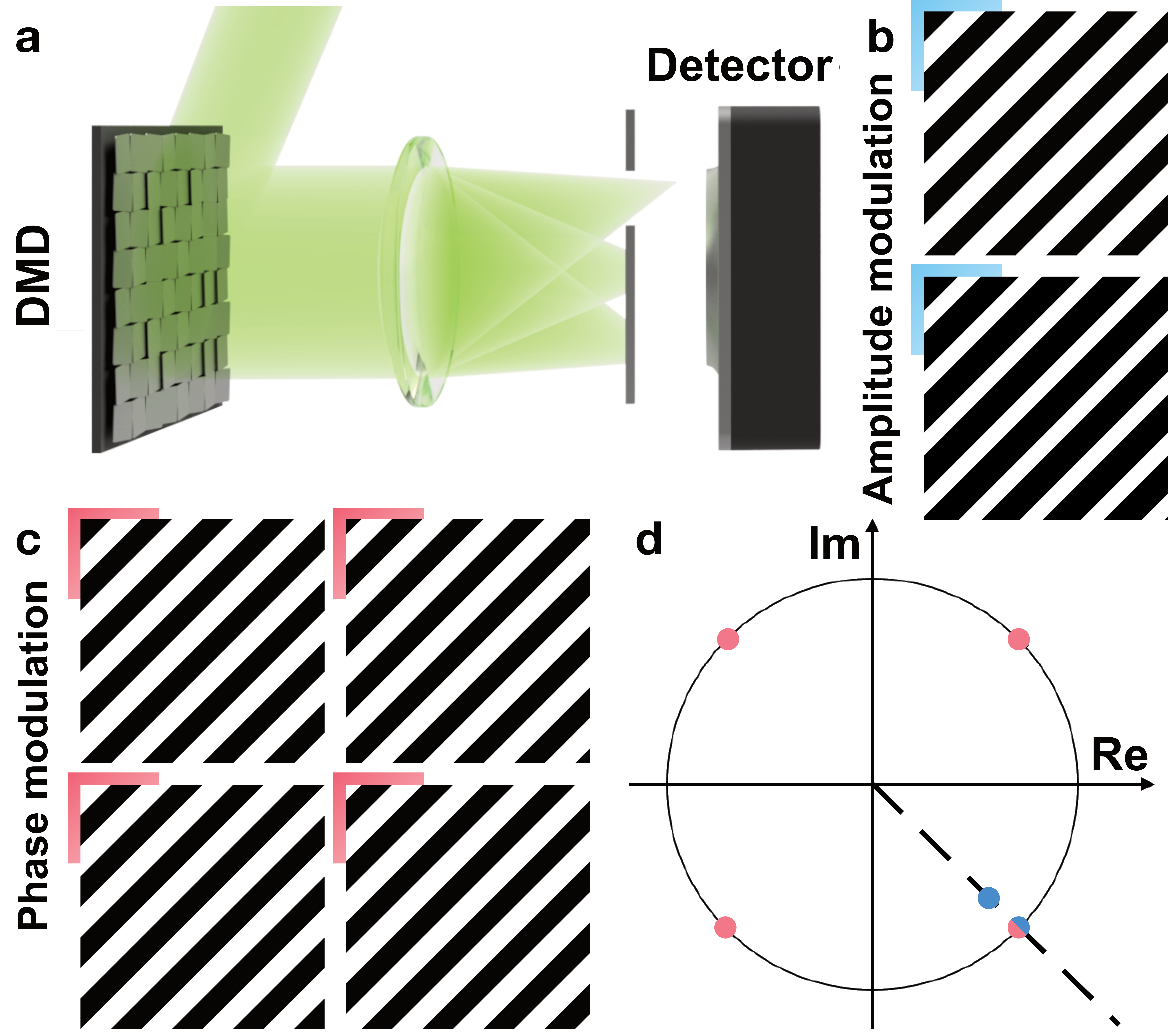}
\caption{(a) Lee hologram method utilizing binary amplitude gratings for off-axis complex modulation. (b) Amplitude modulation is performed via variation of a duty cycle of the grating. (c) Phase modulation is achieved via a lateral shift of the grating. (d) Far-field (Fourier) plane schematically illustrates the modulations caused by the gratings presented in cases b (blue dots) and c (red dots).
}
\label{fig:Lee_hologram}
\end{figure}

The most utilized modulation technique with DMD, known as the Lee hologram \cite{lee1979binary}, allows for tailoring the desired complex field in the first order of diffraction formed by superposition of binary amplitude gratings. The simplest case of a single grating displayed on a DMD is illustrated in Fig. \ref{fig:Lee_hologram}. Here, a grating pitch defines the angle of the first order of diffraction and, as a result, the position of the beam focused on the detector. The lateral shifting of the grating allows modulating the phase of the diffracted light, while a duty cycle controls the amount of light redirected towards a first order, allowing for amplitude modulation.

The main competitive advantage of such modulators comes with the same trade-off in refresh rate and diffraction efficiency as ferroelectric devices mentioned above. While reaching 10s of kHz in the scanning speed, wavefront modulation systems based on DMD usually provide only a few percent of power efficiency in the first diffraction order when employed off-axis \cite{liu2017focusing}. Offering two orders of magnitude faster frame rates, these modulators significantly promote practical applications of the MMF-based endoscopes, allowing to capture dynamic scenes \cite{stellinga2021time}, monitor neuronal activity \cite{ ohayon2018minimally, vasquez2018subcellular} or extend sampling of acquired images to a level comparable to modern video endoscopes \cite{leite2021observing}.
Rather weak power efficiency of DMD in an off-axis regime can be significantly improved by a double-pass scheme, where displayed pattern will be relayed back to the same DMD in a way that light reflected of ON and OFF mirror states will acquire relative $\pi$-phase shift, enabling more efficient binary phase modulation \cite{hoffmann2018kilohertz}. Moreover, such double-pass configuration automatically corrects for pronounced spatial dispersion, inherent for DMD, extending wavefront shaping to broadband and short-pulsed light sources.
Another perspective MEMS modulator, Grating Light Valve (GLV), is based on reflective movable ribbons mounted on a silicon base that can dynamically form diffraction gratings \cite{amm1998}. Recently, GLV technology found its application in beam shaping through the complex medium, demonstrating an outstanding speed level of 100s kHz \cite{tzang2019wavefront}. An efficient way of coupling the light reflected off this 1D modulator into the multimode fiber has to be addressed to benefit from a high refresh rate.

Rapidly growing industrial interest in holographic displays could become a new mass-market driver for further developing cost-efficient phase-modulation SLMs with high pixel count in the upcoming years. One of the most awaited solutions is a MEMS modulator based on a pistonlike micromirror array. Although its design is similar to the DMD, micromirrors do not tilt but produce a vertical stroke, enabling direct phase-only modulation of the incoming wavefront. The availability of such SLMs, simultaneously offering high-resolution, high-speed and low spatial dispersion, will accelerate the practical adoption of methods relying on rapid beam steering and shaping through a multimode fiber, e.g. additive manufacturing \cite{morales2017three}, optical ablation\cite{kakkava2019selective}, volumetric\cite{stellinga2021time} and nonlinear imaging\cite{tragaardh2022cars}. According to recent reports \cite{ketchum2021diffraction, deng2022diffraction, dauderstadt2021analog, witvrouw200911}, such devices are currently under intensive development and already available to the first users.

\subsection{Monochromatic transmission}
\label{sec:monochromatic_transmission}

\begin{figure}[htbp]\centering
\includegraphics[width=82mm]{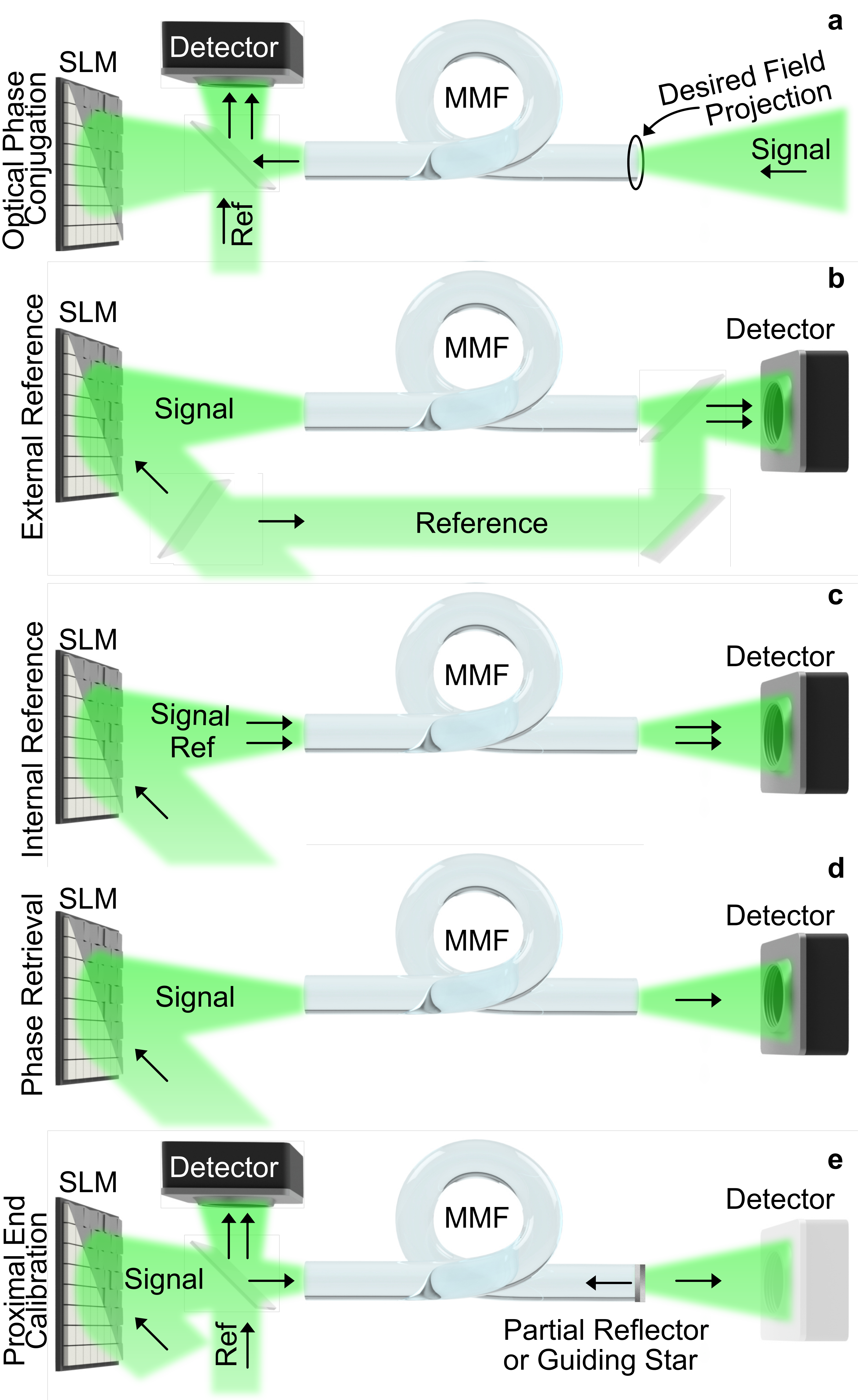}
\caption{Simplified representation of typical optical schemes employed for wavefront shaping through the MMF. (a) Illustration of optical phase conjugation system for wavefront shaping through the MMF. Desired optical wavefront projected onto the distal end and propagated through the fibre is recorded interferometrically at the proximal end. A phase-conjugated copy of the recorded field created by SLM and coupled back to the fiber would result in the original beam shape at the distal end. 
Experimental approaches for measuring the transmission matrix of a multimode fiber. (b) Using a separate reference arm for obtaining the phase distribution of the output fields. (c) Using SLM for both probing signal and phase reference generation. (d) Relying on intensity-only measurements and phase retrieval algorithms. (e) utilizing partial reflectors or guiding starts to allow measurements to be performed at the proximal end.
}
\label{fig:Refer}
\end{figure}

Coherent light coupled into MMFs rapidly transforms into a complex and seemingly random speckled pattern with no resemblance to its original field distribution. Optical wavefront shaping is used to characterize and compensate for such extreme cases of optical aberrations. 
Iteration-based or direct-search feedback algorithms, previously applied to experiments with turbid media, were successfully adapted to enable the first instance of wavefront shaping through a MMF \cite{di2011hologram,bianchi2012multi,cizmar2011shaping}.
Another approach lead to the first holographic experiment with MMFs and is built on the principle of digital phase conjugation \cite{papadopoulos2012focusing, papadopoulos2013high}. The desired light distribution, for example, a focal spot, is initially projected onto the distal fiber facet, and the field propagating through the fiber is measured interferometrically at the proximal end (see Fig. \ref{fig:Refer}a). Finally, the phase conjugated field is generated using SLM, also located at the proximal end, so that the light propagated back along the fiber would form the focal spot at the original position of the output end. Although the alignment is difficult, this approach is inherently fast, requiring a single image to calculate the correct wavefront. Moreover, the digital phase conjugation scheme could be modified, by linking optical fields at SLM and camera planes via transmission matrix, allowing to relax on the pixel-to-pixel pre-alignment step and minimising general aberrations in the system, caused by components or assembling imperfections \cite{mididoddi2020high}.

Probably the most popular method of gaining control over light transmission through MMFs is based on the powerful concept of the transmission matrix (TM) \cite{cizmar2012exploiting, choi2012scanner}. The TM characterizes the optical response of the complex medium and expresses it as a linear operator linking a selected set of input fields coupled to the fiber to another set of transmitted output fields. Once characterized, the TM allows tracking the transformation of every input field that is utilized a priori for probing the fiber response; most importantly, however, it can be used to determine what these inputs must be to obtain the desired output field.

\subsubsection{Transmission matrix measurement}
\label{sec:TM_measurements}

In the experiment, the typical way of retrieving the monochromatic TM of a MMF involves projecting a set of probing optical fields generated by a spatial light modulator, while measuring the interference of the output speckle field with a reference beam in a phase-shifted or off-axis manner. This procedure does not depend on the particular set of input fields chosen, provided the field patterns can be generated in sequence. Phase shifting the input field or reference beam results in a corresponding harmonic intensity evolution in every spatial position across the speckled output field. Recording such evolution for at least 3 phase steps over the $2\pi$-range allows one to retrieve both the amplitude and phase information of the output field. With the output fields usually being conjugated to the camera chip, the interferometric response can be recorded by every individual pixel of a camera simultaneously, such that an entire column of the TM can be measured at once. 

The reference beam can be delivered externally via a separate optical path to avoid scrambling by the fiber, or it can co-propagate through the fiber with the probing fields \cite{cizmar2011shaping}. The first scheme (Fig. 12b) benefits from a uniform reference field for precise measurements across the whole distal facet of the fiber. In contrast, the second scheme  (Fig. 12c) provides outstanding measuring stability at the cost of blind spots in the acquired output fields caused by the speckled reference. These blind spots can, however, be eliminated via repeated measurements with different internal references \cite{bianchi2013focusing, jakl2022endoscopic}. Moreover, using phase retrieval algorithms, inherently complex TMs can be computationally estimated without any phase reference (Fig. 12d), using instead only real-valued intensity measurements and multiple 
calibration procedures \cite{dremeau2015reference,zhao2018bayes,n2018mode}.

\begin{figure}[htbp]\centering
\includegraphics[width=130mm]{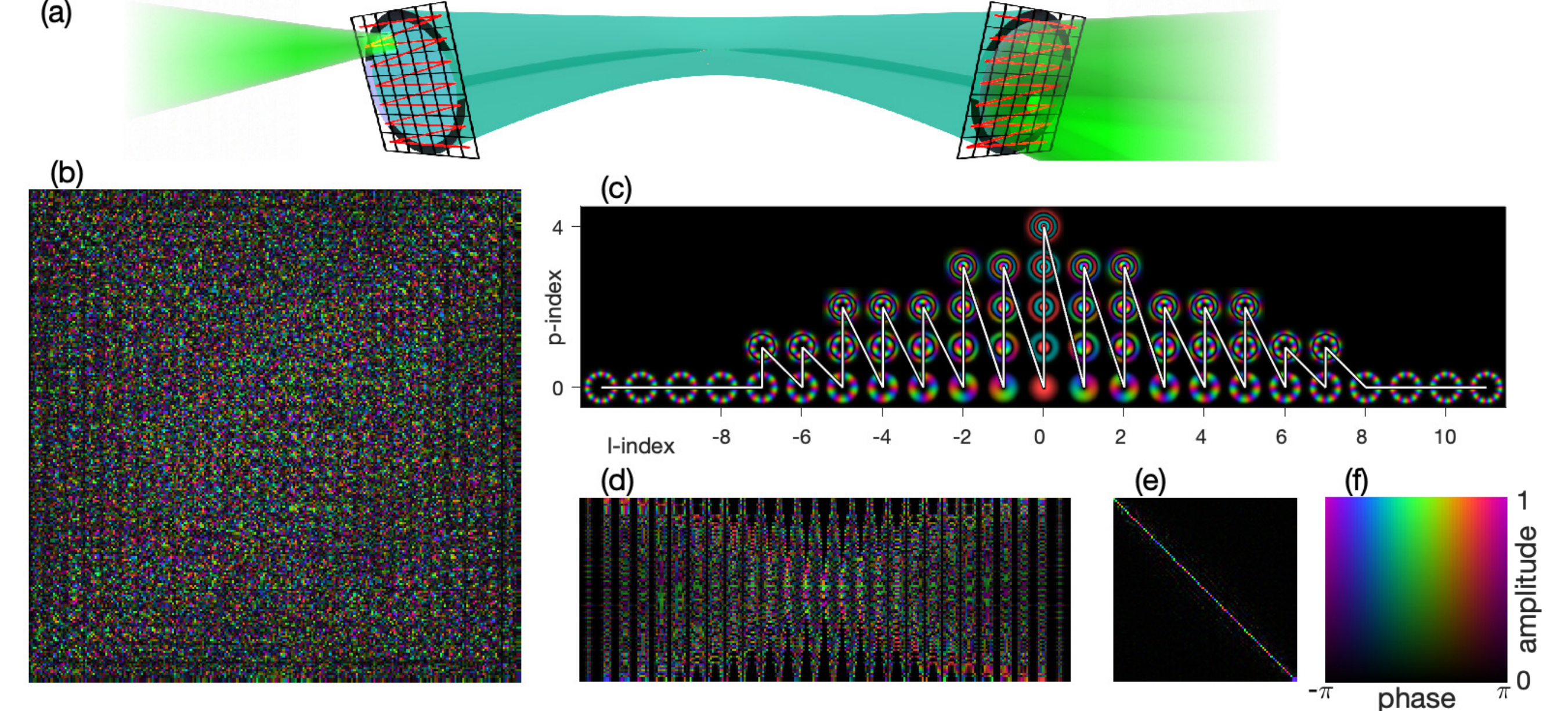}
\caption{(a) The set of diffraction-limited spots at the proximal and distal ends of the fiber, and (b) transmission matrix represented by these states. (c) The basis of the PIMs of the fiber, (d) the conversion matrix between the DLS set and the PIM basis, (e) the diagonal transmission matrix in the PIM basis, and (f) the color map used. Figure adapted from Ref.~\cite{ploschner2015} with permission.
}
\label{fig:TM}
\end{figure}

The set of input and output fields, forming the representation for TM expression, is another essential aspect for TM measurements. One possible choice is the orthonormal basis of the flux-normalized propagation invariant modes of the fiber for both the input and output state, see Fig.~\ref{fig:TM}~(c); in this basis, the TM is diagonal (when fiber loss and mode coupling are negligible) and it contains the phase factors of the individual modes in the diagonal elements, see Fig.~\ref{fig:TM}~(e). However, the PIM basis is not directly experimentally accessible because the exact fiber parameters and consequently the modes are not {\em a priori} known: on the contrary, it is the TM itself obtained by measurement that enables to find the modes as well as the fiber parameters~\cite{ploschner2015}. This way, the basis of PIMs is not convenient for measuring the TM. Instead, the choice of input and output states is dictated by the pixelated nature of the spatial light modulator as a source of the input fields and the camera utilized for recording the speckled outputs. Therefore, a square grid of diffraction-limited points [see Fig.~\ref{fig:TM}~(a)], or truncated plane waves with different propagation angles are frequently used for TM measurements. In this case, both focal points, as well as plane waves (points in the Fourier plane) can be directly associated with pixels of a camera detector or spatial light modulator.

The density of the diffraction-limited points has to be sufficiently large to faithfully sample the input and output states; slightly oversampled sets are usually used, and the input and output states do not form orthogonal systems. In such a representation, the dimension of the TM is larger than the number of the transmitted PIMs, and the TM has a complicated structure that can be seen in Fig.~\ref{fig:TM}~(b). It is important to emphasize that the sets of the input and output states can be different; if the number of input states differs from the number of output states, the TM is a non-square matrix in this representation. Another common set of orthogonal input fields is based on the Hadamard matrices, which is capable of higher signal to noise ratio (SNR) during TM measurement in certain schemes \cite{conkey2012high}.  

Once the TM is measured, the basis in which it is expressed can be changed for the convenience of further operations \cite{ploschner2015seeing,Pai_Opex21}. An excellent example of such a case relevant to this chapter is the compressed sensing of the TM, which promises a significant speed-up compared to the above-mentioned TM measurement methods \cite{li2021compressively}. Relying on the sparsity and highly-diagonal nature of TM in PIMs representation as priors, the compressed sampling approach can estimate the TM of a MMF with high fidelity, using as little as 5\% of the sampling required to complete a TM measurement. 

The above procedure of measuring the fiber transmission matrix requires access to fields at both ends of the fiber, which is not always possible in practical applications. Moreover, changes of fiber configuration or external perturbations may require re-calibration of the transmission matrix {\it in situ} right before imaging. To measure the TM with access only to the proximal end of the fiber (Fig. 12e), a partial reflector or known calibration element is added to the distal end of the fiber~\cite{kahn2015, gordon2019characterizing, lee2020reciprocity, chen2020remote}. For example, a thin stack of structured metasurface reflectors at the distal facet of the fiber introduce wavelength-dependent, spatially heterogeneous reflectance profiles, and the TM can be recovered from  the reflected light arriving at the proximal end of the MMF~\cite{gordon2019characterizing}. Another elegant approach, utilizing the PIMs as a TM basis together with memory effects inherent to MMFs due to their waveguiding nature, allows the TM measurement without accessing the distal fiber facet~\cite{Li2021memory}.

Recently, deep learning methods emerged as a viable alternative to the TM approach, which rely on learning rather than measuring the relationship between coherent fields coupled to the fiber and the resulting output speckled intensity patterns \cite{borhani2018learning, rahmani2018multimode, caramazza2019transmission, fan2019deep, xiong2020deep, liu2020bending, tang2022learning, zhang2022learning}. Demonstrated mostly in image transmission and sensing experiments, these novel approaches showed a pronounced resilience against fiber perturbations, after trained for variety of fiber contortions.

\subsubsection{Spatial control of transmitted light}

\begin{figure}[htbp]\centering
\includegraphics[width=82mm]{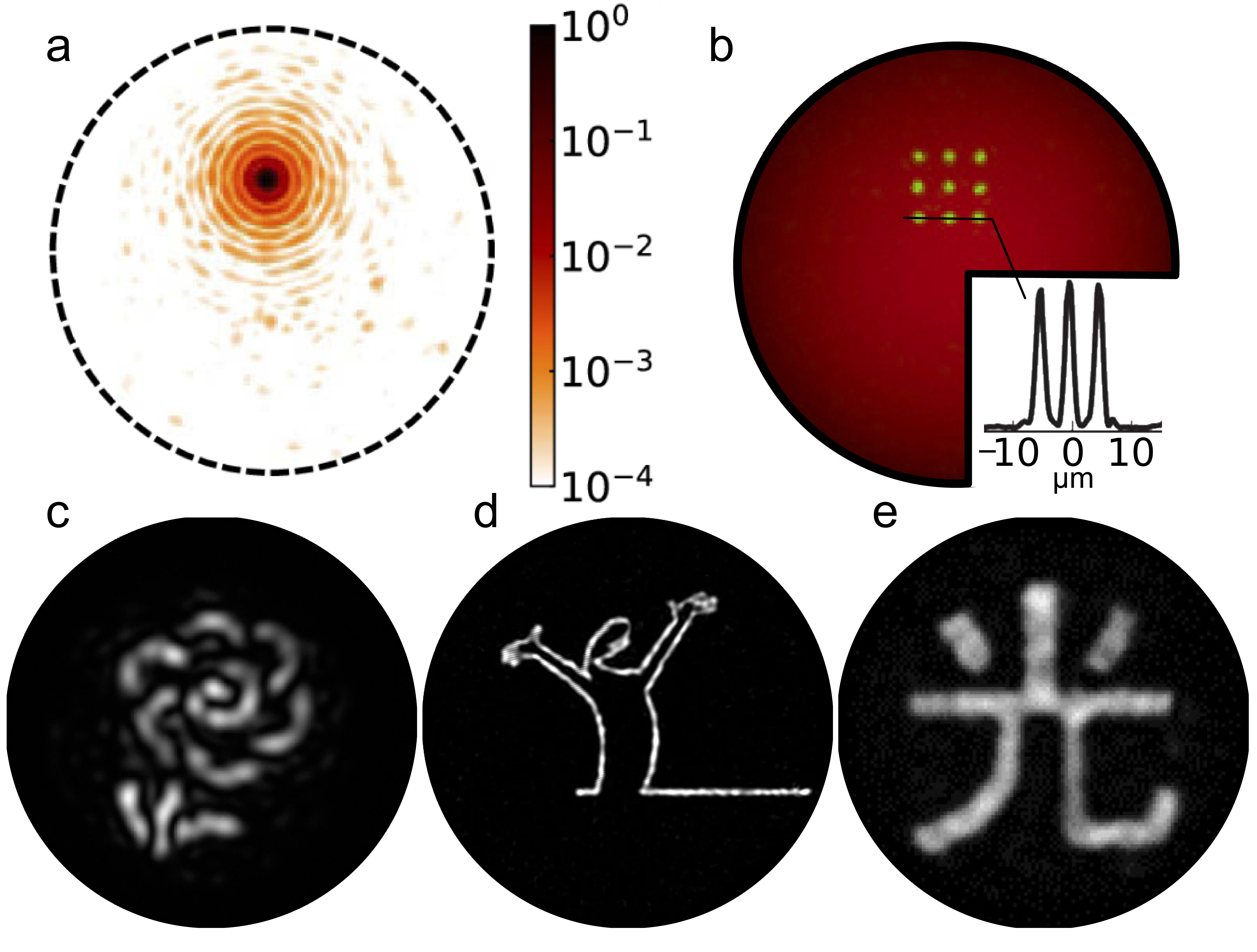}
\caption{(a) Single focal point generated at the tip of a multimode fiber using the measured TM \cite{gomes2022near}. a - Copyright 2022 Optical Society of America. (b,c) Array 3x3 of focal points and the flower were generated using the transmission matrix approach and further optimized by Gerchberg–Saxton (GS) iterative algorithm for uniform amplitude distribution \cite{bianchi2012multi,cizmar2011shaping}. b - Copyright 2011 Optical Society of America. c - Figure adapted from Ref.~\cite{bianchi2012multi} with permission. The red background illumination is provided by a LED source coupled to the same fiber.
(d) Screenshot of a cartoon being projected at the tip of the multimode fiber. Wavefront shaping was achieved by the optical phase conjugation technique using LC-SLM \cite{loterie2015digital}.d - Copyright 2015 Optical Society of America. (e) The Chinese character for light is shaped using the measured TM in a DMD-based wavefront shaping system \cite{li2021compressively}. e - Figure reprinted from Ref.~\cite{li2021compressively}, licensed under a Creative Commons Attribution (CC BY) license.
}
\label{fig:shaping}
\end{figure}

Over the past decade, spatial control of light transmitted through MMFs has been successfully realized using wavefront shaping feedback algorithms first developed for random scattering media \cite{bertolotti2022imaging}, or the measurement of the transmission matrix, as well as principles of digital phase conjugation.

The simplest example of wavefront formation, regardless of the method used, is focusing. In this case, the input wavefront is optimized in such a way as to make all modes constructively interfere only at a single point on the output side, as depicted in Fig. \ref{fig:shaping}a. The focus is a high-intensity spot at the desired location surrounded by faint speckles across the entire field of view. Despite the simplicity and even primitiveness of the approach, many applications are based precisely on the formation of a single focus and raster scanning. The suppression level of residual speckle background formed by a portion of uncontrolled light often defines the quality of the result for such raster scanning applications. The widely used metric for focusing fidelity is the power ratio (PR), defined as the fraction of the optical power carried by the desired focus with respect to the total amount of power transmitted through the fiber \textcolor{black}{(see \cite{gomes2022near} for more details)}. The PR can be linked to the other common complex photonics metric, such as enhancement $\eta$, which describes the ratio between the intensity of the focus and the mean intensity of the background:
$PR= {\eta}/({N+\eta})$, where $\eta = I_{focus}/ \langle I_{background} \rangle$ and N is a number of guided modes in the fiber at a given wavelength \cite{turtaev2018high}.

In principle, any optical field, which can be linearly decomposed into PIMs, can be generated at the output of the fiber. In the case of a focal point, near unity PR can be achieved \cite{tuckova2021computational}.
In practice, the PR depends on the sufficiency of input sampling to address all fiber's degrees of freedom \cite{turtaev2017comparison} and the precision of input fields' amplitude, phase, and polarization control. The PR of over 97\% can be reached experimentally when these and a few minor factors are handled \cite{gomes2022near}.

Light focusing using any direct search algorithm to optimize a input fields for single output channel is an equivalent of measuring a single raw of the TM, which set of output fields conveniently organized as a set of point conjugated to he individual camera pixels. Measurement of the full transmission matrix $T(\omega)$ as described above, allow us to predict, first of all, how any input state $E_{\rm in}$ will be transmitted at the (single) frequency $\omega$ determined by the input laser, to an output state $E_{\rm out}=T(\omega)E_{\rm in}$. (We will use a scalar notation for the in- and output states here just for simplicity.) For this input-output relation to hold, we implicitly assume, of course, that the fiber that hasn't changed its shape since $T(\omega)$ was determined. 

Once the TM is known, one can achieve a desired state at the fiber output by launching the corresponding state at the input; this state can be found by applying the inverse of the transmission matrix to the output state $E_{\rm in}=T(\omega)^{-1}E_{\rm out}$ (see Fig. \ref{fig:shaping}e). The inverse of the TM  can be calculated simply by the matrix inversion if one works in the PIM basis; however, if the TM is represented in terms of the non-orthogonal sets of states as explained above, matrix inversion should be replaced by Hermite conjugation, i.e., transposition and complex conjugation $T(\omega)^{-1} = T(\omega)^{\dagger}$.

\subsubsection{Transmission eigenchannels}
	\label{sec:transchannel}
	The transmission matrix $T(\omega)$ at frequency $\omega$ allows us to determine states with unique transmission characteristics: consider here, e.g., the so-called ``transmission eigenchannels'', whose field patterns are the same at the input and output. This property follows directly from the fact that these transmission eigenchannels are determined (at the input) by the eigenvectors of the transmission matrix, $T(\omega)E^{\rm TM}_{n}=\xi_n E^{\rm TM}_{n}$ (assuming here that $T$ is a square and normal matrix for a MMF). With the eigen-decomposition of  the transmission matrix, 
 \begin{equation}\label{eq:eigendecomp}
 T=U\Xi U^\dagger, 
 \end{equation}
 these right eigenvectors, contained in the columns of $U$, are at the same time the eigenvectors of the Hermitian matrix product $T^\dagger T=U(\Xi^\dagger \Xi)U^\dagger$. The real eigenvalues of $T^\dagger T$ are called ``transmission eigenvalues'' and given as: $\tau_n=|\xi_n|^2$. In the absence of any amplifying mechanism (optical gain), the transmission eigenvalues fall within the interval $\tau_n\in [0,1]$.  For the case of a perfectly cylindrical fiber without any bending or mode-coupling, \textcolor{black}{the eigenvectors of the transmission matrix, which have the same transverse profile at the input and output, are equivalent to the propagation-invariant modes (PIMs), which maintain this profile throughout propagation along the entire fibre.} 
	
	Transmission eigenvalues of $\tau_n$ below unity, i.e., the $\xi_n$ do not lie on, but inside the unit circle in the complex plane, indicate fiber absorption, unwanted scattering, reflection or transmission into the fiber cladding. 
	In all these cases, the matrix $T^\dagger (\omega) T(\omega)$ remains Hermitian per construction. Losses are mainly produced by the overlap with the fiber cladding and by the dissipation in the fiber core. Quite generally, fiber modes corresponding to a larger angle with respect to the fiber axis will suffer more losses. This is because these higher-order modes both have a stronger overlap with the fiber cladding and an increased propagation time through the fiber, simply because their optical path length increases for larger injection angles. This \textit{mode-dependent loss} \cite{ho2011statistics,ho2012exact,chiarawongse2018statistical} lifts the degeneracy among the transmission eigenvalues of unity (in the loss-free and perfectly straight fiber). \textcolor{black}{For the extreme case that this loss is so strong that some higher-order modes are entirely lost during propagation, the transmission matrix ceases to be a square and normal matrix. It is then more appropriate to represent the transmission in terms of a singular value decomposition rather than in terms of the eigen-decomposition shown in Eq.~(\ref{eq:eigendecomp}).} For increasing mode-coupling in the fiber, the angle with respect to the fiber axis is less and less preserved during propagation, leading to a reduction in the overall modal dispersion \cite{ho2011statistics}. As a result, both the losses and the transmission eigenvalues are again more equally distributed among all available modes \cite{ho2011statistics,Ho:11,ho2012exact,chiarawongse2018statistical}, a property that is strongly desirable for practical applications such as for improving the channel capacity \cite{Ho:11}. As we will see in the subsequent section, concepts like the propagation time through a MMF that determine the degree of loss, can be formally defined not only in the ray picture, but also in the wave description of fiber transmission.


	\subsection{Spectrally and temporally resolved transmission}
	\label{sec:spectempTM}

In previous section, we consider  time-harmonic fiber modes that are characterized by a well-defined angular frequency $\omega$. In reality, any light that is launched into a MMF has a finite spectral width. It can be considered as a monochromatic light propagating through a MMF, when its spectral bandwidth is narrower than the spectral correlation width of the fiber, which will be introduced below.  

\subsubsection{Spectral decorrelation}

Let's assume that we scan the (angular) frequency $\omega$ of light injected to a MMF, while an (arbitrary) spatial profile of the incident wavefront remains fixed during the frequency scanning. To understand how the output profile at the distal end of the fiber changes, we first note that both the transverse field profile ${\bf \Psi}_m(r, \varphi)$ of different fiber modes and their propagation constants $\beta_m$ vary with $\omega$. Whereas the frequency dependence of ${\bf \Psi}_m(r, \varphi)$ is weak and will be neglected below, the change of $\beta_m$ with $\omega$ lets the fiber modes accumulate different phase delays during their propagation. As a result, the field at the output end ($z=L$) changes with $\omega$. 

Assuming that the incident field excites all fiber modes, the transmitted field at $\omega$ is expressed as: 
\begin{equation}
{\bf E}(r, \varphi; \omega) = \sum_{m=1}^M A_m \, e^{i \phi_m} \, {\bf \Psi}_m(r, \varphi; \omega) \, e^{i [\beta_m(\omega) \, L - \omega \, t]} \, ,
\label{eq:field}
\end{equation}
where $M$ is the total number of fiber modes, $A_m$ and $\phi_m$ represent the amplitude and phase of the incident field in the $m$-th fiber mode.

The spectral field correlation function of the output field is defined as 
\begin{equation}
C_E(\Delta \omega) \equiv \frac{\langle {\bf E}^*(r, \varphi; \omega) \cdot {\bf E}(r, \varphi; \omega + \Delta \omega) \rangle_{r, \varphi, \omega}} {\langle |{\bf E}(r, \varphi; \omega)|^2 \rangle_{r, \varphi; \omega}} \, ,
\label{eq:C_E}
\end{equation}
where $\langle ... \rangle_{r, \varphi, \omega}$ represents an average over transverse position $(r, \varphi)$ on the fiber cross-section and frequency $\omega$. With increasing frequency detuning $\Delta \omega$, the correlation magnitude $|C_E(\Delta \omega; L)|$ decays monotonically, and eventually approaches 0 for large $\Delta \omega$. 

The spectral correlation width $\Delta \omega_c$ of a MMF is usually defined as the full-width-at-half-maximum (FWHM) of the field correlation function, namely, $|C_E(\Delta \omega_c/2)| = |C_E(0)|/2 $. It is inversely proportional to the temporal spread that light experiences while propagating through the fiber. The latter is characterized by the width of optical path-length distribution $\Delta \mathcal{L}$ in the fiber \cite{redding2016evanescently}. In a MMF, different fiber modes have varying group delays. The modal dispersion will broaden the path-length distribution of transmitted light through a MMF \cite{ho2011statistics}. Consider all modes are excited at the fiber input. When mode coupling is negligible, $\Delta \mathcal{L}$ scales with the fiber length $L$, thus $\Delta \omega_c \propto 1/L$ \cite{redding2012using, redding2013all}. In the presence of strong mode coupling, light is scattered back and forth among the fiber modes with different group velocities, and the optical path-length distribution is narrowed. Such process can be considered as light diffusion or random walk in fiber mode space. The width of path-length distribution scales as $\Delta \mathcal{L} \propto \sqrt{L}$,  and $\Delta \omega_c \propto 1/\sqrt{L}$ \cite{redding2016evanescently}.

Using graded index fibers, whose transverse index profile is designed to reduce the difference in the phase velocities, is a viable strategy to counter-act modal dispersion. In most cases refractive index profiles decrease nearly parabolically from the central fiber axis to the cladding, achieving a strong (albeit not perfect) reduction in modal dispersion and a periodic refocusing of incident light along the propagation direction \cite{ghatak_introduction_1998}. On top of the modal dispersion, the fiber will also suffer from ``chromatic'' or ``material dispersion'' due to the frequency-dependence of the refractive index $n(\omega)$ that the fiber is made of. \textcolor{black}{In highly-multimode fibers especially the step-index fibers, the material dispersion is usually weaker than the modal dispersion.} 

\subsubsection{Multi-spectral and time-gated transmission matrices}
\label{sec:TM_measurements_nonmono}

For an optical pulse of frequency bandwidth $\Delta \omega_i$ exceeding the fiber spectral correlation width $\Delta \omega_c$, the number of uncorrelated spectral channels is $M_s \simeq \Delta \omega_i / \Delta \omega_c$. As the field transmission matrix becomes decorrelated after the input frequency shifts by $\omega_c$, one needs to measure $M_s$ transmission matrices at frequency spacing of $\omega_c$. The frequency-resolved or ``multi-spectral'' transmission matrices make it possible to predict (through a simple Fourier transform) how a pulse of spectral bandwidth $\Delta \omega_i$ arrives at the distal end of the fiber both in its spatial and temporal shape. The multi-spectral transmission matrices are measured using a frequency-tunable laser. The relative phases between different $T(\omega)$ are acquired with a reference beam \cite{carpenter_observation_2015, xiong2016spatiotemporal}. The reference arm length matches the MMF length, to avoid errors from phase drift of the laser during the measurement \cite{xiong2017principal}. 
  
The Fourier reciprocity between frequency and time degrees of freedom suggests that, instead of measuring a frequency-resolved transmission matrix $T(\omega)$, one can also measure directly a time-resolved (or ``time-gated'') transmission matrix $T(t)$. Indeed, such measurements are possible and typically being carried out by sending an input pulse through the MMF and through a delay line in parallel. Tuning the delay then allows one to scan the time at which both parts of the pulse jointly reach the detector. Through the corresponding interference process, the time-resolved transmission matrix $T(t)$ can be determined \cite{mounaix_deterministic_2016}.

\subsection{Temporal control of fiber transmission}
\label{sec:tempcontrol}

When a short pulse launched into a MMF, the modal dispersion typically broadens and breaks up this pulse into several temporal components. For controlling and mitigating such distortions, a number of different techniques have been developed that will be reviewed in this subsection. 

\subsubsection{Spatio-temporal focusing}	
	
\begin{figure}\centering
	\includegraphics[width=0.95\textwidth]{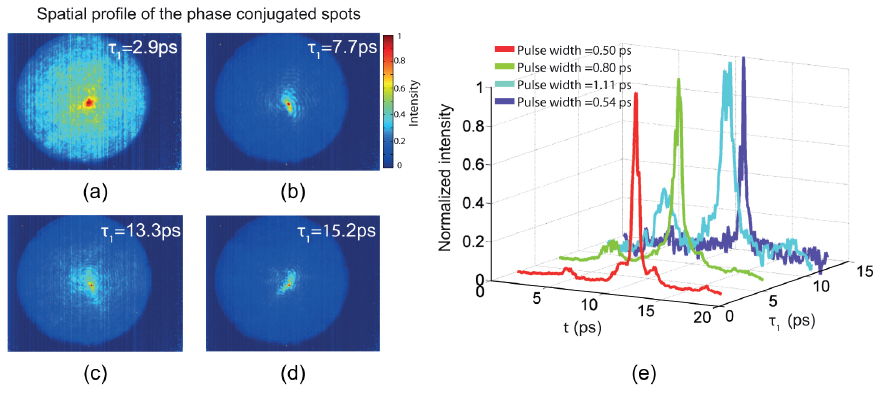}
	\caption{Spatio-temporal focusing through a multimode fiber. Spatio-temporal characterization of the reconstructed phase conjugated spot. (a)-(d) Intensity of the spatial profile and (e) temporal profile of the phase conjugated spots generated from the reconstructed holograms recorded at preselected times. Figure adapted from \cite{morales2015delivery} with permission. Copyright 2015 Optical Society of America}
	\label{fig:STfocus} 
\end{figure}

 The simultaneous spatial and temporal focusing of ultrashort pulses has been achieved in MMFs by employing time-gated interferometry and subsequent digital phase conjugation \cite{morales2015delivery}, as shown in Fig.~\Ref{fig:STfocus}. Specifically, only modes with similar group delays have been excited through selective phase conjugation. These modes only follow those nearby paths in the fiber that interfere constructively at the distal end of the fiber, both in a focused spot (spatially) and with minimal distortions (temporally) \cite{morales2015delivery}. In this way, temporal broadening due to modal dispersion could be reduced by a factor of around 30 (the effects of chromatic dispersion remain). An alternative technique to mitigate modal dispersion that has already been successfully employed for the case of disordered media, is the optimization of a non-linear signal (like two-photon fluorescence) behind the complex medium \cite{katz_focusing_2011}. In this case, both spatial and temporal distortions can be corrected by manipulating only the spatial degrees of freedom of the incident wavefront.

 Instead of spatial wavefront shaping, spectral or temporal shaping of broadband femotosecond pulses has been employed for spatio-temporal focusing through multimode fibers \cite{liu2018space, velsink2021spatiotemporal}. Since the temporal speckle fields at different MMF output
locations are uncorrelated, spectral pulse shaping at the fiber input can focus a short pulse at a specific output spatial location, while keeping the field at other output locations speckle-like \cite{liu2018space}. While the input pulse is in a single spatial mode, light scrambling in a MMF, temporal shaping of the input pulse, and nonlinear optical feedback are utilized to focus the output pulse to a predefined location \cite{velsink2021spatiotemporal}.

\subsubsection{Total temporal impulse response}

While the last subsection describes temporal focusing of a pulse to a single spatial channel, here we show how to control the total temporal impulse response of a multimode fiber. The global enhancement or suppression of the total light intensity exiting a multimode fiber at arbitrary delay times can be achieved with the time-resolved transmission matrix $T(t)$ of a MMF \cite{xiong_long-range_2019,mounaix2019control}.  

The maximum (or minimum) power that can be delivered at a given time $t$ is determined by the largest (or smallest) eigenvalue of the Hermitian ``temporal focusing matrix'' $T(t)^\dagger T(t)$ and the corresponding spatial wavefront of the pulse that achieves this, is given by the corresponding eigenvector. The enhancement (or suppression) of power delivery at targeted times is significantly improved by the presence of long-range spatio-temporal correlations in the transmission matrix of a MMF \cite{xiong_long-range_2019}. 

The positive correlations among spatial channels enable a global enhancement of transmitted energy at a selected arrival time by shaping the incident wavefront. Experimentally, a higher enhancement is obtained when the target time is before or after the mean arrival time, as a result of stronger long-range correlations. Theoretically, a quantitative relation between spatio-temporal correlations and the time-dependent enhancement of transmitted power is established in a MMF with strong mode coupling \cite{xiong_long-range_2019}. 

\subsubsection{Principal modes}
	Since the time-resolved transmission matrix $T(t)$ is just the Fourier transform of the frequency-resolved transmission matrix $T(\omega)$, the same amount of control is also possible through $T(\omega)$ directly. Quite remarkably, however, the availability of $T(\omega)$ at two neighboring frequency values already enables efficient temporal control. Consider here the situation, when we launch a time-harmonic input field $E_{\rm in}$ into a MMF and change the input frequency from $\omega$ to $\omega+d\omega$. 
	An interesting question to ask is whether modal dispersion that leads to the change of the output pattern ($E_{\rm out}$) at the distal end of the fiber with input frequency can be compensated through a proper choice of the input field $E_{\rm in}$. Similar to the transmission eigenchannels, that are invariant with respect to propagation through the fiber, we are looking here for states that are invariant in their output pattern with respect to a change of the input frequency (the input field stays the same). 
	
	\begin{figure}\centering
		\includegraphics[width=0.8\linewidth]{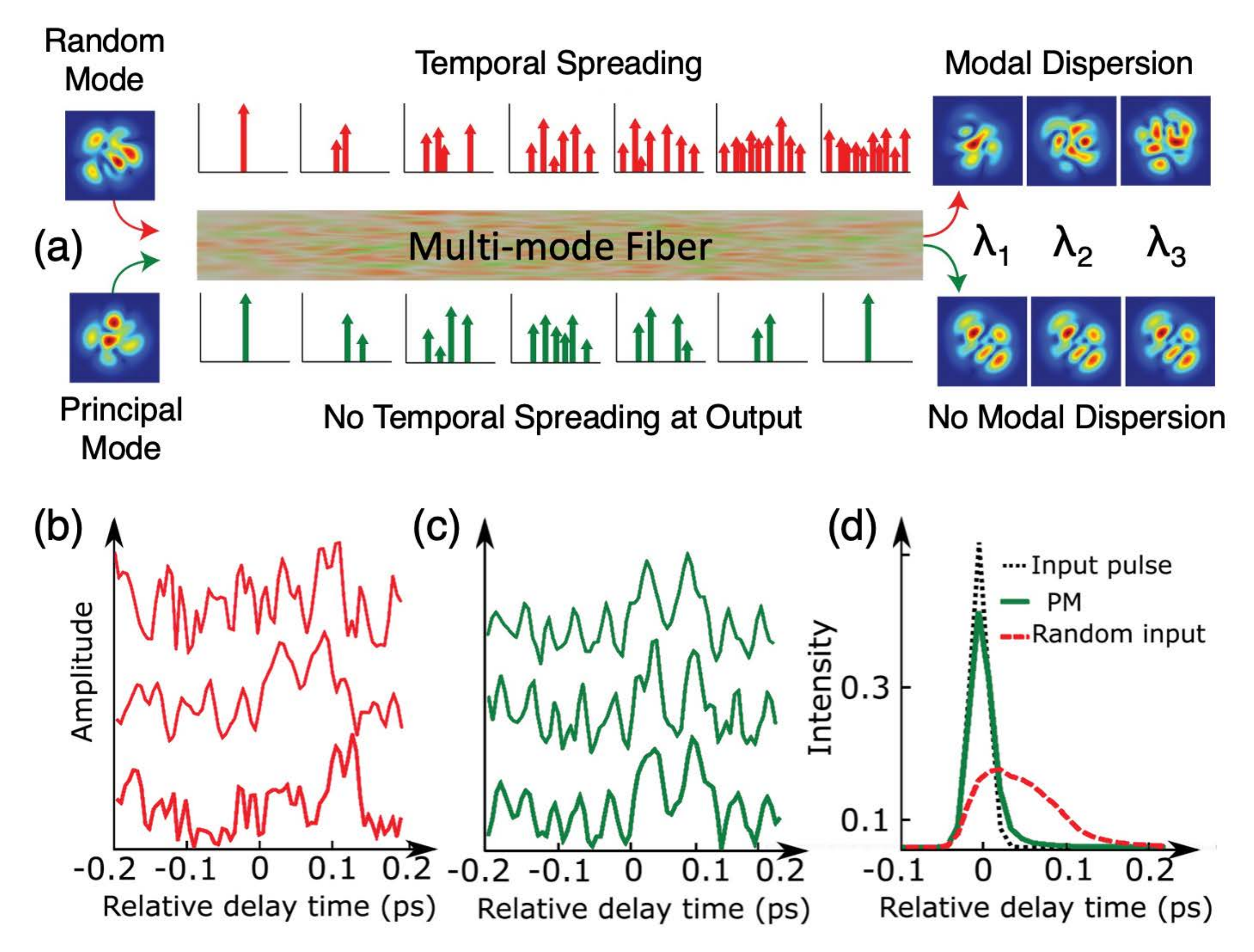}
		\caption{Principal modes in MMFs. (a) Illustration of the concept: an arbitrary input pulse launched into a MMF will spread in time such that a stretched pulse or even several of them will arrive at the output. The reason for this spreading is modal dispersion, resulting in the indicated wavelength-dependent output speckle pattern. Principal modes, on the contrary, are designed to be free of modal dispersion with a wavelength-independent output speckle pattern. A single input pulse launched into them will also arrive as a single output pulse with negligible broadening and shape distortions. Figure adapted from \cite{carpenter2015observation} with permission from Springer Nature. (b),(c) Time-dependent variation of the output field in three independent speckle grains when a pulse is launched into a random mode (b) or in a principal mode (c), respectively. The temporal traces of these individual spatial channels differ significantly from each other for the random input, but are almost identical for the case of a principal mode. (d) Intensity of a Gaussian input pulse integrated over the entire fiber cross-section (black dotted line) and that of the corresponding output pulses when they propagate in a random spatial profile (red dashed line) or in a PM (green solid line). The reduced dispersion of the PM is clearly visible, even though a MMF with strong mode coupling is used here. Figure adapted from \cite{xiong2016spatiotemporal}.}
		\label{fig:pms} 
	\end{figure}
	
	Formalizing this property results in the following defining equation for the so-called ``principal modes'' (PMs) of a MMF \cite{fan_principal_2005}: \begin{equation}
    \label{eq:PM}
    T(\omega+d\omega)E_{n}^{\rm PM}=\alpha_n T(\omega)E_{n}^{\rm PM}.
    \end{equation} 
    The complex eigenvalue $\alpha_n$ in this equation is a global factor, indicating that the frequency shift $\omega\rightarrow\omega+d\omega$ may change the output field in overall phase and brightness, but not in its spatial structure, see Fig.~\ref{fig:pms} (a). Quite instructively, a Taylor expansion of $T(\omega+d\omega)$ in the above equation shows that the PMs are the eigenstates of the following operator: 
    \begin{equation}
    \label{eq:Q}
    Q=-iT(\omega)^{-1}\partial_\omega T(\omega).
    \end{equation}
    Already the derivative with respect to frequency indicates that this operator must have something to do with time. Indeed, a rigorous scattering theory analysis carried out by Eisenbud, Wigner and Smith (EWS) \cite{eisenbud_notitle_1948,wigner_lower_1955,smith_lifetime_1960} shows that the eigenvalues of this ``time-delay operator'' measure the times associated with the operator's eigenstates. More specifically, these ``time-delay'' eigenvalues  correspond to the time lag between the envelopes of incoming and outgoing wavepackets launched in such a time-delay eigenstate at the center frequency $\omega$. (This correspondence only holds for wavepackets that are spectrally narrower than $\omega_c$.) Provided that the transmission matrix $T(\omega)$ is a unitary matrix, the EWS time-delay operator $Q$ is Hermitian, resulting in an orthogonal set of PMs as eigenstates and in real eigenvalues. Since these real time-delay eigenvalues are inversely proportional to the wave packets' group velocity, they are also referred to as ``group delays''. While arbitrary input states in a MMF are characterized by several group delays, the PMs are associated with a single group delay only and are thus robust with respect to a small change in the input frequency: $\omega\rightarrow\omega+d\omega$. 
	
	This spectral robustness of principal modes \cite{shemirani2009principal}, which comes with several advantageous properties for  multiplexing in data transmission \cite{ho2011statistics,juarez2012perspectives,franz2012experimental,prasad2018impact}, was first demonstrated experimentally in MMFs with weak mode coupling \cite{carpenter2015observation}, where PMs may exceed random input modes by several orders of magnitude in terms of their frequency stability. When the input wavefront is set to one of the PMs, the output field pattern decorrelates very slowly with frequency detuning. The spectral correlation width of a PM, given by the full-width-at-half-maximum of the transmitted field correlation function, is much broader than that of random inputs ($\Delta \omega_c$).  PMs also survive in the more challenging case of MMFs with strong mode coupling \cite{xiong2016spatiotemporal}, where their spectral correlation width remains about a factor of 2 larger than that of random input fields. Quite remarkably, even in the presence of mode coupling, a pulse with a spectral width exceeding the spectral correlation width $\Delta\omega_c$ will arrive at the distal end of the fiber not only without significant  broadening and shape distortions, but with a spatial output pattern that stays the same throughout the entire pulse duration, see Fig.~\ref{fig:pms} (b-d). This feature illustrates the decoupling of spatial and temporal variations of a PM's output field which enables a full spatio-temporal control of pulse transmission \cite{xiong2016spatiotemporal}.  

    \textcolor{black}{Principal modes have meanwhile also been used as a basis in which the multi-spectral transmission matrix $T(\omega)$ can be conveniently parametrized. Using the principal modes' spectral stability, it has been shown that only a few measurements at selected frequencies are sufficient to approximate $T(\omega)$ inside a spectral window that approaches that of the spectrally most stable principal modes in bandwidth \cite{lee2023efficient}. }
 
	The fact, that both the temporal focusing operator $T(t)^\dagger T(t)$ and the EWS time-delay operator $Q(\omega)$ feature eigenstates that are associated with a well-defined arrival time at the fiber output, suggests that these two operators are closely connected. Indeed, under the assumption of input pulses that are centered around the frequency $\omega$ and that have a spectral width $\Delta \omega$ that is shorter than the spectral correlation width $\Delta \omega_c$, it can be shown \cite{melchard2020time} that the following matrix identity holds:
    \begin{equation}
    T(t)^\dagger T(t)\approx 1-(\Delta\omega/2)^2[1-Q(\omega)]^2.
    \end{equation}
    This relation demonstrates that $T(t)^\dagger T(t)$ and $Q(\omega)$ share a common eigenbasis and that the EWS time-delay operator, indeed, acts as a temporal focusing operator for spectrally narrow pulses.
\subsubsection{Super- and Anti-Principal Modes}	
	A further increase of the spectral stability of these time-delay eigenstates can be obtained through the concept of ``super-principal modes'', which outperform the stability of conventional PMs \cite{ambichl_super-_2017}. The idea here is to use, instead of the derivative $\partial_\omega$ in the time-delay operator $Q(\omega)=-iT(\omega)^{-1}\partial_\omega T(\omega)$, a finite difference quotient: 
     \begin{equation}
     \tilde{Q}(\omega,\Delta\omega)=-iT(\omega)^{-1} \frac{T(\omega+\Delta\omega)-T(\omega)} {\Delta\omega}
 \end{equation} 
 which ensures that the eigenstates of this matrix remain perfectly correlated in their output patterns when shifting the frequency from $\omega$ to $\omega+\Delta\omega$. As it turns out, however, when this shift $\Delta\omega$ is larger than the spectral correlation width $\Delta\omega_c$, the spectral field correlation $C_E(\Delta\omega)$, see Eq.~(\ref{eq:C_E}), 
 between the normalized output fields cannot be perfect in the entire frequency interval  $[\omega,\omega+\Delta\omega]$. The reason is simply that, to maintain the correlation, an input state would have to be a simultaneous eigenstate of all operators $\tilde{Q}(\omega,\Delta\omega)$ within this frequency interval. Since the corresponding operators do, however, not commute, such joint eigenstates do in general not exist. Instead, one can use a numerical optimization routine to find input states that maximize the frequency-integrated correlation, $\int |C_E(\Delta\omega)| \, d\omega$,
within a given spectral interval. Such states are called super-PMs as they perform better than conventional PMs in terms of the spectral stability of their output spatial field profiles \cite{ambichl_super-_2017}. 

Inverting this concept also allows one to create anti-PMs, which have a minimal spectral correlation width and the output field profiles are considerably more sensitive to a frequency change than typical input wave fronts. Decomposing them in the PM basis reveals that the super-PMs are formed via interference of principal modes with close delay times, whereas the anti-PMs are a superposition of principal modes with the most-different delay times available in the fiber. Therefore, PMs constitute the natural basis not only to capture the dynamical aspects of light scattering, but also for synthesizing novel types of states with unique spatial, temporal and spectral characteristics.

In the same way as PMs are invariant with respect to a frequency shift, also states in a fiber can be found that are invariant with respect to a shift in another parameter that the transmission matrix depends on – such as those parameters $\alpha$ that describe potential deformations of the fiber when pressing or bending it. As it turns out, such ``deformation principal modes'' \cite{matthes_learning_2021} are the eigenstates of the generalized EWS operator  $Q_\alpha=-iT(\omega,\alpha)^{-1}\partial_\alpha T(\omega,\alpha)$ in which the frequency derivative $\partial_\omega$ from the conventional EWS time-delay operator is replaced by a derivative with respect to the deformation parameter $\partial_\alpha$ \cite{ambichl_super-_2017,horodynski_optimal_2020}. The perturbation-insensitive channels that can be obtained as the eigenvectors of this operator have meanwhile also been implemented experimentally  \cite{matthes_learning_2021}.

	\subsection{Polarization of transmitted light}
	\label{sec:vectorTM}

  	The vectorial nature of electromagnetic waves plays an indispensable role in light-matter interaction, optical transmission and imaging. A control over the polarization state of light has been widely exploited in single molecule detection, nanoplasmonics, optical tweezers, nonlinear microscopy and optical coherence tomography. However, a well-prepared state of polarization can be easily scrambled when transmitting through a fiber. Polarization maintaining single-mode fibers have been widely used, but they operate with a fixed polarization state. An arbitrary polarization state can be obtained for transmitted light by either modifying the fiber configuration or tailoring the input polarization state. 

 In a multimode fiber, the modes generally do not maintain a linear polarization state due to spin-orbit interaction. Yet the circular polarization state can be maintained in most modes of ideal cylindrical step-index fibers. However, fiber deformation, residual birefringence or strain lead to polarization scrambling. When light is launched into an individual mode of the MMF, it will spread to other modes, each of which will experience distinct polarization change. Thus the output polarization state varies from one mode to another \cite{xiong2018complete}. 

One way to avoid depolarization is to use polarization-maintaining MMFs \cite{ahmad2017polarization, chen2018design}. However, if a fiber has several hundreds of guided modes, maintaining the polarization states for all of them is challenging. Moreover, the output light from all modes of a polarization-maintaining MMF will have an identical polarization state, e.g., linear or circular polarization. It is difficult to have arbitrary polarization states for the transmitted light, especially making the polarization state vary from one mode to another. It is also difficult to control the polarization states of all modes by adjusting the fiber configuration, which would require a lot of bending and twisting of the MMF.  Below we will show how to control the polarization state of light transmitted through a MMF by wavefront shaping. 

\subsubsection{Full polarization control}

\begin{figure}\centering
		\includegraphics[width=0.8\textwidth]{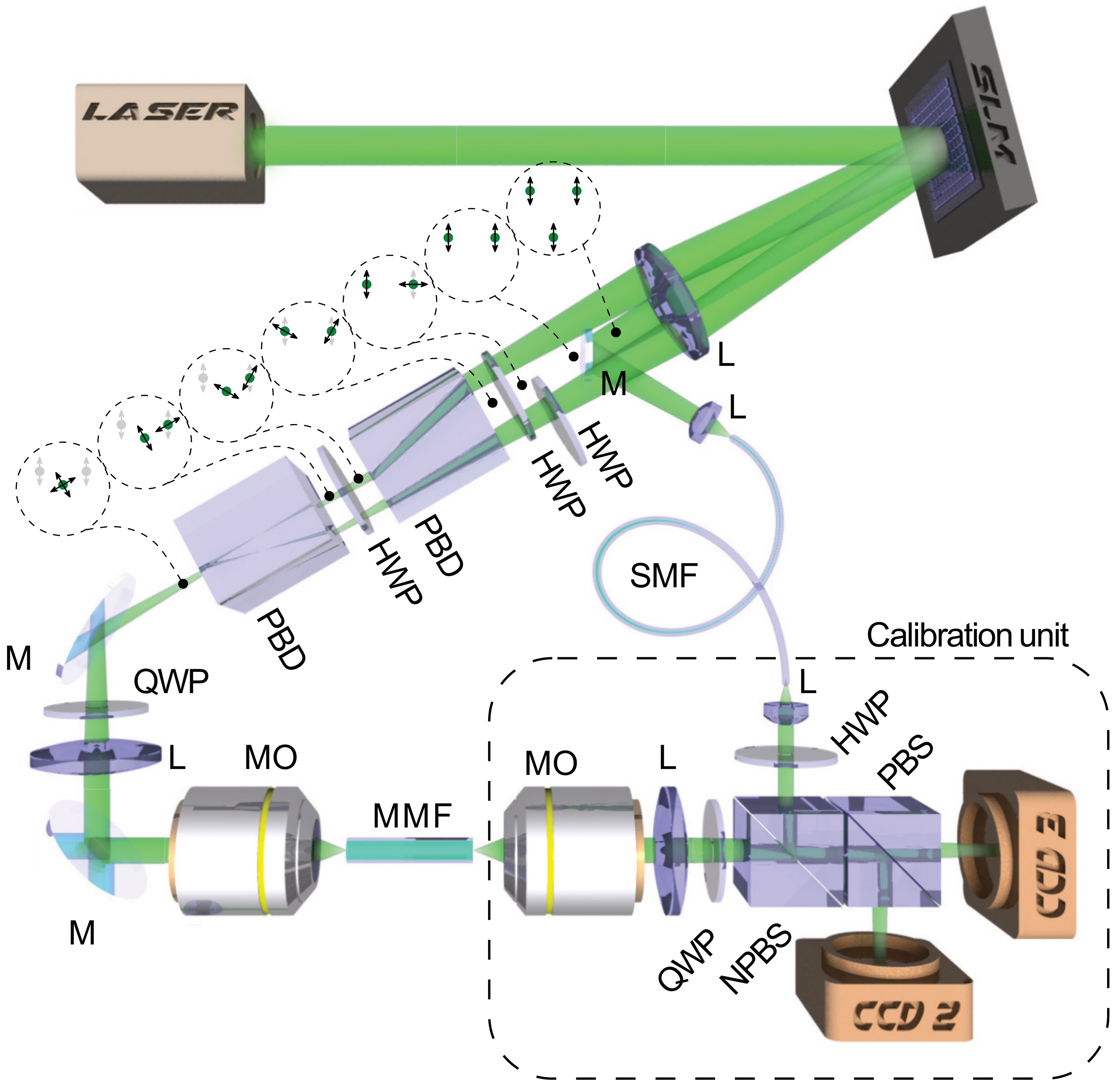}
	\caption{\label{fig: Full_polar_control} Experimental geometry for polarization complete TM measurements. Relying on off-axis configuration, it utilizes different carrier frequencies as a simple method for multiplexing the complex fields in the farfield region of the SLM. Two linearly-polarized beams are diffractedoff the SLM beams, and one of them has a polarization rotation before the two beams are merged again, giving an independent control for both input polarization states coupled to the fiber. The third beam is chosen to be a static reference for the phase-shifting TM measurements and therefore directed to the calibration module by the single-mode fiber. Output optical fields at the distal end of the fiber are split by a polarization beam splitter and recorded with two cameras, enabling polarization-resolved TM measurements. Such a scheme employed for wavefront shaping can gain full control of complex field amplitude and polarization of the output light and allow the study of polarization effects in multimode fibers. The following abbreviations are used: L-lens, M-mirror, MO-microscope objective, HWP - half-wave plate, QWP - quarter-wave plate, PBS-polarizing beam splitter, \textcolor{black}{PBD - 
 polarization-dependent beam displacer,} NPBS-non-polarizing beam splitter, SMF - single-mode fiber. Figure adapted from Ref.~\cite{ploschner2015} with permission.
	}
    \label{full_polarization_scheme}
\end{figure}

The simplified yet highly practical scalar representation of the waveguiding modes of optical fibers, described in section 2.A, implies polarization degeneracy.
Every mode can be defined in two orthogonal polarization states, thus the TM is complete only when both states are measured. Such measurements require access to orthogonal polarization states at both the proximal and distal end of the fiber, allowing for generating input fields and recording output responses in a polarization-resolved manner. Figure \ref{full_polarization_scheme} represents the optical scheme for wavefront shaping through the MMF based on complete TM acquisition. This setup utilizes standard digital holography technique of superimposing the desired hologram with a phase grating on SLM, which allows redirecting the beam off the optical axis and therefore center the Fourier plane towards the carrier frequency given by the modulation period of the applied grating. Relying on such an off-axis configuration, it utilizes different carrier frequencies as a simple method for multiplexing the complex fields in the modulator's farfield region.  

The single-frequency, linearly polarized laser light illuminates the SLM. The displayed pattern, formed as the superposition of three carrier frequencies, diffracts light into distinct directions, propagating at different angles from the SLM. Focused by the lens, the beamlets form focal points in different regions of the Fourier plane. One of the spots is chosen to serve as a static reference for phase-shifting measurements of TM coefficients and transferred to the calibration module via the single-mode fiber, the other two are utilized for wavefront shaping. Using a half-wave plate, the polarization state of one of the remaining beams is then rotated by 90°. Two beamlets, now orthogonally polarized, can be merged via beam displacers and coupled to the proximal fiber end through the microscope objective with a tube lens. The distal facet of the fiber is imaged on two cameras placed behind the ports of the polarizing beamsplitter so that output fields carried by both polarization states can be measured independently. 

The TM measurement procedure in this geometry can be performed in the representation of diffraction-limited spots for the input fields, since the focused spots can be easily generated experimentally using an SLM or DMD as depicted in Fig.~\ref{fig:Lee_hologram}. During the measurement, focused spots are generated at different positions of the input fiber facet. The resulting fields at the output facet are superimposed with the reference beam, delivered via the single mode fiber, and recorded by the camera to provide both amplitude and phase information \cite{turtaev2017comparison}.  In this case, however, TM measurements have to be performed for two input sets of fields corresponding to both orthogonal polarization states.

Complete, polarization-resolved TM measurements, linking amplitude, phase and polarization state for the sets of input and output fields, enables one to generate any complex vector beam allowed to propagate in the optical fiber \cite{ploschner2015seeing}. Moreover, control of both input polarization states allows for near-perfect fidelity of the generated beams \cite{gomes2022near}.

Such a scheme for polarization control can be utilized with both LC-SLMs \cite{ploschner2015seeing, cizmar2012exploiting} as well as DMDs \cite{mitchell2016high}. Moreover, this exemplary case can be simplified to using just one beam displacer and one camera \cite{gomes2022near} for practical convenience in applied experiments.

In addition to the aforementioned TM, including two orthogonal polarizations, the TM based on the Stokes parameters for transmitted light is introduced  to describe depolarization effects in a MMF  \cite{fan2020polarization}. Such TMs have been used for complete control of polarization states of light at one or multiple focal spots at the distal end of a MMF.

\subsubsection{Polarization conversion}

For particular applications, right and left circular polarization states of input optical field may be used to prevent or minimize polarization change through short pieces the fiber. In contrary, the polarization mixing in MMFs due to pronounced spin-orbit interaction when working with linearly polarized modes, can be intentionally utilized to promote cross-talk between polarization states, and, gain the complete control of the output polarization of the light without the need of independent control of two polarization components at the proximal end. 

Furthermore, random mode coupling and polarization scrambling, caused by fiber imperfections and external perturbations, further couple the spatial degrees of freedom and the polarization degrees of freedom, allowing one to control the polarization state of the transmitted light by shaping the spatial wavefront of an incident beam. For illustration, we consider a fiber with only two guided modes, and the incident light is linearly polarized in the horizontal (H) direction. Without mode coupling, the relative phase of input fields in the two modes does not affect the output polarization state of either mode. However, with mode coupling, the output field of one mode also depends on the input field of the other. For example, the vertical (V) polarization of mode 1 has contributions both from the input field in mode 1, which is converted to the vertical polarization, and from the input field in mode 2, that is coupled to the vertically polarized mode 1. The interference of these two contributions can be made constructive or destructive by varying the relative phase of input fields to the two modes, which will modify the vertically polarized output field in mode 1. This degree of control is effective only when there is mode mixing in the fiber. 
 
    Beyond a qualitative description above, a quantitative evaluation of the degree of control with strong mode and polarization coupling is given below. Let us consider a fiber with $M$ guided modes at frequency $\omega$, each mode has a two-fold degeneracy which corresponds to two orthogonal polarizations. Without loss of generality, the horizontal (H) and vertical (V) linear polarizations are used below as the basis to describe the vector transmission matrix (TM) of a MMF at a single frequency $\omega$ 
	\[
	T=
	\left[ {\begin{array}{cc}
			T_{\rm HH} & T_{\rm HV} \\
			T_{\rm VH} & T_{\rm VV} \\
	\end{array} } \right]
	\] where $T_{\rm HH}$ (or $T_{\rm VH}$) is a $M \times M$ matrix that represents the horizontal (or vertical) components of transmitted fields when the input light is horizontally polarized. 
	
	A concatenated fiber model was built to calculate the TM with random mode and polarization coupling in a MMF \cite{ho2011statistics}. In this model, the fiber is divided into many small segments. Within each one of them, light propagates without polarization or mode coupling such that the transmission matrix of each segment is a diagonal matrix in mode basis. Each matrix element accounts for the phase delay and the loss of an individual mode. Between adjacent segments, fiber modes with different spatial profiles and polarization states are randomly coupled. Adjacent modes with similar propagation constants tend to couple more easily, thus the coupling can be described by a banded random matrix. The bandwidth gives the range of fiber modes that are coupled, and the magnitude of matrix elements within the band reflects the strength of coupling. In the case of strong mode coupling, where each segment is already shorter than a transport mean free path, every mode is coupled to all other modes at the interface between two adjacent segments. The random unitary coupling matrices are drawn from the circular unitary ensemble (CUE). The total TM is the product of the propagation matrices for all segments and the coupling matrices in between them.  
	
	\begin{figure}[htbp]
		\centering
		\includegraphics[width=0.7\linewidth]{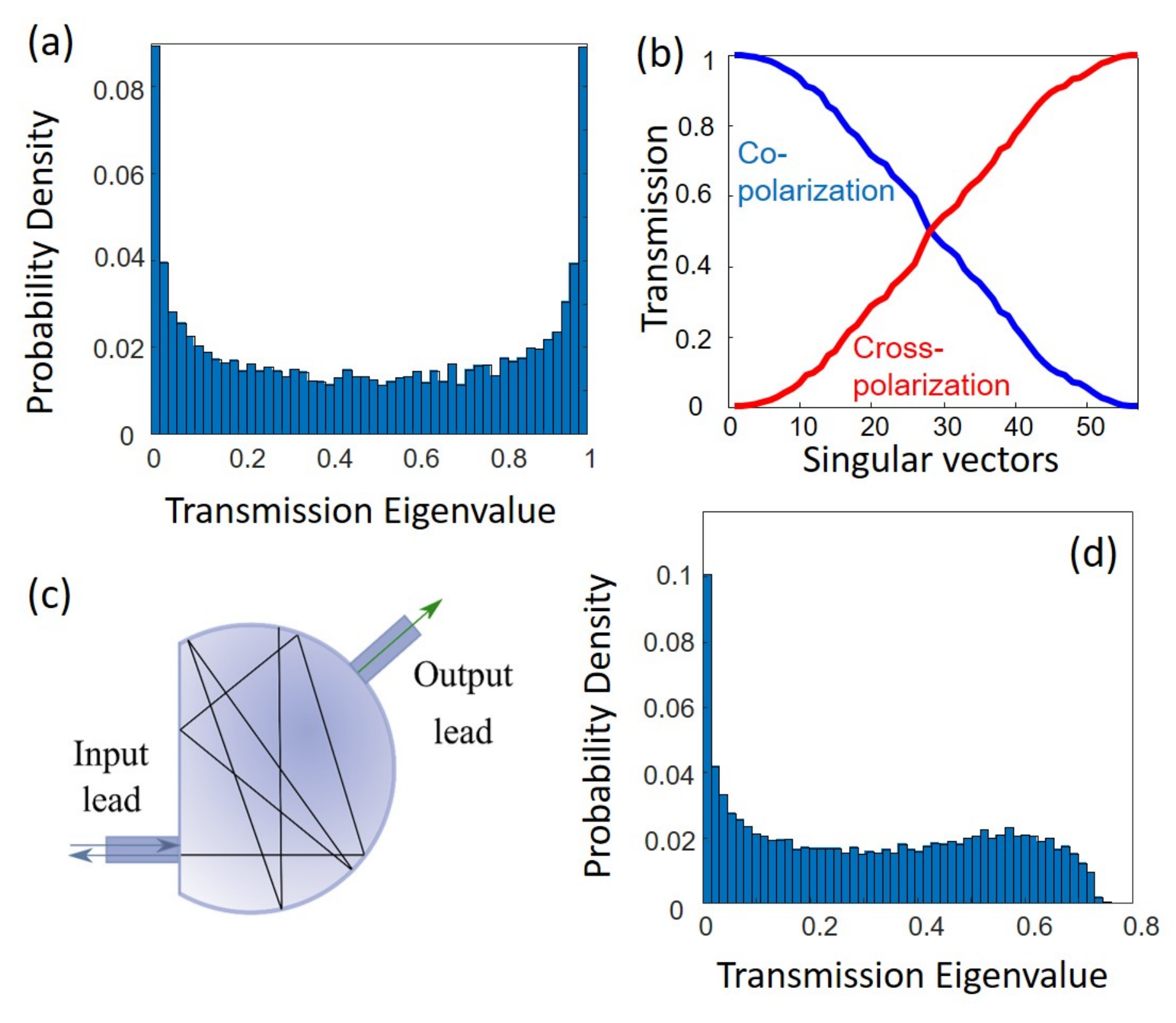}
		\caption{Polarization of transmission through a multimode fiber. The fiber has strong mode coupling and polarization mixing. The number of guided mode is much larger than 1.
			(a) Probability density of transmission eigenvalues for horizontally polarized input and output light  has a bimodal distribution when the fiber loss is negligible. 
			(b) Transmission of horizontal (H) and vertical (V) polarization components for individual eigenvectors, which are numbered by their eigenvalues from high to low. The decrease of H is accompanied by an increase of V, and their sum remains 1. 
			(c) Eigenvalue distribution in (a) is identical to that of reflection eigenvalue in a chaotic cavity with two leads, drawn schematically.  
			(d) Eigenvalue distribution is modified by fiber loss. Figure adapted from Ref.~\cite{xiong2018complete}, licensed under a Creative Commons Attribution (CC BY) license. }
		\label{fig:polarization}
	\end{figure}
	
	Let us first ignore the loss in a MMF which has strong strong mode and polarization coupling. The Hermitian transmission matrix product $T_{\rm HH}^{\dagger}\, T_{\rm HH}$ has $M$ real transmission eigenvalues $\tau_{\rm HH}$ in the range of 0 and 1. The largest eigenvalue determines the maximum energy that can be retained in the horizontal polarization after propagating through the fiber. Due to strong mode and polarization mixing, the eigenvalue density $P(\tau_{\rm HH})$ has a bimodal distribution for $M \gg 1$, as shown in Fig.~\ref{fig:polarization}(a). The peak at $\tau_{\rm HH} \simeq 1$ corresponds to the transmitted light retaining the input polarization (H), although the spatial mode contents at the output differ from those of the input. The other peak at $\tau_{\rm HH} \simeq  0$ corresponds to 100$\%$ conversion to the orthogonal polarization (V), in spite of strong polarization mixing in the fiber. As $\tau_{\rm HH}$ decreases from 1 to 0, the percentage of transmission in the horizontal polarization drops, while that in the vertical polarization rises, as seen in Fig.~\ref{fig:polarization}(b). 
	
	In a MMF with strong spatial- and polarization-mode coupling, the analytical expression for the probability density function (PDF) is $P(\tau_{\rm HH}) = 1/\pi \sqrt{\tau_{\rm HH} \ (1 - \tau_{\rm HH})}$. It is identical to the expression found already earlier \cite{baranger1994mesoscopic, jalabert1994universal, rotter2017light} for wave transmission in a lossless chaotic cavity [see the symbols and lines in Fig.~\ref{fig:polarization}(a)], revealing the analogy between a MMF with random spatial- and polarization-mode coupling and the scattering of scalar waves through a chaotic cavity with two leads, as drawn schematically in Fig.~\ref{fig:polarization}(c). The fiber transmission in the same polarization as the input is analogous to the reflection from a chaotic cavity, in the sense that light exits the cavity via the same lead. Similarly, the fiber transmission to the polarization that is orthogonal to the input polarization is analogous to the transmission into the other lead of the chaotic cavity. The PDF $P(\tau_{\rm VH})$ for the eigenvalues $\tau_{\rm VH}$ of $T_{\rm VH}^{\dagger}\, T_{\rm VH}$ has the same functional form as $P(\tau_{\rm HH})$. One peak at $\tau_{\rm VH} \simeq 1$  means $100\%$ conversion of input horizontal polarization to output vertical polarization, the other peak at $\tau_{\rm VH} \simeq 0$ indicates the input and output polarizations are identical.
	
	Therefore, depolarization of light transmitting through a MMF with strong polarization mixing can be suppressed by coupling light into the eigenvector of $T_{\rm HH}^{\dagger} \, T_{\rm HH}$ with the maximum eigenvalue $\tau^{\rm max}_{\rm HH}$. The ensemble average of maximum eigenvalue, $\big \langle\tau^{\rm max}_{\rm HH} \big \rangle = 1 - 1 / (M^2 + 1)$, approaches unity rapidly with the increase of fiber modes $M$. The polarization extinction ratio (PER) of transmitted light is  $\big \langle \tau_{\rm max} \big \rangle /(1 -  \big \langle \tau_{\rm max} \big \rangle) = M^2$.  
	
	If loss in the fiber is non-negligible, the bimodal distribution of eigenvalue density will be modified. As shown in  Fig.~\ref{fig:polarization}(d), the peak at eigenvalue close to 1 is reduced and shifts to smaller values \cite{chiarawongse2018statistical}. However, the peak near 0 remains in the presence of strong loss. If the input light is horizontally polarized, by coupling it to the eigenvector of $T_{\rm VH}^{\dagger}\, T_{\rm VH}$ with eigenvalue close to 0, the transmitted light has a vanishing vertical component. Hence, depolarization is avoided, even though part of incident light is lost. Similarly, all transmitted light can become vertically polarized by exciting the eigenvector of $T_{\rm HH}^{\dagger} \, T_{\rm HH}$ with eigenvalue close to 0.

	\begin{figure}[htbp]
		\centering
		\includegraphics[width=0.8\linewidth]{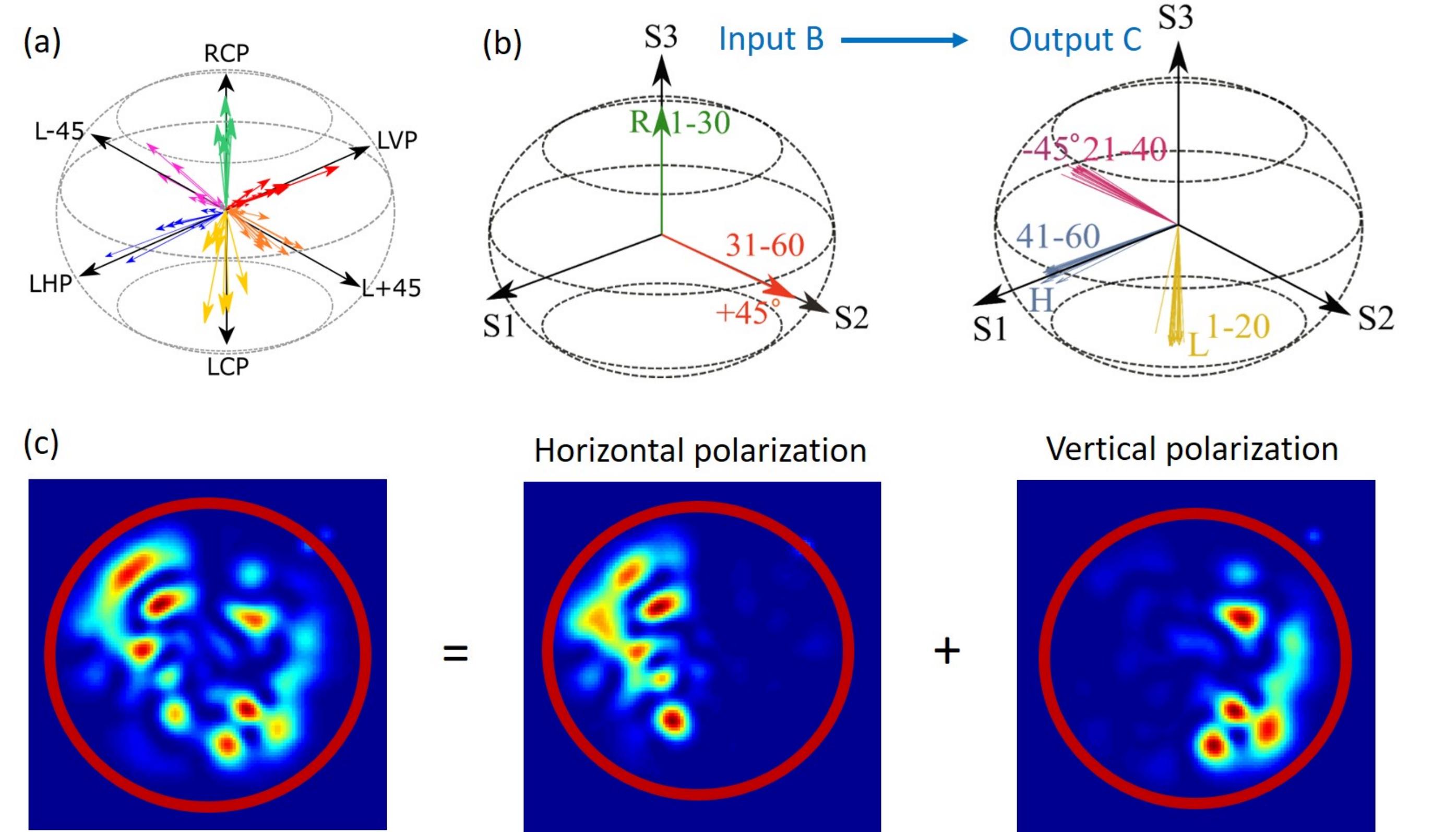}
		\caption{Polarization conversion of transmission through a multimode fiber with strong mode and polarization mixing. 
			(a) Wavefront shaping of an incident monochromatic light transforms the input linear horizontal polarization (LHP) to various output polarizations such as LHP, LVP (linear vertical polarization),  L$\pm$45 (linear polarization in $\pm 45^{\circ}$ direction), RCP (right-hand circular polarization), LCP (left-hand circular polarization). Arrows in the Poincar{\' e} sphere represent the polarizations of individual spatial modes at the fiber output, the arrow length indicates the intensity of each mode.
            (b)  Transformation of an input polarization state B (left) to the output polarization state (C) by exciting the maximum-transmission eigenchannel of $T_{\rm BC}$ 
			(c) Output intensity pattern is the sum of its horizontal and vertical polarization components, revealing the transmitted field in the left half of the fiber facet is horizontally polarized, and the right half vertically polarized. Figure adapted from \cite{xiong2018complete},  licensed under a Creative Commons Attribution (CC BY) license. }
		\label{fig:polcontrol}
	\end{figure}
	
	Strong mode and polarization mixing in a MMF makes it possible to create arbitrary polarization states for all output modes by modulating the spatial wavefront of a linearly polarized beam \cite{xiong2018complete}. For example, complete transformation from horizontal linear (H) polarization to right-hand circular (R) polarization is possible by resorting to the transmission eigenchannels of $T_{\rm RH} = \frac{1}{\sqrt{2}} (T_{\rm HH} - i\, T_{\rm VH})$ for H input and R output. The eigenvalue density $P(\tau_{\rm RH})$ is identical to that of $P(\tau_{\rm HH})$, as long as the fiber completely scrambles light polarization, making all polarization states equivalent. With strong spatial- and polarization-mode coupling and negligible loss, $P(\tau_{\rm RH})$ has a bimodal distribution, and the peak at $\tau_{\rm RH} = 1$ ($\tau_{\rm RH} = 0$) allows a full conversion of horizontal polarization to right (left) circular polarization. As shown by  green arrows in Fig.~\ref{fig:polcontrol}(a), all output modes are right-hand circularly polarized. 
	
	Instead of having the same polarization for all output modes, it is possible to acquire different polarizations for individual modes. Beyond the fiber mode basis, the spatial channels can be represented in real space. For example, the output polarization state A is designed to have the horizontal (H) linear polarization for the spatial channels within the left half of the fiber cross-section, and the vertical (V) polarization in the right half. The transmission matrix $T_{\rm AH}$ is constructed by concatenating one half of $T_{\rm HH}$ and the other half of $T_{\rm VH}$. The conversion of inpu polarization H to output A is realized by exciting the maximum eigenvector  of $T^\dagger_{\rm AH} T_{\rm AH}$, as shown in Fig.~\ref{fig:polcontrol}(c)
	
	Thanks to strong spatial- and polarization-mode  coupling, wavefront shaping can transform any input polarization state (e.g., individual spatial modes with distinct polarizations) to arbitrary output polarization state with nearly $100\%$ efficiency.  One example is given in Fig.~\ref{fig:polcontrol}(b) for a MMF with 60 modes. The input polarization state B has right-hand circular polarization (R) for fiber modes 1–30 and linear $+45^{\rm o}$ polarization for modes 31–60. The output polarization state C has left-hand circular polarization (L) for modes 1–20, linear $135^{\rm o}$ polarization for modes 21–40, and horizontal polarization (H) for modes 41–60. The field transmission matrix from input state B to output C is given by $T_{\rm BC}$. If the fiber loss is negligible, the eigenvector of $T_{\rm BC}^{\dagger} T_{\rm BC}$ with the near-unity eigenvalue will provide the input wavefront for perfect conversion of input polarization B to output polarization C. Hence, controlling the spatial degrees of freedom alone can transform a multimode fiber into a highly efficient reconfigurable matrix of waveplates.
	
\begin{figure}[htbp]
		\centering
		\includegraphics[width=0.9\linewidth]{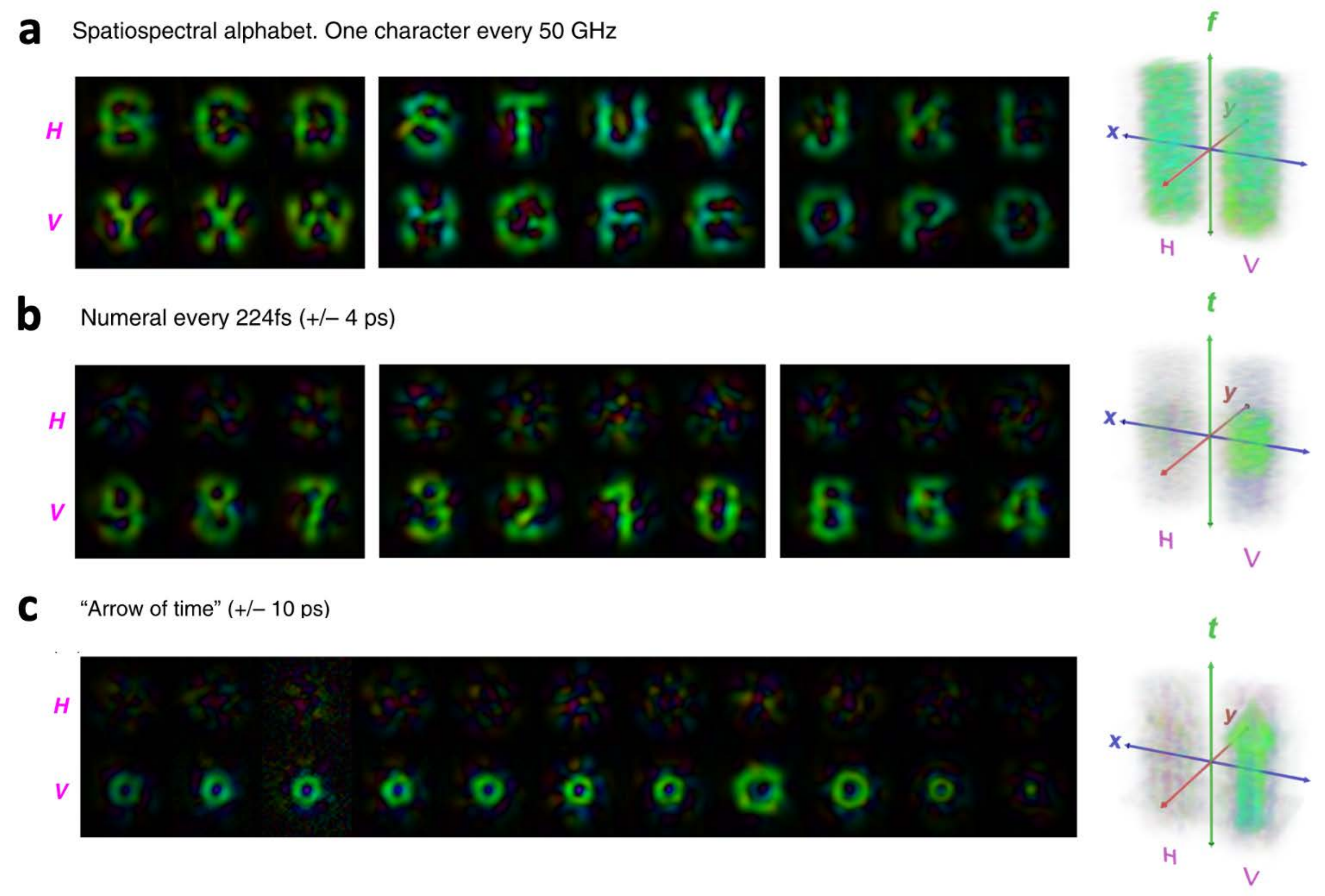}
		\caption{Examples for the simultaneous control of the spatial, spectral/temporal and polarization degrees of freedom of a light field recorded at the distal end of a MMF. (a) A vector spatiospectral state is shown that displays different letters of the alphabet in the horizontal ($H$) and vertical ($V$) polarization component at every 50 GHz. In the right panel the letters are stacked on top of each other. (b,c) Spatiotemporal  states with numerals and an ``arrow of time'' pointing upwards in the time-axis. Figure adapted from \cite{mounaix2020time}, licensed under a Creative Commons Attribution 4.0 International License. }
		\label{fig:arrowoftime}
	\end{figure}

	The above discussion of polarization control is restricted to monochromatic light. For broadband optical pulses, polarization control can be combined with temporal control by wavefront shaping. This has been achieved by measuring the time-resolved vector transmission matrix \cite{mounaix2019control}. The total transmission is enhanced for arbitrary delays and polarization states. These ideas have also been carried further to achieve a full control of all of the classical degrees of freedom of a light field simultaneously \cite{Mounaix2020}. Using a multi-plane light converter in combination with a polarization-resolved multi-port spectral pulse shaper, both the two spatial components as well as the spectral/temporal degrees of freedom have been controlled for both polarisations at the distal end of a MMF. Examples of such volumetrically shaped light fields both in the spatio-spectral and the spatio-temporal domain are shown in Fig.~\ref{fig:arrowoftime}.

	\section{Applications}
	\label{sec:application}

This section covers emerging applications of multimode fibers (MMFs). Wavefront shaping enables new applications of MMFs in imaging and endoscopy (sec. 4.A), optical trapping (sec. 4.B), and microfabrication (sec. 4.C). Furthermore, an output speckle pattern of a MMF encodes the information of the input spectrum, temporal pulse shape, polarization state. Hence, a MMF can function as a high-precision spectrometer and snap-shot hyperspectral imager, as described in section 4.D. The MMF has also been used for full-field measurement of an optical signal (sec. 4.E), and serve as a compressive radio-frequency receiver (sec. 4.F). Finally, random fluctuation of light transmission through a MMF is utilized for remote key generation and establishment in section 4.G. 

	\subsection{Imaging and endoscopy}
	Having the possibility to generate a spectrum of desired optical fields at the output of a MMF opens a variety of routes to perform imaging. The most straightforward one can be found in raster-scanning objects point by point with a large number of diffraction-limited foci. 
 Although in the remaining part of this section we focus mainly on such point-scanning approaches, there are other available choices. 
 Imaging can, for example,  be achieved by exposing the sample to an ensemble of completely random speckle intensity distributions, which have been recorded prior imaging has taken place~\cite{Mahalati:13}. The records are then employed in a computational algorithms essentially providing the pseudo-inversion of the linear optical system. Practical implementations of this concept are virtually analogous to that of computational ghost imaging \cite{PhysRevA.78.061802,PhysRevA.79.053840}. Under the no-noise conditions, the algorithms provide images free from background and with the highest contrast of all available spatial frequencies of the image. Moreover, these approaches are compatible with the exploitation of compressive sensing, allowing for reduction of image acquisition time and, with the use of sparsity priors, resolution enhancement~\cite{AmitonovaLSA2020}.  
 In presence of noise, the speckle pattern illumination approaches however result in severe numerical artifacts, which often render resulting images unreliable. 
 Imaging using diffraction-limited foci are known to offer the best resilience to noise in low-photon regimes~\cite{gu2014noise}. 

	The illumination foci are typically distributed across an orthogonal grid and constrained to a single axial plane of choice as well as the zone which the MMF outputs can illuminate. In most cases, the group of the scanning foci is identical to the basis of the output modes which were used for the preceding TM acquisition, hence the selection of the focal plane (the working distance measured from the output facet) is carried over. In cases, where the phase relations between the output modes are known, i.e., the TM has been measured using a reference signal with a flat wavefront, it is straightforward to computationally shift the focal plane by any desired distance\cite{ploschner2015seeing}. The selection of the working distance however affects the available field of view (FOV) as well as the size of the focus, and thereby the spatial resolution of imaging. Directly at the MMF output facet, the resolution is dictated purely by the numerical aperture of the MMF, while the FOV is governed by the size of the MMF's core. Moving the focal plane away from the fiber, these relations first mix and ultimately, at infinite distance, they reverse.   
	The freedom of choice in the focal plane distance brings by itself the possibility to refocus within the sample volume but also to extend the raster-scanning into three-dimensional paths, thereby achieving volumetric imaging. 
	
	Each focus, and therefore a pixel of the acquired image, requires a specific modulation to be applied onto the light coupled into the MMF. The speed of the acquisition (the pixel rate) then derives from the refresh rate of the used light modulator. Depending on the chosen contrast mechanism, the objects residing within the grid of the scanning foci then interact with the light signal producing some measurable response to be detected and expressed as an image once the scanning focus visits all positions of the desired sequence. 
	
	This imaging mechanism is also associated with a true random access feature. In contrast to many implementations of scanning approaches, no moving mechanical parts, e.g., galvo- or piezo-driven mirrors, are involved. In order to change the focus position, one has to apply new settings to each pixel of the used light modulator, inflicting identical delay, regardless of how remote the fresh position form the expired one is. One has therefore always a choice which individual foci of the available sequence will be used and in what order. 

\textcolor{black}{Although this review does not cover fibre bundles and multi-core fibres, they are frequently compared for pros and cons with the MMF-based approaches when considered for a specific application, especially in the domain of imaging. Arguably, MMFs are hard to beat in reaching the most minuscule footprint of the resulting instrument and in many cases also the photon efficiency. There are therefore uniquely suited in photon-hungry imaging modalities in hardly accessible locations. Multi-core fibres and fibre bundles, however offer several important benefits. As the power exchange between the cores is frequently negligible, the transmission matrix (in representation derived from the waveguide's structure) is close to diagonal. The element can therefore be treated as a planar random aberration, thereby offering very robust memory effects, allowing for calibration-free and single-shot image acquisition and much simpler solutions to the problems related to bending. Further, as multi-core fibres are almost free from modal dispersion, they lend themselves to non-linear and mutiphoton imaging modalities without the need of complex and costly technological solutions. These considerations and further prospects of imaging using multi-core fibres can be found in \cite{gigan2022roadmap}.}

	In the following parts we will turn our attention to the specificities and the most promising applications of the most commonly used contrast mechanisms. 
	
	\subsubsection{Fluorescence imaging}
	Fluorescence is a process of energy exchange between light and matter, whereby the absorption of radiation is almost instantly re-emitted at  
	longer wavelengths. Assuming that photobleaching and saturation do not play significant roles, the emission of fluorescent signals from fluorophores (fluorescent chemical compounds) can be considered as being directly proportional to the intensity of the excitation signal.
	Fluorescence is an immensely powerful instrument for life sciences, as through labelling with fluorescent stains and techniques exploiting expressions of florescent proteins, one can separately visualize constituents of cells including genetic molecules, cytoskeleton, membranes and various vesicles.  Further one may selectively visualize specific cell types, e.g., excitatory or inhibitory neurones while monitoring their metabolism and signalling. The development of powerful, more energy efficient and versatile fluorescent labels and proteins goes hand in hand with the development of imaging technologies, together pushing the boundaries of biological research. 
	Endoscopy, based on MMFs is nowadays recognised as a strategy to achieve detailed observations deep inside even the most sensitive tissues such as that of the brain, inside living and fully functioning organism, without inflicting unnecessary damage and affecting its natural behavior. 
	
\begin{figure}[htbp]\centering
			\includegraphics[width=.8\textwidth]{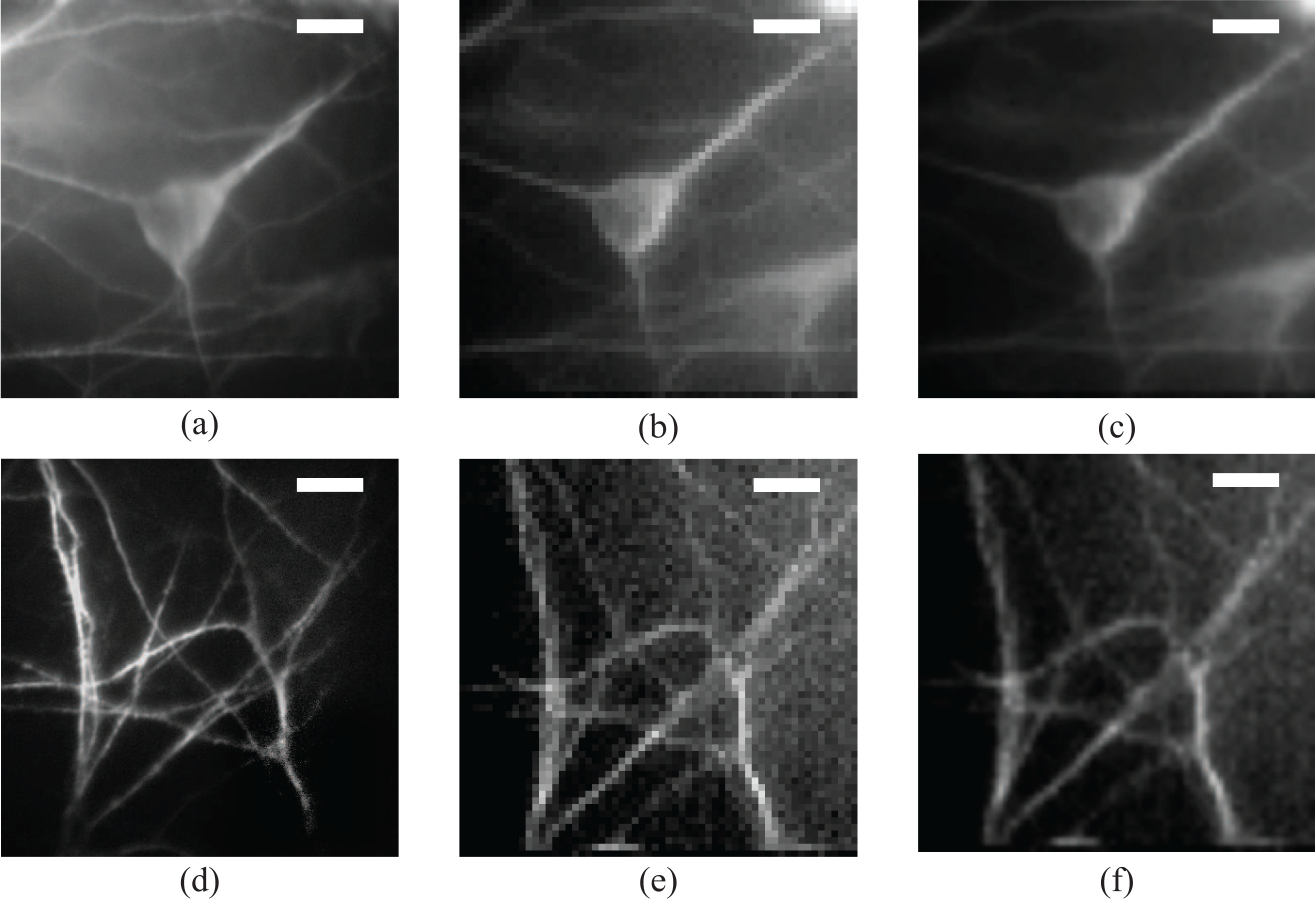}
		\caption{Images of fluorescently stained neuronal cells acquired with the multimode fiber endoscope and compared against conventional images acquired with a microscope objective. Left column, (a) widefield fluorescent image of a single neuron soma and (d) detail of dendrites. Middle column: (b) and (e) direct stitched image as acquired from the fiber. Right column: image from the fiber resampled and filtered so that the pixelation induced by the scanning acquisition is overcome. Highly detailed images of the neuronal soma and the dendritic network can be resolved by the fiber imaging system. The working distance is \SI{200}{\micro\metre} to compensate for the coverslip that separates the cells from the fiber facet. Field of view is \SI{60}{\micro\metre} by \SI{60}{\micro\metre} and scale bars in all images are \SI{10}{\micro\metre}. Image adapted from \cite{papadopoulos2013high}. Copyright 2013 Optical Society of America.}
		\label{fig:neuroimg}
	\end{figure}

	This enormous potential has been identified shortly after the methods for controlling light propagation through optical fibers have been established. Although fluorescent micro-particles have been the first objects imaged with SLM-controlled MMF endoscope geometries \cite{bianchi2012multi,cizmar2012exploiting}, imaging fluorescently labelled neuronal cells have followed shortly afterwards \cite{papadopoulos2013high} (see Fig. \ref{fig:neuroimg}). 

\begin{figure*}[htbp]\centering
			\includegraphics[width=1.05\textwidth]{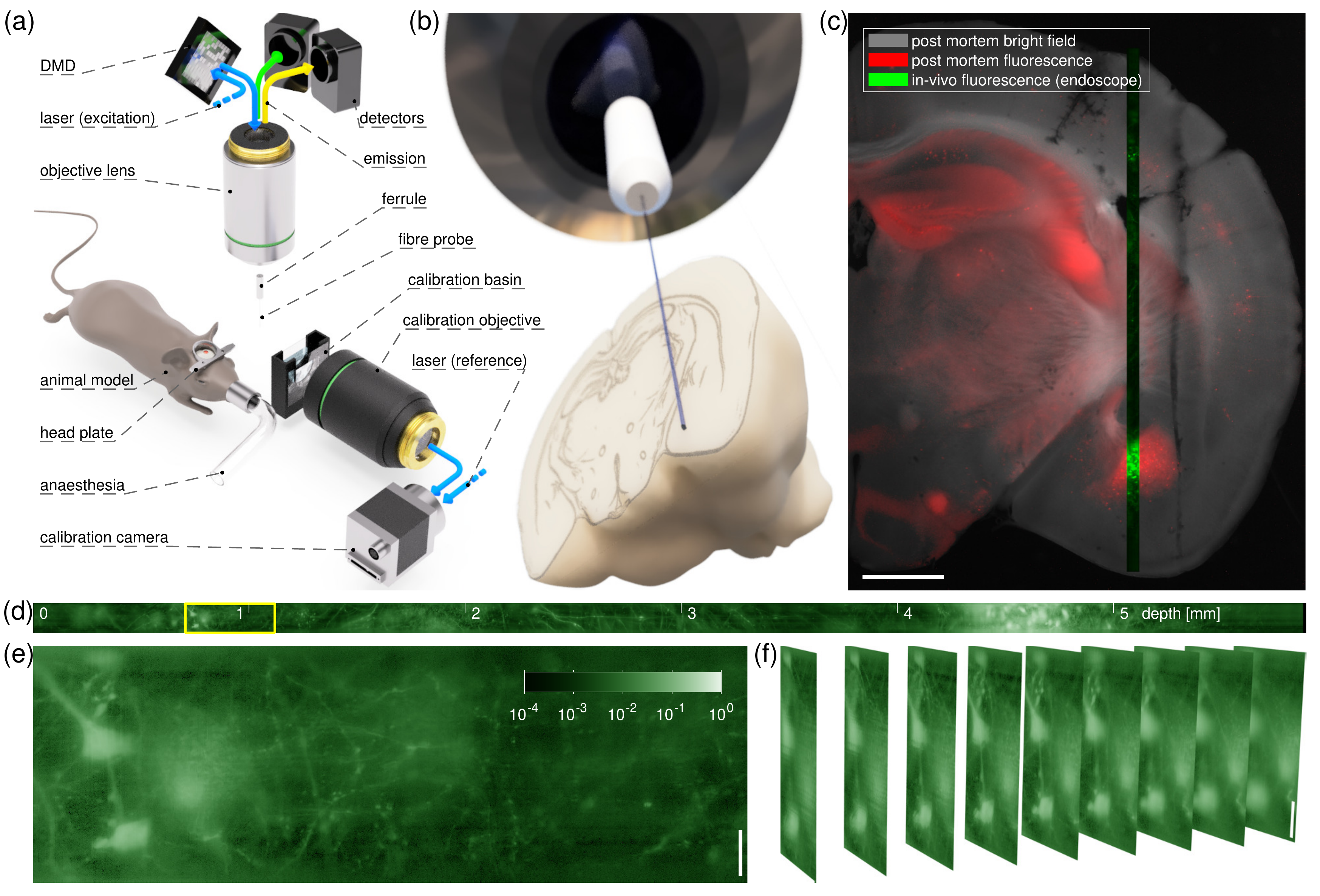}
		\caption{\emph{In-vivo} fluorescence imaging through a MMF in the living brain. 
  (a) Simplified experimental scheme. 
  (b) To-scale fiber probe and mouse brain.
  (c) Overlay \emph{in-vivo} MMF endo-microscope record detailed in (d)--(f) with post-mortem bright-field and confocal fluorescence microscopy. 
  (d) Record of the 'side-view' \cite{silveira2021side} endoscope progression throughout the whole brain depth of a Thy1-GFP line M mouse. 
  (e) Detail of the same record from the location of cortex, shown in full resolution. 
  (f) Volumetric data corresponding to left part of (e) organised in 9 parallel planes away from the probe tip, displaced by \SI{2.5}{\micro\metre} from one another.
The horizontal scale bar in (c) corresponds to the length of \SI{1}{\milli\metre}. Vertical scale bars in (d) and (e) correspond to the length of \SI{20}{\micro\metre}.
  Fluorescence intensity data in (d) and (e) are shown in logarithmic scale as indicated by a colour bar in (d).
  Figure adapted from Ref.~\cite{Stiburek2023} with permission.
  }
		\label{fig:invivo}
	\end{figure*}

 In 2018, several studies have exemplified this concept in \emph{in-vivo} imaging of neurones inside brains of small animal models, also pioneering the concepts of volumetric imaging and random access, while enjoying the boosted recording speed due to the transition from liquid crystal-based SLMs to more than two orders of magnitude faster DMD modulators \cite{ohayon2018minimally,turtaev2018high,vasquez2018subcellular}. Details of the structural connectivity within neuronal circuits (dendritic spines and axonal boutons) have been successfully visualized and the activity of individual neurones has been recorded.   
	The current state-of-the art is visualized in Fig. \ref{fig:invivo}.
	
	Separate efforts have been focused on implementation of two-photon microscopy via MMFs. Two photon excitation requires simultaneous absorption of two photons carried by a femtosecond laser pulse, which together provide the necessary energy for a fluorophore molecule to transit into the excited state. Each of the photons therefore carries only half of the excitation energy and its wavelength is thereby double of that for the standard, single-photon excitation. As at longer wavelengths the tissue scattering weakens significantly, one can achieve much larger penetration depths. In addition, the excitation is proportional to the square of the excitation beam, thereby exciting the sample much less outside the focal plane when compared to the single-photon case. As the excitation is very well localised to the selected focal plane,  this results in elegant means of sectioning. These considerations make two photon excitation a very desirable modality for MMF endoscopes. Although very short pulses have been successfully realized through MMFs \cite{morales2015delivery} and imaging possibility has been verified \cite{morales2015two}, these results have been achieved with relative slow scanning rates allowed by liquid crystal SLMs. The DMD technology does not allow for power-efficient wavefront control of broadband signals \cite{hoffmann2018kilohertz}, so the practical relevance of multi-photon excitation in biological application will likely emerge once the means of much faster phase-only modulation become widely accessible.

	\subsubsection{Reflectance imaging}
	
	Reflectance imaging, i.e., compiling an image from the intensity of light signals which have reflected off the sample, is arguably the most straightforward imaging modality, which has been exploited in numerous pioneering studies in this domain \cite{mahalati2012adaptive, choi2012scanner, Mahalati:13, gu2014noise, Loterie:15, ploschner2015seeing, kahn2015}. In MMF implementations, the reflected signal is ideally collected and delivered towards the detector backwards, through the same fiber, which has provided the illumination signals. \textcolor{black}{As the same wavelength is used both ways through the fibre, the TM for the returning light is just the inverse of that determined in the calibration procedure. This can be efficiently utilized in implementations of confocal, phase contrast and other microscopy modalities \cite{loterie2015confocal,lee2022confocal}.}
 The desired signals can however be mixed with other returning signals including reflections off the MMF's facets or power back-scattered inside the fiber. While studying the principles of light  transport through MMFs and various aspects of their potential applicability, researchers often utilize the transmittance alternative, i.e. collect the signals transmitted through semitransparent objects of known properties (e.g., the 1951-USAF resolution target) serving as the ground truth, although such geometries do not provide viable technological solutions for endoscopy \cite{ploschner2015seeing,turtaev2017comparison,li2021compressively,Li2021memory}. 
	
	One area, where reflectance MMF endoscopes stand a solid chance of applicability, emerged from translating their image plane to very large distances away from the fiber facet, and due to the minuscule aperture size given by the size of the MMF core, they essentially operate as infinity focus cameras with giant depth of focus, starting only a few centimetres behind the facet. They can be readily switched into high resolution microscopes when the observed objects are brought the vicinity of the distal MMF facet, purely by scheduling a different set of modulation patterns for the used light modulator. Combining these geometries with highly precise measurement of the photons' time-of-flight, it has been shown that reconstructing the depth of the object is possible, with the precision of a few millimetres\cite{stellinga2021time}. 
	
	\begin{figure}[htbp]\centering
			\includegraphics[width=\textwidth]{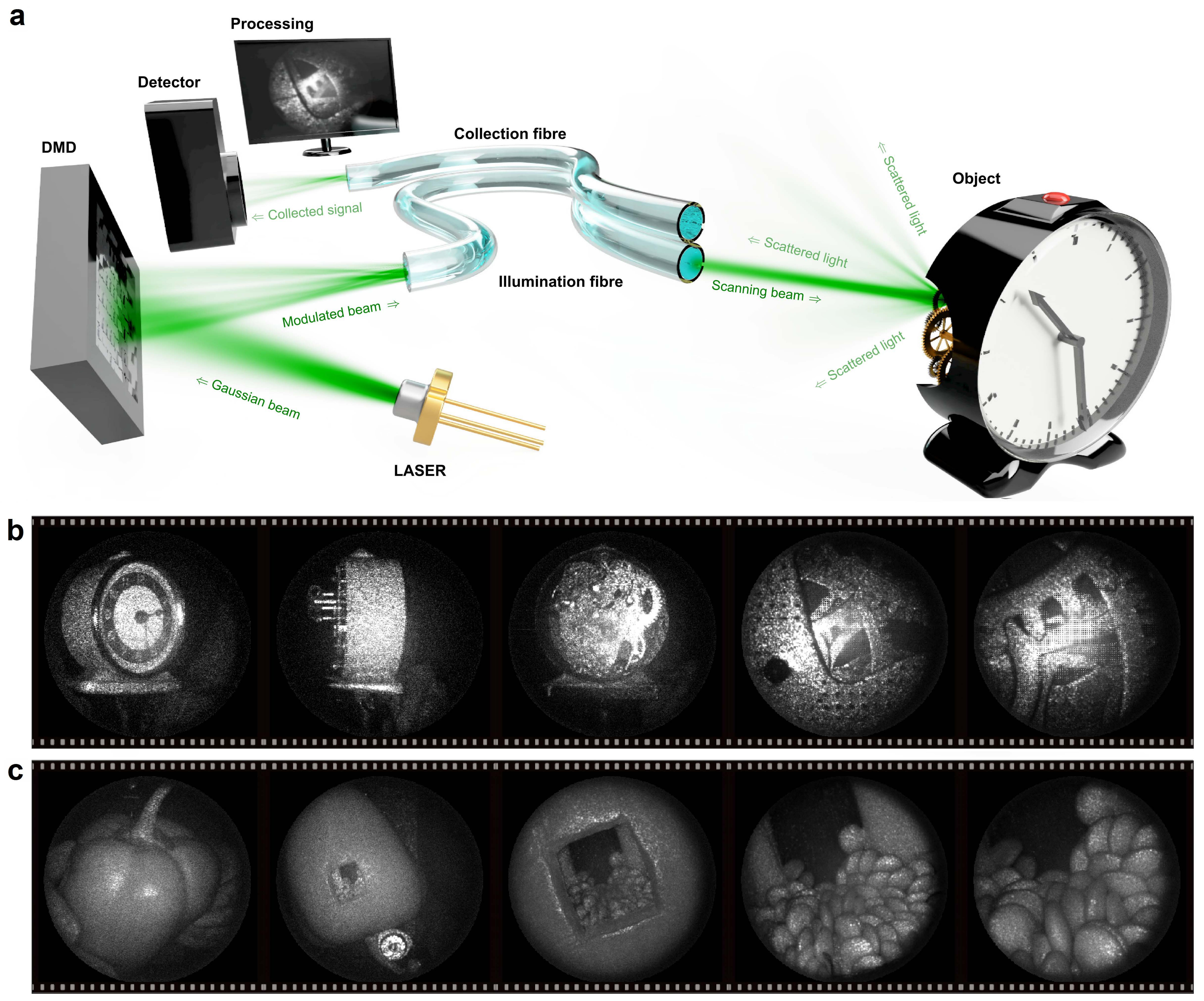}
		\caption{(a) A sequence of holograms displayed by a digital micro-mirror device (DMD) spatially shapes the wavefronts coupled into a multimode optical fiber, in such way that a far-field focus scans the distal field of view. The light signals back-scattered by an object are partially captured by a collection fiber, allowing real-time image reconstruction at the proximal side.
			(b)  Imaging a mechanical clock as an example of a dynamic object. 
			(c) Demonstration of endoscopic imaging of a bell pepper via a small opening. 
			Figure reprinted from \cite{leite2021observing}, licensed under a Creative Commons Attribution (CC BY) license.}
		\label{fig:farfiled}
	\end{figure}
	Nevertheless, when light focused on a distant object reflects back towards the MMF, the tiny fiber core and thereby the minuscule collection efficiency makes the collected light signal returning towards the detector feeble with respect to the strength of the undesired signals from facet reflections and MMF scattering. Although more elaborate solutions (e.g. time-gating) are possible, practical implementations have sacrificed on the instruments footprint and implemented separate collection fiber, free from the parasite light signals. Further, issues originate from the coherent nature of the illuminating light, which is essential for the very enabling principles of the instruments' operation. Upon reflection off matte distant objects, light returns towards the distal facet as a randomised speckle, with grain size comparable to the diameter of the core. It is therefore a question of chance and statistics, what will be the brightness of a pixel, which corresponds to a certain location of a diffusive object. With the influence growing as the object moves away from the distal facet, the speckle manifestation of coherent light is therefore carried over as a random intensity modulation of the object's image (see Fig. \ref{fig:farfiled}).  
	Even with the present limitations, however, these far-field MMF-based micro-endoscopes are already exploitable in various clinical and diagnostics applications, remote inspection, and security systems.
	
	\subsubsection{Chemical contrast imaging} 
	Label-free methods enabling instant imaging of chemical contrast represent a greatly desired alternative to post-mortem histopathology, the current gold-standard in clinical diagnostics of potentially cancerous tissues which, due to the remote and time-consuming procedures involved, delay decision-making and frequently require repeated surgical interventions. 
	A potential answer is the concept of light biopsy, utilizing methods derived from vibrational, particularly Raman spectroscopy. Here, the inelastic scattering of light by atoms and molecules leads to alternations of the scattered photons' frequencies thereby leaving characteristic spectral signatures for different materials.  
	Although Raman scattering has been shown to have great relevance to medical diagnostics \cite{C5CS00564G}, its main limitation is in the sparsity of the inelastically scattered photons (only 1 in $\approx10^6$ collisions is inelastic), necessitating long collection times while keeping the sample exposure intensities at safe levels. Next to spontaneous Raman scattering, several non-linear chemical imaging techniques, which derive from Raman scattering, particularly coherent anti-Stokes Raman scattering (CARS), and stimulated Raman scattering (SRS) have been introduced \cite{doi:10.1126/science.aaa8870}. 
	As they are tuneable to image at a specific line of the Raman spectrum, they are much faster and safer to the sample, yet their implementation is considerably more elaborate as they require two synchronous pico- or femto-second laser pulses to illuminate the sample point by point.   
	Before these techniques can be exploited in clinical settings, such label-free chemical contrast techniques must be allowed to operate in extended depths of highly scattering tissues, thereby necessitating the development of minimally invasive endoscopic solutions \cite{Lombardini:2018cs}. 
 
\begin{figure}[htbp]\centering
			\includegraphics[width=.9\textwidth]{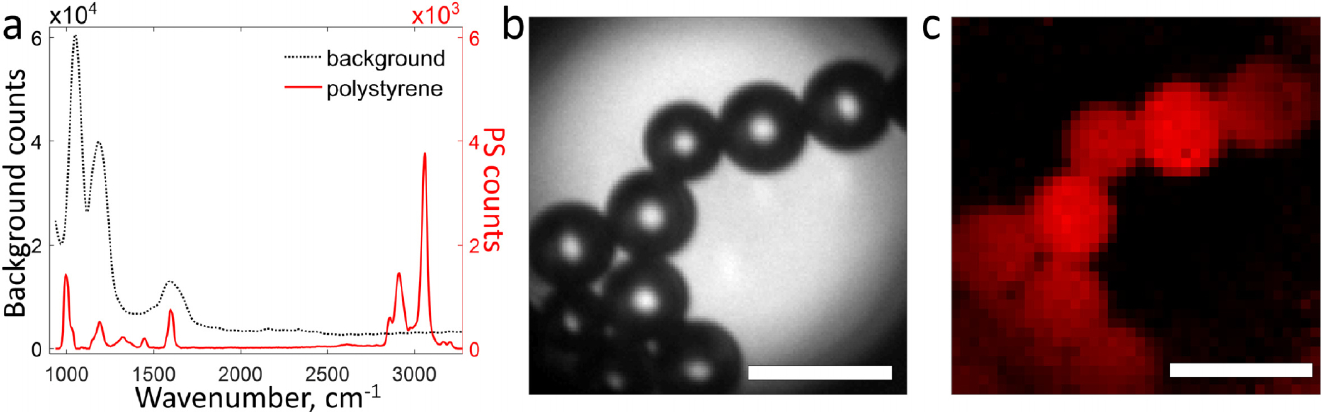}
		\caption{Raman imaging through a multimode fiber of polystyrene particles dried on a glass coverslip. (a) Background (black dotted) and polystyrene (red solid) spectral information. (b) Bright field image of the particles. (c) Weights for polystyrene spectral components, showing a Raman image of the particle distribution. Scale bars are \SI{20}{\micro\metre}.
  Figure adapted from \cite{Gusachenko:17}. Copyright 2017 Optical Society of America.}
		\label{fig:ramanpic}
	\end{figure}

	MMF endoscopes feature many of the properties desired for such advancements, particularly their minuscule footprint and high spatial resolution. 
 They have joined this race by implementing spontaneous Raman imaging modality in 2017 \cite{Gusachenko:17}, recording the complete Raman spectrum of various specimens for each pixel of the field of view with 5-20 second acquisition time per pixel (see Fig. \ref{fig:ramanpic}).
 
\begin{figure}[htbp]\centering
			\includegraphics[width=.9\textwidth]{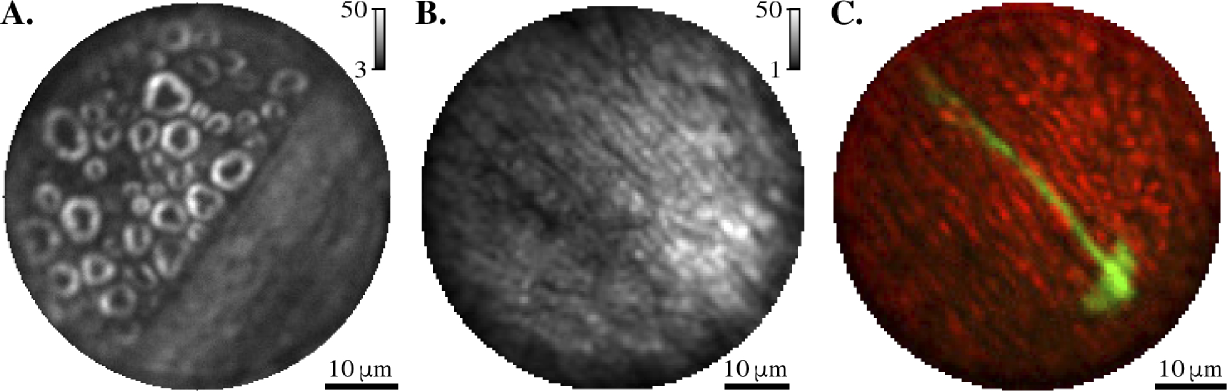}
		\caption{A. CARS image of myelin sheaths in a transversal cut of a sciatic nerve taken on the surface. B. CARS image of myelin sheaths in the corpus callosum of mouse brain taken with the probe inserted 1.5mm
			inside the tissue. C. Multimodal CARS (red) and TPEF (green) image of myelinated nerve fibers in the cerebellum.
			Figure adapted from\cite{Pikalek:22}. Copyright 2022 Optical Society of America.}
		\label{fig:chemim}
	\end{figure}

 A potential limitation was found in the strong Raman and fluorescence background originating from the fiber itself, at shorter wavenumbers (< 500 --1700 cm$^{-1}$). This has been addressed by employing sapphire MMFs, which enabled collecting Raman spectra at wavenumber near 1000 cm$^{-1}$, thereby opening the possibility to image within the most prominent window for clinical applications (Raman shifts of aromatic amino acid phenylalanine, unsaturated fatty acids, CH2 and CH3 deformation vibrations, amide bands of proteins) \cite{Deng:19}. 
	MMF endoscopes have further been demonstrated to offer even non-linear chemical imaging modalities, particularly CARS\cite{tragaardh2019label}. The most advanced version \cite{Pikalek:22} utilizes femtosecond laser pulses with spectral focusing modality \cite{doi:10.1063/1.1768312} and composite fiber probes for suppression of strong background signals originating in non-linear processes inside the fiber. Importantly, this concept enabled reduction of the pixel dwell time to millisecond intervals (see Fig. \ref{fig:chemim} for performance examples). Similarly as in the case of multi-photon imaging, the bottle-neck for higher imaging speeds is here the slow operation of liquid crustal SLMs and further progress will be enabled through the availability of faster light modulators capable of directly manipulating the phase.  
	
\subsubsection{Photoacoustic imaging} 	
	Photoacoustic imaging provides a contrast based on optical absorption. It can be used as a label-free method to image highly absorbing tissues locations, where hemoglobin, melanin or lipids occur in high concentrations. The contrast can be further enhanced by compounds targeting desired structures. Upon illumination by nanosecond pulses delivered through the MMF in the shape of difraction-limited foci, light absorption and resulting thermoelasticity causes acoustic waves to emerge. The strengths of this acoustic signal is detected and used as contrast for the resulting image~\cite{zhao2021towards,mezil2020single,caravaca2019hybrid}. In 2020, it has been shown, that the generated acoustic waves can be efficiently detected from the vicinity of the sample also with the use of an optical fiber (separate from the MMF for light delivery), thus making for a narrow all-fibre photoacoustic endoscope to be used in large depths of tissues, which can further be combined with other modalities described above, particularly the fluorescence \cite{mezil2020single,caravaca2019hybrid}.

	\subsubsection{Wide-field approaches} 
	
	A na\"{i}ve idea of an ideal microscope is the use of two lenses sharing one focal plane. The optical fields found in the remaining focal planes are essentially related by a physical realization of two Fourier transforms between the spatial domain and that of the spatial frequencies. The first lens converts the field of the observed object into the spatial spectrum, the second reverses the conversion and forms the field of the object's image to be observed by a naked eye or acquired by a camera.  Due to the mathematical equivalence of the Fourier transform and its inverse the image is upside down, and conveniently scaled by a factor given by focal lengths of both lenses. Considering ideal lenses, this principle applies regardless of the spatial and temporal coherence of the light relaying the object with its image. As oppose to scanning approaches, all pixels of an image are not formed sequentially, but simultaneously, e.g., from a single camera acquisition. 
	
	A TM of a MMF is similarly an operator, which also has its mathematical inverse. Finding solutions for implementing computational or even physical inverse of a TM are greatly desired as they can significantly speed-up image acquisition and enable powerful wide-field methods of modern microscopy such as localisation super-resolution techniques such as PALM and STORM~\cite{betzig2006imaging,rust2006sub}. 
	
\begin{figure}[htbp]\centering
			\includegraphics[width=\textwidth]{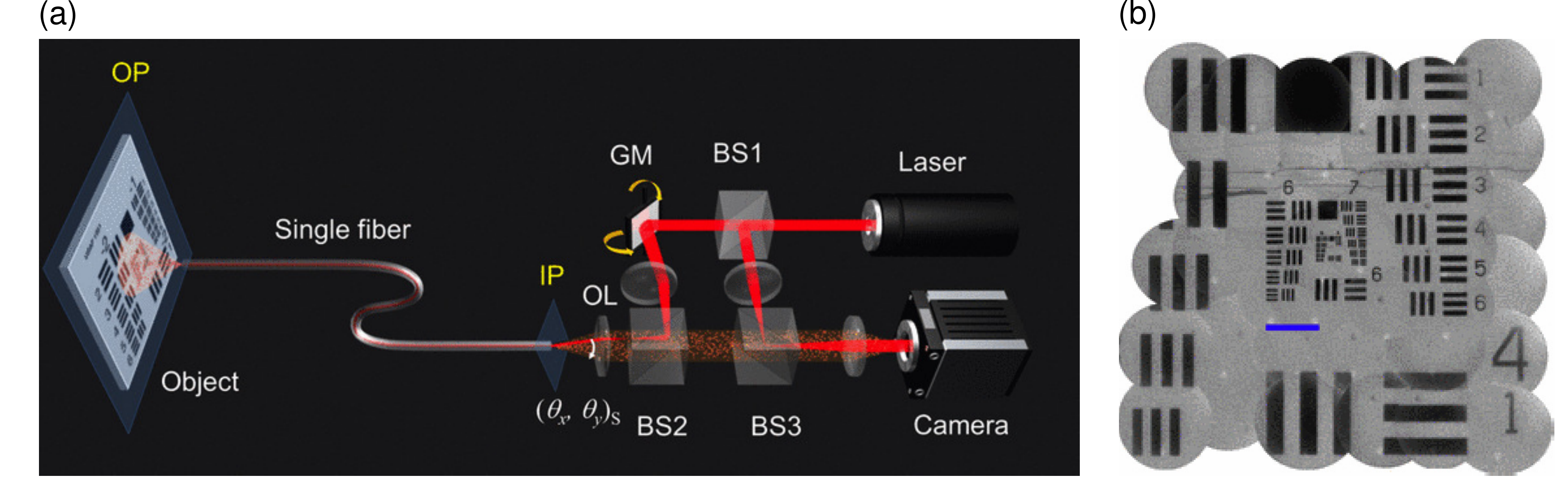}
		\caption{Wide-field imaging using coherent signals. 
(a) The setup is based on an interferometric phase microscope. The interference between the reflected light from the object located at the opposite side of a single multimode fiber and the reference light is recorded by a camera. A two-axis galvanometer scanning mirror (GM) controls the incident angle 
to the fiber. BS1, BS2, and BS3: beam splitters. OL: objective lens. IP: input plane of a multimode optical fiber (0.48 NA, 
\SI{200}{\micro\metre} core diameter). OP: object plane. (b) 
Reconstructed images are stitched to enlarge the field of view. Scale bar indicates  
\SI{100}{\micro\metre}
			Figure adapted from Ref.~\cite{choi2012scanner} with permission.}
		\label{fig:scannerfree}
	\end{figure}

	When fully (spatially and temporarily) coherent signals are considered, the knowledge of the TM offers a direct route to implement wide-field imaging \cite{choi2012scanner} (see Fig. \ref{fig:scannerfree}).
 In this concept, light reflected of an object and transformed by the propagation through the MMF is captured in both amplitude and phase using off-axis interferometry. A mathematical inverse of the TM is then used to compute the object field corresponding to the received signal. Due to the coherence of the light used, the reconstructed image field however remains modulated by speckle, therefore averaging the result over numerous realizations is necessary to obtain a speckle-free image. 
	
	Further it has been shown, that single-plane phase modulation can be used to acquire images with over 100 image pixels from a single camera acquisition, utilizing temporary coherent yet spatially incoherent signal, resulting in wide-field (bright-field and dark-field) imaging modalities. Although there is no speed bottle-neck other than the camera detector itself, this approach was associated with immense power losses, which scale proportionally to the amount of pixels in the image~\cite{cizmar2012exploiting}. The efficiency and the number of pixels can be greatly improved if the concept of mode convertors was implemented~\cite{Fontaine:2019ho}. These devices, featuring a series of single-plane phase modulations, decompose a propagating signal into the basis of propagation-invariant modes and direct the optical power carried by each of them into spatially separated regions. In a recent theoretical study \cite{https://doi.org/10.48550/arxiv.2204.02865} a pair of 15-plane convertors of 400 propagation-invariant modes of a step-index MMF
 were considered and one additional corrector plane, all together forming a physical inverter of the TM. The light propagating through an optical fiber is decomposed into its PIM constituents by the first convertor, the corrector plane removes the phase shifts they acquired while propagating through the MMF, and finally, they are combined into the image by the second convertor, thus forming a physical copy of the field found in front of the MMF. Even-though this scheme will be immensely challenging to realize experimentally and it would limit the use of broadband signals for larger segments of fibers (>10nm bandwidth is achievable with < 1cm long fiber), it may render any form of wide-field microscopy at the tip of a hair thin segment of MMF possible. 
	
	Finally, when fluorescence signals of multiple emitters are coupled into a MMF, the output intensity distribution is essentially a linear superposition of individual intensity landscapes, which would emerge if each emitter was present in the sample plane on its own.  Acquiring such individual MMF output intensity distributions can be achieved for example by a single fluorescent particle anchored to a microscope slide. 
The particle is densely position stepped across the desired object plane in front of the distal MMF facet and the corresponding fluorescence intensity distributions are recorded for each position. These can further be used when reconstructing the image of more complex objects algorithmically \cite{Kim:2017jv}. This technique is simple and elegant, yet highly sensitive to noise and prone to numerical artifacts, especially in low-photon-number regimes, e.g., when observing details of structural connectivity of neurones. It can be efficiently used in imaging of sparse fluorescent objects, possibly even for signalling activity of neurones within the brain tissue.  

\subsubsection{Image transmission} 
\begin{figure}[htbp]\centering
			\includegraphics[width=.8\textwidth]{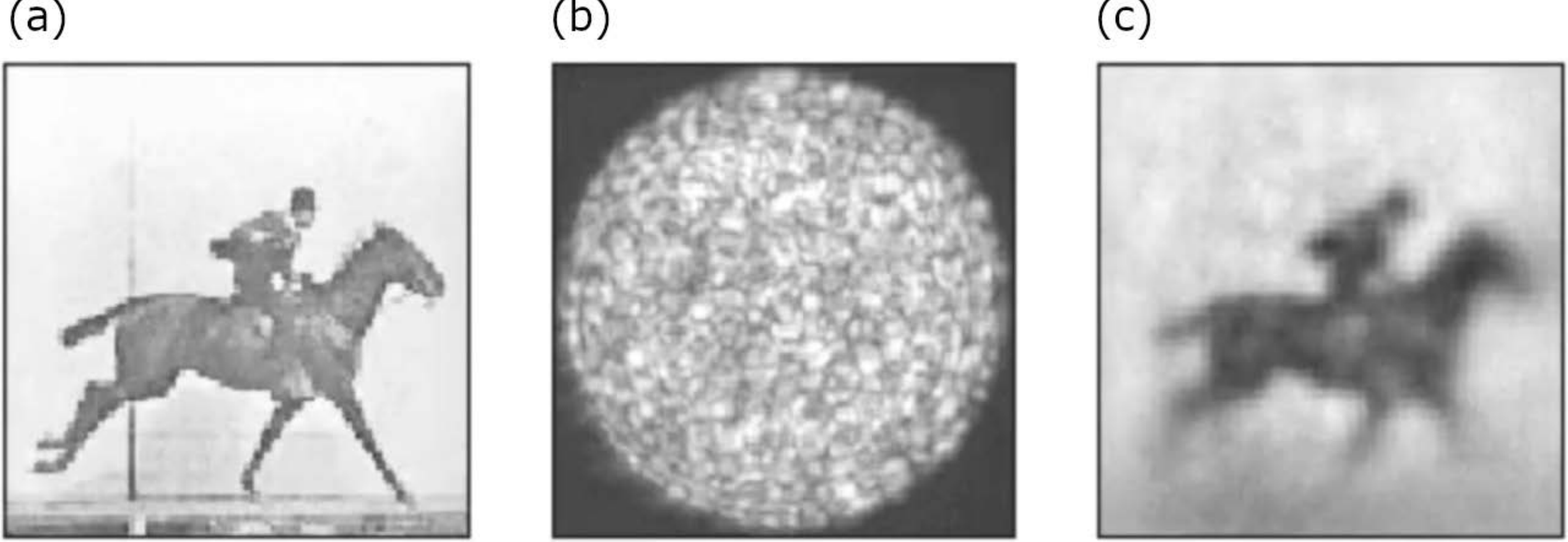}
		\caption{Transmitted image reconstruction using TM obtained by an artificial neural network. (a) Scene coupled into the MMF. (b) Output speckle. (c) Reconstructed image.  
			Figure adapted from ~\cite{caramazza2019transmission} with permission.}
		\label{fig:CNN}
	\end{figure}

Image transmission is a technique, which utilizes a monochromatic beam with usually a flat wavefront and the intensity spatially modulated to the shape of an image. This field is then sent  through a MMF and the image is recovered from the speckle distribution.  Unlike in imaging applications, the light does not interact with an object to give a contrast. The approach is frequently used as a test bed for advanced computing algorithms, particularly artificial neural networks~\cite{RahmaniOguzTeginHsiehPsaltisMoser+2022+1071+1082}. Using a convolutional network, Rahmani et al. \cite{rahmani2018multimode} reconstructed the input intensity (or phase) randomised by propagation through a \SI{0.75}{\metre} segment of step index fiber having \SI{50}{\micro\metre} in diameter. After training, the network has been used to reconstruct images of different classes. Although the network could transfer its knowledge for retrieval of untrained image classes, the retrieval performance deteriorated significantly. This has been addressed by an alternative method, designed to statistically reconstructs the inverse transformation matrix of the MMF, thereby enabling reconstruction of untrained images with significantly higher structural complexity~\cite{caramazza2019transmission} (see Fig. \ref{fig:CNN}).

Further, neural networks have been used to reconstruct image information in fibers which are subjected to bending \cite{fan2019deep, liu2020bending, resisi2021, zhao2021high, zhu2021image,  tang2022learning, wang2022upconversion, xuhigh, song2022deep} and spectral correlation \cite{kakkava2020deep}.

	\subsection{Optical trapping}
	
	Micro-objects, which find themselves in an optical field, experience mutual exchange of the momentum with the light, which manifests itself by optical forces. One component of this force related to the idea of radiation pressure, so called scattering force, accelerates the particles in the direction of the beam propagation.  Further, such particles whose refractive index is higher than that of the surrounding medium are attracted towards the locations of high intensity as a result of so called gradient force. Optical tweezers  are three-dimensional optical traps formed by tightly focused laser beams, where both force components are balanced.      
	A number of exciting applications of this concept have been introduced, particularly due to their ability to exert and measure subpiconewton forces on single molecules or molecular assays. Optical tweezers have greatly shaped our understanding of molecular motors and the functionality of biological polymers driving the mechanics of cells. Holographic optical tweezers (HOT) are arguably the most powerful and versatile embodiments of optical tweezers, which exploits wavefront shaping to facilitate a large number of trapping sites that can be simultaneously manoeuvred in three dimensions at will. HOT requires the use of high numerical aperture focusing element (NA > 0.8), most commonly a specialised microscope objective. Further, its performance degrades steeply in the presence of aberrations, thereby precluding its exploitation deep in scattering tissues.   
	
	Due to the close mutual relevance of wavefront shaping and optical trapping methodologies, the potential to exploit shaped structured fields emerging from MMFs for optical manipulation has been identified very early\cite{cizmar2011shaping,bianchi2012multi}. Forming HOT at the tip of a hair-thin fiber endoscope would enable acting and measuring minuscule optical forces into optically complex environments beyond the reach of bulky focusing optics and would allow for studies of important biological processes directly inside complex  living organisms. 
	
	As the numerical aperture of the commercial MMFs does not reach sufficiently high values for stable three-dimensional confinement of micro-objects, the early achievements of optical manipulation though MMFs have been reduced to two-dimensional, lateral trapping whereby the particles were prevented from escaping the optical trap in the axial direction by the boundary of the sample chamber. 

\begin{figure}[htbp]
\centering
			\includegraphics[width=0.9\textwidth]{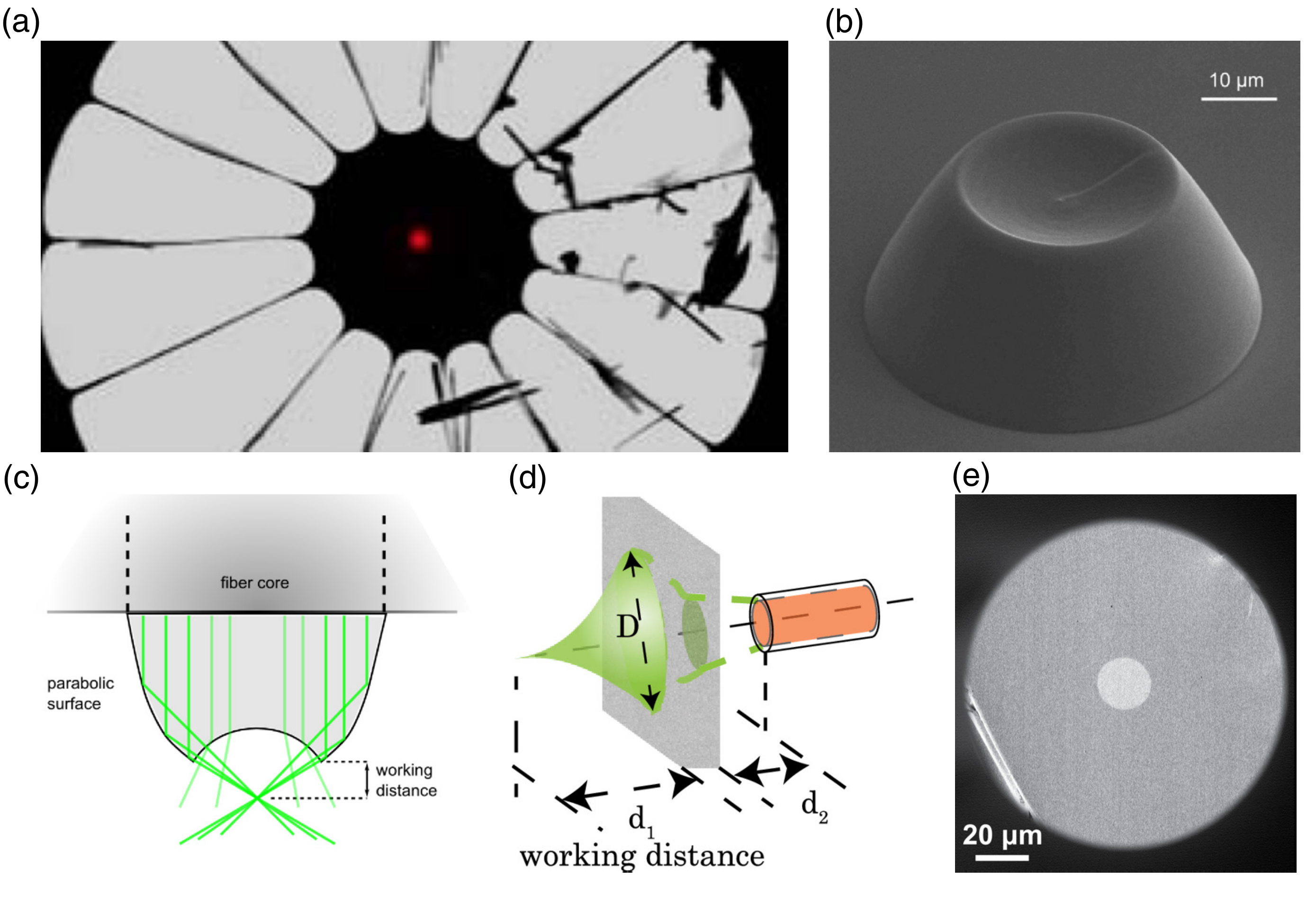}
		\caption{Strategies for tight focusing of light at the distal end of a multimode fiber. 
  (a) Air-cladding photonic crystal fiber with generated focus.   
			Figure adapted from~\cite{Amitonova:16}. Copyright 2016 Optical Society of America.
   (b) and (c) Fiber with micro-fabricated optics. Figure adapted from~\cite{Bianchi:13}. Copyright 2013 Optical Society of America.
   (d) Exploiting the properties of highly scattering layer close to the distal facet of a MMF.
   Figure adapted from~\cite{Papadopoulos:13}. Copyright 2013 Optical Society of America.
   (e) All solid step index fiber from soft-glass materials.
   Figure adapted from ~\cite{Leite2018} with permission.
   }
		\label{fig:NA}
	\end{figure}

	In order to increase the numerical aperture to the required level, several possibilities have emerged recently and although they have not found use in optical trapping applications, it is worth mentioning them for their benefits in other applications including imaging. 
	Placing a highly scattering element in the close vicinity of the distal MMF facet \cite{Papadopoulos:13,Choi:13} is probably the simplest route towards achieving tight focusing (see Fig. \ref{fig:NA}d). This solution only allows a fraction of the optical power to form the desired foci, with the rest forming a background speckle signal. This would be problematic particularly when forming a large amounts of optical traps, with which the background signal can undesirably interfere. Further options involve micro-fabrication of focusing element, directly at the distal facet \cite{Bianchi:13} (see Fig. \ref{fig:NA} b and c), with which it is possible to achieve tight and high-purity foci, yet the field of view is no longer as large as the instrument's footprint, instead it is reduced by the same factor with which the NA is enhanced. Air-cladding photonic crystal fibers also offer the formation of tight and pure  foci \cite{Amitonova:16} (see Fig. \ref{fig:NA}a), yet due to the necessity of the outer layer, holding the core in place via narrow bridges, they have rather large footprint. Moreover it is unclear whether aqueous media would not invade the air cavities and alter the MMF's TM during experiments.  
	
	An implementable solution was however found in manufacturing all-solid fibers formed by soft-glass materials \cite{Leite2018} (see Fig. \ref{fig:NA}e). 
	\begin{figure}[htbp]\centering
			\includegraphics[width=.9\textwidth]{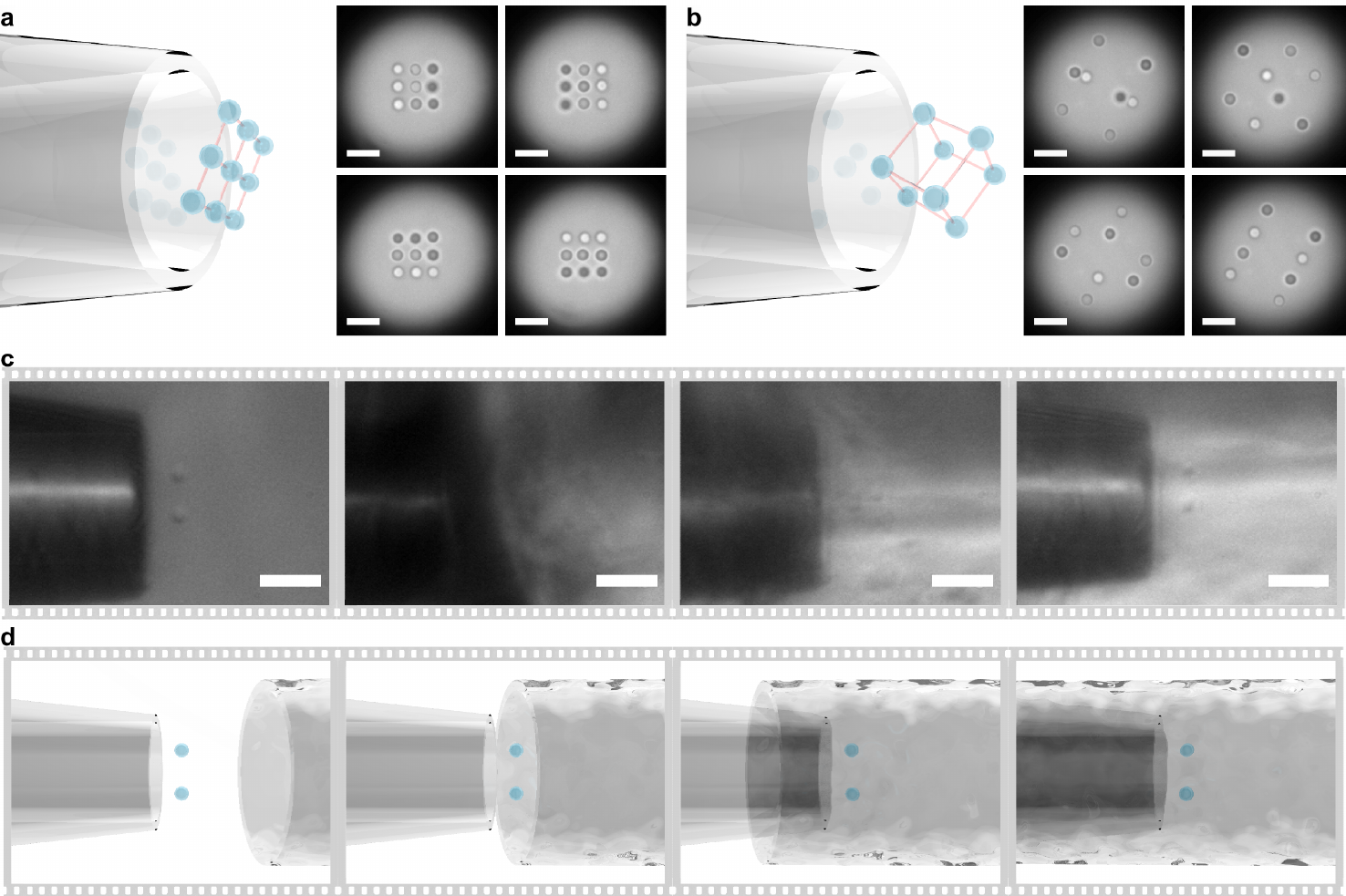}
		\caption{Multiple holographic tweezers delivered through a lensless multimode fiber.
			(a,b) Multiple holographic tweezers trapping of (a) nine particles in a square-grid arrangement, and (b) eight particles forming a rotating three-dimensional cube. The scale-bars correspond to \SI{5}{\micro\metre}.
			(c,d) Optical manipulation of two particles inside a turbid cavity comprising a complex, hard-to-access environment. The scale-bars correspond to \SI{10}{\micro\metre}.
			In (a-d), all the particles are \SI{1.5}{\micro\metre}-diameter silica microspheres in a water suspension, the fiber length is \SI{65}{\milli\metre}, and the trapping wavelength is \SI{1064}{\nano\metre} in vacuum. Figure adapted from \cite{Leite2018} with permission. 
		}
		\label{fig:trap}
	\end{figure}
	The fibers were successfully employed in confining and manipulating up to 9 particles simultaneously, with position sensitivity as small as 5 nm (see fig. \ref{fig:trap}).

	\subsection{Microfabrication}
	
	\label{sec:Microfabrication}
	
	The formation of 3D microstructures brings novel advanced methods in several disciplines, including biophotonics, plasmonics, metamaterials, and micro-electromechanical systems \cite{bernardeschi2021review}. One of the 3D printing processes that makes it possible to create high-resolution microstructures is direct laser writing. Most precise systems rely on the two-photon polymerisation principle, which due to its nonlinear nature, utilizes ultrashort pulses to trigger polymerisation only in the highly confined volume of the photoresin. It recently attracted a lot of attention for creating structures of highly complex architecture with feature sizes below \SI{100}{\nano\metre} \cite{maruo2008recent}. 
	
	Sub-micron resolution implies using high-NA objectives, which limits  3D printing ability to a small volume at close proximity to the bulk optics elements, preventing applications in the non-transparent, hard to reach or simply narrow cavities. Here multimode fiber endoscopy can play its role \cite{morales2017three, delrot2018single, konstantinou2023improved}. 
	
\begin{figure}[htbp]\centering
			\includegraphics[width=.8\textwidth]{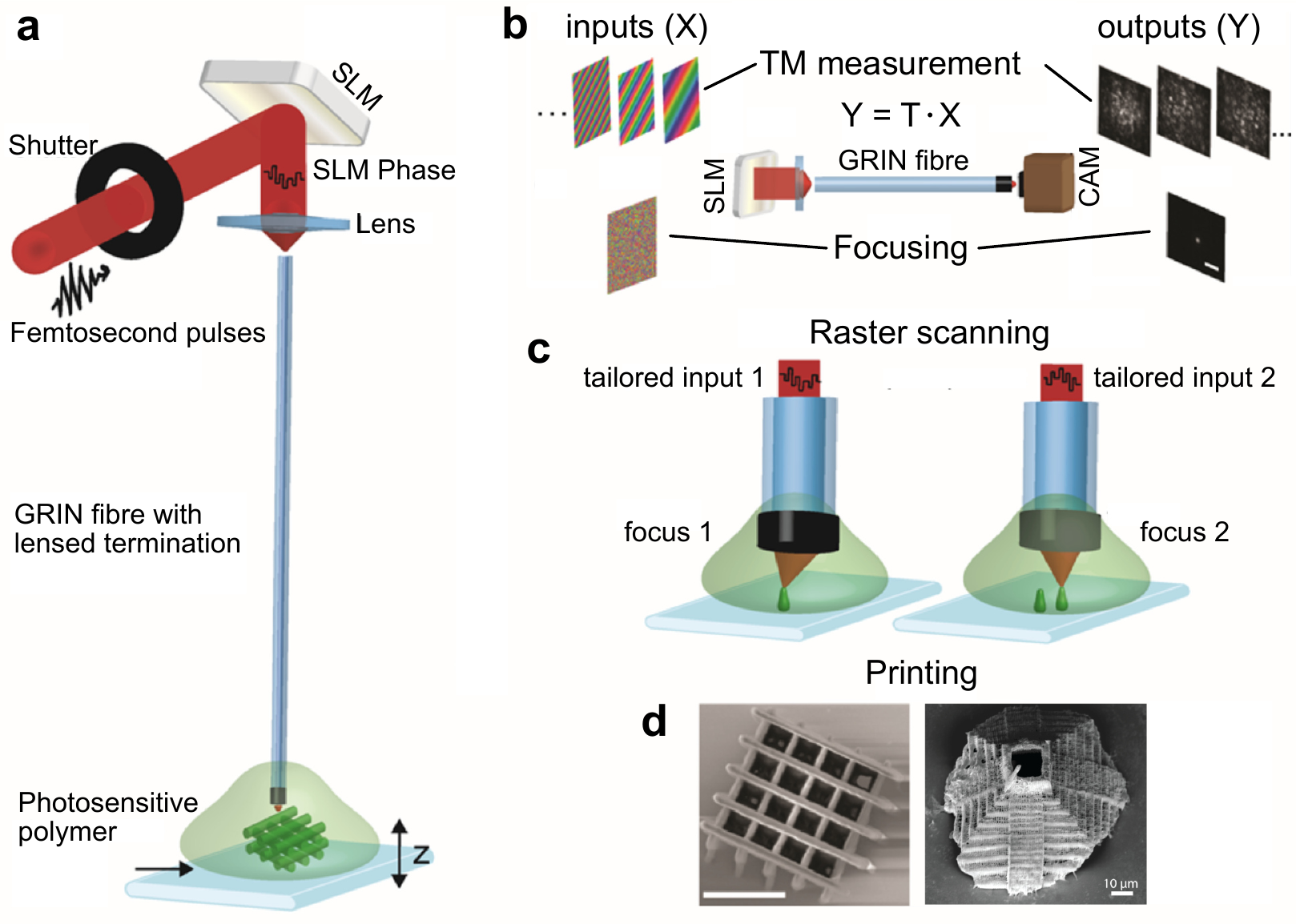}
		\caption{Schematic for the two-photon endofabrication through a multimode fiber.
			(a) A spatial light modulator illuminated by femtosecond laser allows wavefront shaping through a GRIN multimode fiber (400 $\mu$m core diameter, length: 50 mm). The distal end of the fiber is dipped into a photosensitive polymer. Axial (z) positioning is implemented at the sample side via motorized translation stage;
			(b) Calibration procedure based on TM approach allowing light focusing through the probe;
			(c) Raster scanning by precalculated phase patterns;
			(d) SEM image of printed structures of the woodpile and the pyramid of Chichen Itza, both scale bars are \SI{10}{\micro\metre}. Left figure adapted from  \cite{morales2017three}. Copyright 2017 Optical Society of America. Right figure adapted from\cite{konstantinou2023improved} with permission.}
		\label{fig:microfabrication}
	\end{figure}

	In experiment depicted in Fig. \ref{fig:microfabrication}, a graded index fiber terminated with a grin lens, resulting in an NA exceeding 0.5, was utilized to deliver tightly focused laser beams through only 0.5 mm thick probes \cite{morales2017three}. Both optical phase conjugation \cite{morales2015delivery} or TM-based approaches \cite{konstantinou2023improved} were successfully applied in combination with time-gating, resulting in wavefront shaping only for the modes of similar propagation constant, therefore minimising the effect of modal dispersion and significant pulse broadening. This focused pulse can be scanned across the photoresist volume, resulting in highly localised polymerisation.
	
	This raster scanning approach at the tip of the fiber enables the additive manufacturing of complex structures (see Fig. \ref{fig:microfabrication}c) with sub-diffraction limited precision to be performed in hard-to-reach areas. Moreover, the inherently reconfigurable nature of the wavefront shaping approaches allowed us to optimize the beam's PSF for faster printing \cite{konstantinou2023improved}.
	
	Further development of fast and flicker-free spatial light modulators   will significantly reduce printing time and improve quality, enabling practical applications.

	\subsection{Spectroscopy, hyperspectral imaging}
	\label{sec:MMFspec}
	
	
	\subsubsection{Multimode fiber spectrometer}
	
	Optical spectrometers are widely used for chemical and biological sensing, material analysis, and light source characterization. The spectral resolution $\delta \lambda$ sets the smallest difference in wavelength $\lambda$ that can be distinguished. 
	The wavelength range covered by one measurement is given by $M_{\lambda} \, \delta \lambda$, where $M_{\lambda}$ is the number of spectral channels that can be measured in one acquisition. The ultimate limit is the free spectral range (FSR), beyond which a spectrometer may generate the same response at well-separated wavelengths. The sensitivity of a spectrometer depends on the light throughput and the signal to noise ratio (SNR).

	\begin{figure}\centering
		\includegraphics[width=0.9\textwidth]{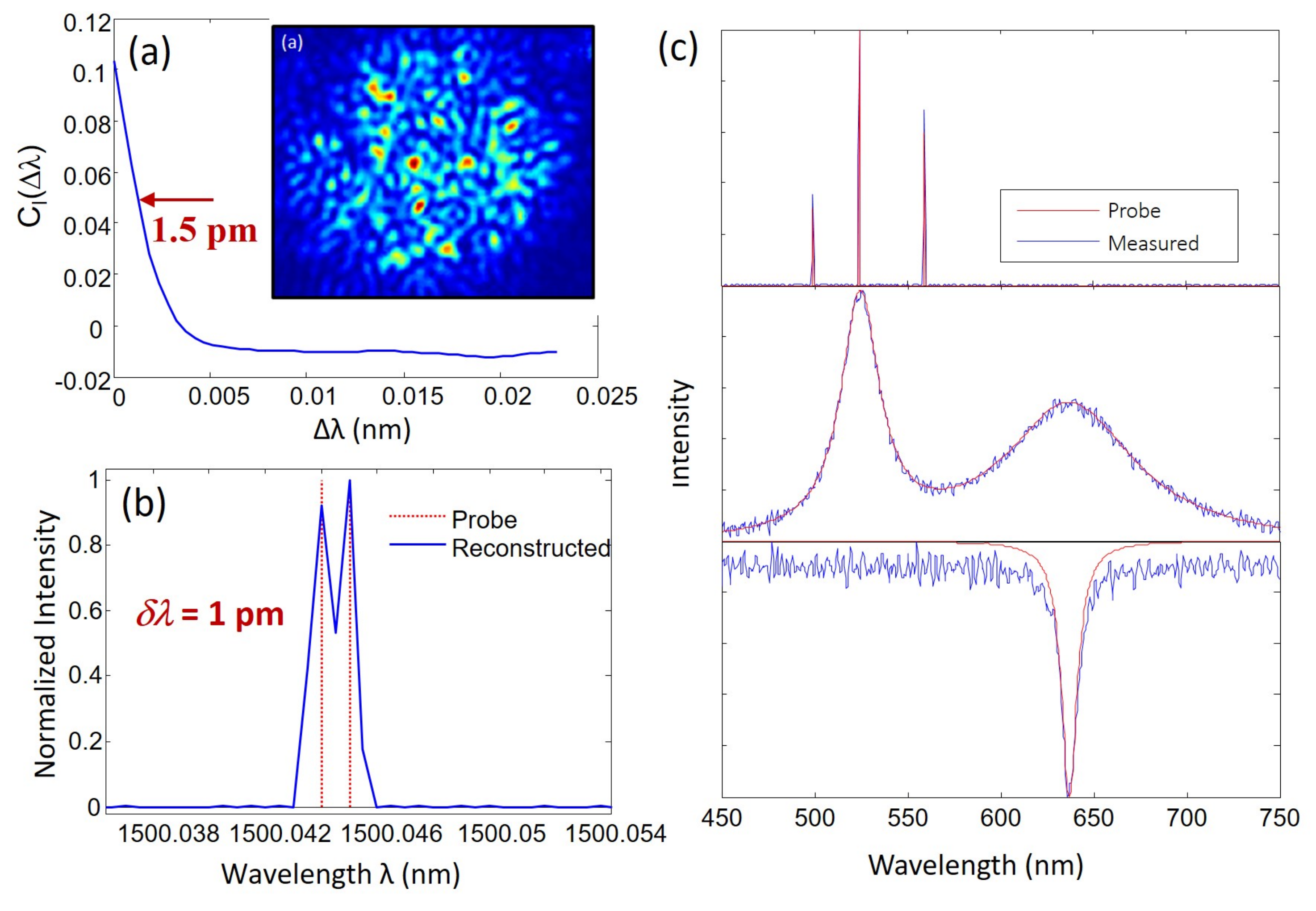}
		\caption{Multimode fiber spectrometer featuring high resolution and broad range.
			(a) Spectral correlation function of the transmitted intensity pattern $C_I(\Delta \lambda)$ through a 100-m-long MMF has a half-width-at-half-maximum (HWHM) of 1.5 pm at $\lambda$ = 1500 nm. Inset: Speckle pattern recorded at fiber output. 
			(b) Reconstructed spectrum (blue line) of two narrow lines separated by 1 pm. The red dotted vertical lines mark the probe wavelengths.
			(c) Three broadband spectra recovered with 1 nm resolution from the speckle patterns of a 4-cm-long MMF. Figure adapted from Ref.~\cite{redding2014high} with permission. Copyright 2014 Optical Society of America. 
			\label{fig:MMFspec}
		}
	\end{figure}
	
	A standard multimode optical fiber can function as a compact spectrometer with high resolution and broad range. Let us consider a monochromatic light of wavelength $\lambda$ is coupled via a polarization-maintaining single-mode fiber to a MMF of length $L$ and with $M$ excited guided modes. The transmitted field is filtered by a linear polarizer, and the spatial distribution is given by Eq.~\ref{eq:field}. Once the spatial wavefront and polarization state of incident light are fixed, the transmitted intensity pattern $I_t(r, \varphi; \lambda) = |E_t(r, \varphi; \lambda)|^2$ is unique for each wavelength $\lambda$, and serves as a fingerprint. The input spectrum $I_i(\lambda)$ can be reconstructed from $I_t$,
	\begin{equation}	
		I_t(r, \varphi) = \int P(r, \varphi; \lambda) \, I_i(\lambda) \, d \lambda \, ,
		\label{eq:OutputIntensity}
	\end{equation}
	where $P(r, \varphi; \lambda)$ represents the output intensity pattern at wavelength $\lambda$. Once $P(r, \varphi; \lambda)$ is calibrated by recording the output speckle patterns at individual wavelengths using a tunable light source, any unknown spectrum $I_i(\lambda)$ can be recovered from the measured $I_t(r, \varphi)$ \cite{redding2012using}. 
	
	The spectral resolution is determined by how small a wavelength shift of the input light can cause a notable change in the output intensity pattern. Such a change is quantified by the spectral intensity correlation function:
	\begin{equation}
		\label{SpetralCorrelation}
		C_I(\Delta \lambda) \equiv \frac{\langle P(r, \varphi; \lambda) \, P(r, \varphi; \lambda + \Delta \lambda) \rangle_{r, \varphi, \lambda}} {\langle P(r, \varphi; \lambda) \rangle_{r, \varphi, \lambda} \, \langle {P(r, \varphi; \lambda + \Delta \lambda) \rangle_{r, \varphi, \lambda}}} -1 \, ,
	\end{equation}
	where $\langle ... \rangle_{r, \varphi, \lambda}$ denotes an average over spatial position $(r, \varphi)$ and wavelength $\lambda$. As $\Delta \lambda$ increases, $C_I(\Delta \lambda)$ gradually decays to zero, as shown in Fig.~\ref{fig:MMFspec}(a).  The half-width-at-half-maximum (HWHM) provides an estimate of wavelength resolution $\delta \lambda$. 
	
	Ignoring material dispersion, the spectral resolution is dictated by modal dispersion that broadens the optical path-length distribution as light propagates through a MMF. If all $M$ fiber modes are equally excited ($A_m = 1$) and their coupling is negligible, the width of the optical path-length distribution scales as $(\dot{\beta}_1 - \dot{\beta}_M) \, L$. Here, $\dot{\beta}_1 - \dot{\beta}_M = d \beta_1 / d \lambda - d \beta_M / d \lambda$ reflects the maximal difference of group velocity, and it scales as NA$^2$ for a step-index MMF of NA $\ll$ 1 \cite{redding2013all}. The spectral resolution $\delta \lambda$ is inversely proportional to $\Delta \mathcal{L}$:  
	\begin{equation}
		\label{MMFresolution}
		\delta \lambda \propto \frac{1}{(\dot{\beta}_1 - \dot{\beta}_M) \, L }
	\end{equation}
	If mode coupling is strong in a MMF, light hops among the fiber modes, and $\delta \lambda \propto 1/\sqrt{L}$.     
	
	Since the optical fiber has been optimized for long-distance transmission with minimal loss, a long MMF can be used to reach ultrahigh resolution without sacrificing light throughput. As shown in Fig.~\ref{fig:MMFspec}(b), a 100-meter-long step-index MMF (core diameter = 105 $\mu$m, NA = 0.22) provides a wavelength resolution of 1 pm at $\lambda$ = 1500 nm \cite{redding2014high}. Such a long fiber is coiled on a small spool, making a compact, light-weight spectrometer with ultrahigh resolution.  
	
	The number of spectral channels $M_{\lambda}$ that can be recovered from a single measurement is determined by the number of speckle grains in the output intensity pattern, which is equal to $M$. Without prior information of the probe spectrum, $M_{\lambda} \simeq M$. 
	Since a MMF maps a 1D spectrum to 2D space, a large number of spectral channels are measured by a 2D camera in a single acquisition. As shown in Fig.~\ref{fig:MMFspec}(c), a 4-cm-long MMF (core diameter = 105 $\mu$m, NA = 0.22) is able to cover the entire visible spectrum with 1 nm wavelength resolution in a single measurement \cite{redding2014high}.

	One advantage of the MMF spectrometer, over the conventional grating spectrometers, is that the spectral channels in one measurement do not need to be contiguous in wavelength. If it is known {\it a priori} that the signals are located in certain wavelength regions that are disconnected, the wavelengths in $P(r, \varphi; \lambda)$ may be assigned only to those regions. In this way, a finite number of spectral channels can effectively cover a broad wavelength range. 
	
	To recover a large number of spectral channels $M_\lambda$, it is important to excite as many modes as possible at the fiber input. This is realized by using a fiber mode scrambler \cite{valley2016multimode} or off-set fusion of fibers \cite{wang2020study}.  The upper bound for $M_\lambda$ is the total number of fiber modes, which is proportional to the fiber core area. While a large-core MMF provides a high $M_\lambda$, the speckle intensity contrast $C_t \simeq 1 / \sqrt{M_{\lambda}}$ is low for a dense spectrum of $M_\lambda \gg 1$. Once the contrast is below the noise level, an accurate spectrum recovery is impossible \cite{redding2014noise}.
	
	To overcome the trade-off between spectral range and resolution, a wavelength division multiplexer (WDM) is integrated with a bundle of MMFs \cite{liew2016broadband}. The WDM divides a broad spectrum to multiple windows, with each window being covered by a single MMF. The output speckle patterns from all MMFs are recorded simultaneously by a large-area camera (with a million pixels). Then they are separately processed in parallel, greatly reducing the complexity of the spectrum reconstruction and increasing the speed of reconstruction. Using five 2-meter-long MMFs, dense spectra centered at $\lambda$ = 1500 nm with 100 nm bandwidth are recovered with 0.03 nm resolution in snapshot measurements. 
	
	Compared to other high-resolution spectroscopy tools such as the scanning Fabry-Perot etalon or the virtually imaged phase array (VIPA), a notable advantage of the MMF is that its spectral range is not limited by the FSR. This is because fiber fabrication imperfections (random refractive index variation) and external perturbations (fiber bending, twisting) make the probability of a highly multimode fiber producing identical speckle patterns at distinct wavelengths extremely low. A MMF spectrometer, combined with an optical frequency comb source, has been adopted in broadband metrology-grade spectroscopy to resolve individual comb lines \cite{coluccelli2016optical}. The wavelength resolution obtained with a 100-meter-long MMF is an order of magnitude higher than that of dual-comb spectroscopy. Using the MMF spectrometer, 500 comb lines are measured simultaneously and 3500 lines sequentially for direct comb spectroscopy. 
	
	The MMF also functions as an ultrahigh-resolution wavemeter that can precisely determine a single wavelength \cite{bruce2019overcoming}. By modulating the input signal with an acoustic optical modulator (AOM) and applying the principal component analysis, attometer resolution at $\lambda \simeq$ 780 nm is obtained with a 18-cm-long step-index MMF (core diameter = 105 $\mu$m, NA = 0.22). Multiple wavelengths may be resolved simultaneously by modulating them at different acoustic frequencies \cite{bruce2020femtometer}. 
	
	The resolving power of a MMF spectrometer/wavemeter can be further improved by increasing the fiber length and/or the differential group delay. However, the speckle patterns are sensitive to environmental changes including mechanical vibration and temperature drift. Recent studies show that a combination of thermal and mechanical stabilization together with software correction can enable robust performance of a high-resolution MMF spectrometer \cite{redding2014high, coluccelli2016optical}. Alternately, the spectral-to-spatial mapping $P(r, \varphi; \lambda)$ may be calibrated as a function of temperature, and the one matched to the current temperature will be used for spectrum recovery.

	Further stabilization has been realized with on-chip implementation of the MMF spectrometer. A silicon multimode waveguide (width = 10 $\mu$m, length = 18 mm) is coiled in an Archimedean spiral \cite{redding2016evanescently}. By introducing evanescent coupling of light between adjacent waveguide arms (separated by small air gap), light will leap forward or backward in time, greatly broadening the  optical path-length distribution. Consequently, the spectral resolution is dramatically enhanced, and such enhancement is nonresonant and broadband. The wavelength resolution of 10 pm is obtained at $ \lambda \simeq$ 1500 nm with a spiral of outer radius 250 $\mu$m. Like the MMF, the spiral waveguide effectively disperses light at any wavelength at which the material absorption is negligible. 
	
	Since the output speckle pattern varies with the launch condition of input light, it is crucial to ensure the consistency between the launch used in the calibration and that in spectrum measurement. On the other hand, the input launch condition may be switched among multiple choices to increase the number of spectral channels $M_\lambda$ \cite{piels2017compact}. The space-division multiplexing scheme has been implemented with a multicore fiber \cite{meng2019multimode} or an on-chip photonic lantern \cite{yi2020integrated}.
	
	\subsubsection{Hyperspectral imager}
	\label{sec:hyperspec}
	
	\begin{figure}\centering
		\includegraphics[width=0.8\textwidth]{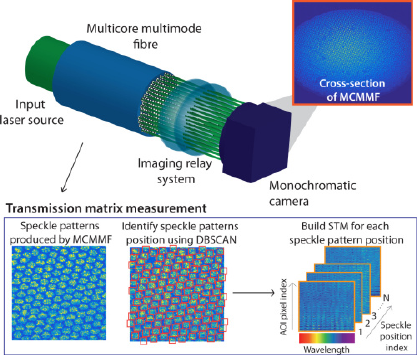}
		\caption{Hyperspectral imaging with multicore multimode fiber. Light travels through a multicore multimode fiber and the cross-section of the end of the fiber is imaged on to a camera. A clustering algorithm detects the coordinates of the speckle patterns produced by each fiber core on the camera. The pattern coordinates are then used to calibrate a spectral intensity transmission matrix by measuring wavelength-dependent speckle patterns at every core position and storing them in a 3-dimensional data cube. Figure reprinted from Ref.~\cite{french2018snapshot} with permission. Copyright 2018 Optical Society of America.  
			\label{fig:hyperspec}
		}
	\end{figure}
	
	Hyperspectral imaging (over a continuous spectrum) and multispectral imaging (at multiple frequencies) are of great importance for acquiring both spatial and spectral information, with applications in remote sensing and threat detection. Scanning-based hyperspectral imaging techniques, such as pushbroom and whiskbroom spectral imaging systems, require performing many sequential measurements to reconstruct an image. A more desirable approach is to acquire both spatial and spectral information in one measurement. Such snapshot hyperspectral imaging has recently been demonstrated with a multicore multimode fiber array in combination with a monochrome CMOS camera \cite{french2018snapshot}. As shown in Fig.~\ref{fig:hyperspec}, 3000 cores in a 30-cm-long fiber bundle produce distinct speckle patterns, from which space-dependent spectra are recovered with sub-nanometer wavelength resolution. A clustering algorithm is employed in combination with $l_1$-minimization to limit data collection at the acquisition stage for the reconstruction of spectral images that are sparse in the wavelength domain. The efficient acquisition of spatial and spectral information in one measurement enables high-resolution, high-throughput snapshot spectral imaging. 
	
	In addition to compressive sensing algorithms, deep learning has been employed for fast and reliable reconstruction of both discrete and continuous spectra from speckle patterns \cite{kurum2019deep}. The artificial neural network is trained with the multispectral datasets numerically constructed from the measured spectral PSFs. Compared to analytical inversion methods and compressive sensing algorithms, the deep learning approach is robust to system drift and measurement noise. Although it takes time to train an artificial neural network, once it is done, the reconstruction is fast enough for real-time recovery of hyperspectral information. 
	
	\subsection{Full-field measurement}
	\label{sec:timeMeasurement}
	

	Full-field (both amplitude and phase) characterization of ultrashort optical pulses with femtosecond to attosecond pulse duration is crucial to ultrafast science. Existing sensors are not fast enough to directly resolve such short pulses, and the pulses are recovered indirectly, often through algorithmic methods. Various schemes have been developed for multi-shot measurements, which require trains of identical pulses. However, in many cases the probed pulses are not reproducible, prompting the development of single-shot characterization methods.
	
	Single-shot measurements are more challenging for several reasons. First, a trade-off between the temporal range $\Delta t$ and resolution $\delta t$ limits the time-bandwidth product $\Delta t/ \delta t$, where $1/\delta t$ is proportional to the spectral bandwidth of detection. Secondly, the single-shot measurement is highly sensitive to noise, as there is no averaging over multiple pulses and the power of an incoming signal is always limited. Thirdly, a full-field characterization requires recovering the amplitude $A(t)$ and phase $\theta(t)$ simultaneously, and both may evolve rapidly in time $t$. 
	
	A Fourier transform of the temporal field $E(t) = A(t) e^{i \theta(t)}$ gives the spectral field $E(\omega) \equiv \mathcal{F}[E(t)] = A(\omega) e^{i \theta(\omega)}$.  Simultaneous reconstruction of spectral amplitude $A(\omega)$ and phase $\theta(\omega)$ will recover $E(\omega)$ and $E(t)$. As shown in section~\ref{sec:MMFspec}, the power spectrum $I(\omega) = |E(\omega)|^2 = [A(\omega)]^2$ can be reconstructed from a time-integrated intensity measurement of optical speckle patterns generated by a multimode fiber. However, such speckle patterns do not contain any information of spectral phase, thus preventing complete pulse recovery. 
	
	Lately two methods have been developed for single-shot full-field measurement of optical pulses using (i) spatio-temporal speckles and (ii) nonlinear speckles. This subsection will introduce these methods. While method (i) employs a reference pulse, (ii) is reference-free or self-referenced.  
	
	\subsubsection{Parallel temporal ghost-imaging}
	
	\begin{figure}\centering
		\includegraphics[width=0.8\linewidth]{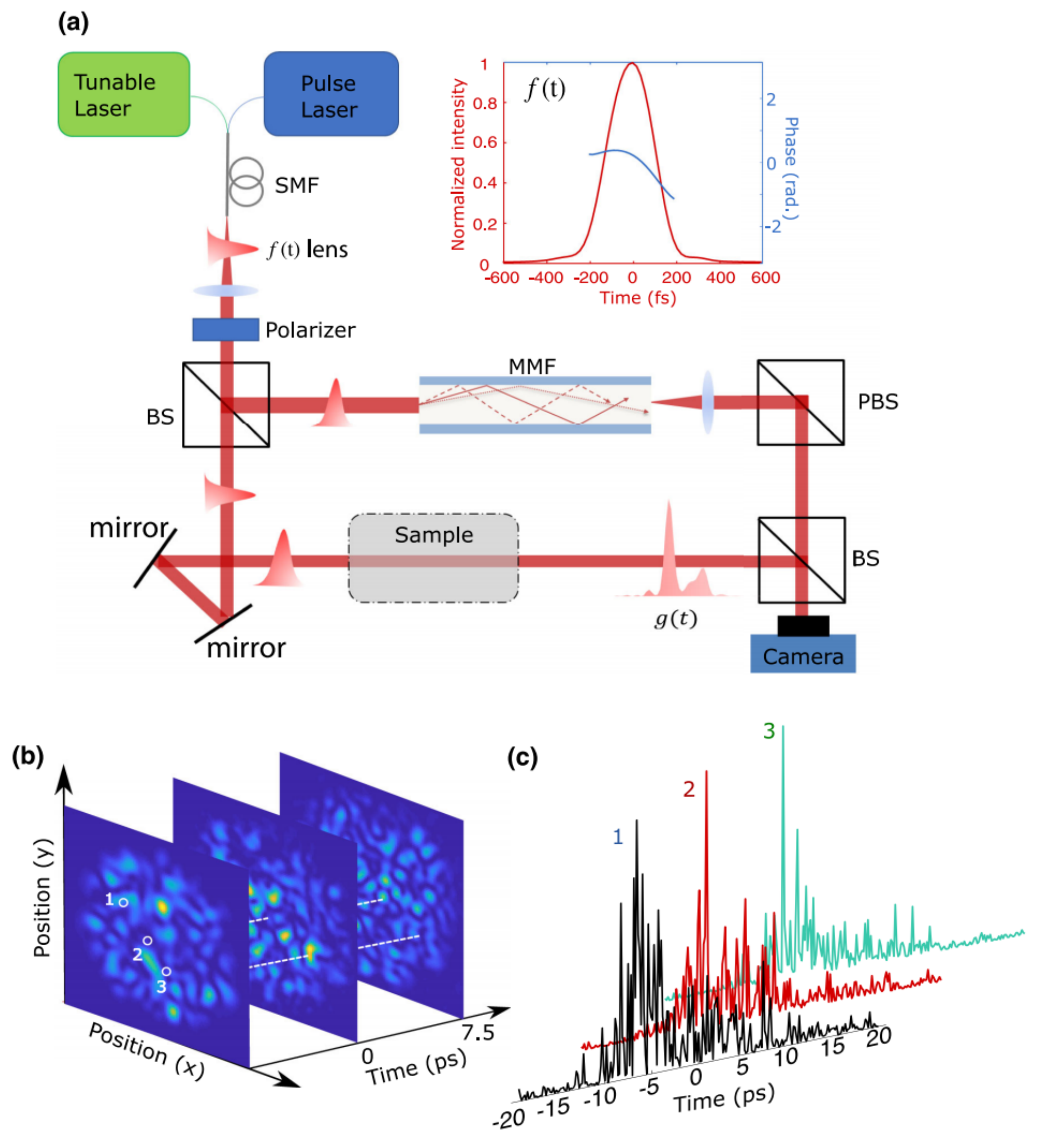}
		\caption{Single-shot pulse characterization using spatio-temporal speckles. 
			(a) Schematic of an interferometric setup for calibration of a spatio-temporal speckle field generated by a reference pulse through a multimode fiber (MMF). Its interference with the signal (pulse transmitted through a sample) is measured without temporal resolution by a camera. The time-integrated measurement enables full field recovery of the signal. Inset: intensity (red solid line) and phase (blue solid line) of the reference pulse launched into the MMF. 
			(b) Spatial field amplitude distribution of the reference pulse transmitted through the MMF evolving with time. (c) Transmitted field amplitudes at different spatial positions of the fiber output exhibit distinct temporal fluctuations. Figure reprinted from Ref.~\cite{xiong2020multimode} with permission. Copyright 2020 Optical Society of America. }
		\label{fig:LinearPulse} 
	\end{figure}
	
	Figure~\ref{fig:LinearPulse}(a) illustrates the single-shot full-field measurement using spatio-temporal speckle pattern out of a multimode fiber (MMF) \cite{xiong2020multimode}. A linearly polarized laser pulse with known field $f(t)$ is split into two arms of a Mach-Zehnder interferometer. In one arm, the pulse is launched into a multimode fiber (MMF) to create spatio-temporal speckle field $E_s(\mathbf{r},t)$ in transmission. In the other arm, the pulse interacts with a sample, and the transmitted/reflected field $g(t)$ interferes with $E_s(\mathbf{r},t)$. The time-integrated intensity pattern $I_p(\mathbf{r}) = \int |E_s(\mathbf{r},t)+g(t)|^2 dt$ is recorded by a slow camera. By applying a Hilbert filter in the Fourier domain of $I_p(\mathbf{r})$, the interference term $I_q(\mathbf{r}) = \int dt [E(\mathbf{r},t) \, g^*(t) +  E(\mathbf{r},t)^* \, g(t)]$ is extracted. 
	
	In the frequency domain, $ E_s(\mathbf{r},\omega) = \mathcal{F}[E_s(\mathbf{r},t)] =  T(\mathbf{r},\omega) \, F(\omega)$, where $T(\mathbf{r},\omega)$ is the frequency-resolved transmission matrix of the MMF, and $F(\omega) = \mathcal{F}[f(t)]$. The interference term can be expressed in the frequency domain as  
	\begin{equation}
		I_q(\mathbf{r}) =  \begin{bmatrix}
			T(\mathbf{r},\omega) F(\omega) &
			T^*(\mathbf{r},\omega) F^*(\omega)
		\end{bmatrix}
		\begin{bmatrix}
			G^*(\omega)\\
			G(\omega)
		\end{bmatrix},
		\label{eq:Id}
	\end{equation}
	where $G(\omega) = \mathcal{F}[g(t)]$. Once $T(\mathbf{r},\omega)$ is calibrated and $F(\omega)$ is known, $G(\omega)$ can be retrieved from $I_q(\mathbf{r})$, and an inverse Fourier transform gives $g(t)$. 
	
	The temporal range  $\Delta t$ of the single-shot measurement is set by the temporal length of $E_s(\mathbf{r},t)$, which is inversely proportional to the spectral correlation width of the MMF. A fiber with longer length and/or stronger modal dispersion has a faster spectral decorrelation, and can cover a longer time window. The temporal resolution $\delta t$ is equal to the temporal speckle size, which is given by the inverse of the spectral bandwidth of the reference pulse. A reference pulse of broader bandwidth (not necessarily transform-limited in time) provides higher temporal resolution. Therefore, the temporal range and resolution can be tuned separately by varying the parameters of the fiber and the reference pulse. Experimentally, single-shot full-field measurement with a 1.8-m-long MMF (105 $\mu$m core diameter, 0.22 NA) provides 230 fs temporal resolution over a window of 35 ps, and the time-bandwidth product is 152 \cite{xiong2020multimode}. 
	
	The time-bandwidth product reflects the number of independent temporal channels that are measured simultaneously. Its upper bound is given by the number of uncorrelated time traces generated by the MMF, which is equal to the number of guided modes in the fiber. The time-bandwidth product for a MMF with a large core and a high NA may well exceed 1000, and a further increase is possible with a bundle of MMFs \cite{liew2016broadband}. 
	
	This single-shot measurement scheme is equivalent to parallel ghost imaging in time \cite{ryczkowski2016ghost}. While conventional ghost imaging relies on sequential generation of different temporal waveforms for sampling, the MMF simultaneously creates many distinct time traces in different spatial channels to sample the signal [Fig.~\ref{fig:LinearPulse}(c)]. The complex yet deterministic spatio-temporal speckle field enables parallel sampling in a single shot for a full-field recovery, eliminating the requirement for repetitive signals.
	
	The spatio-temporal coupling of broadband light in a MMF has also been employed for axial reflectivity profiling \cite{lee2020single}. The closely spaced yet distinct propagation constants of various spatio-temporal modes in a MMF yield a set of spatially distributed functions that are distinct for varying delay-times (path-lengths). Such encoding functions can be used for depth referencing. A signal that is delayed by reflection at a specific sample depth only interferes with the path-length-matched encoding function [Fig.~\ref{fig:LinearPulse}(b)], and creates a distinct speckle pattern that is unique to this depth. An arbitrary sample reflectivity profile produces a linear superposition of the associated speckle patterns. By calibrating the random encoding functions, the 1D sample depth profile is reconstructed from a single recording of the interference pattern without the need for any mechanical or optical scanning. Such configuration can be considered as spatially multiplexed low-coherence interferometry, and is a parallel implementation of ghost optical coherence tomography \cite{amiot2019ghost}. Experimentally, axial depth profiling is demonstrated with bandwidth-limited resolution of 13.4 $\mu$m over a range of 13.4 mm \cite{lee2020single}.  By varying the MMF length, the depth range is scaled from several millimeters to well beyond one centimeter, relaxing the challenging hardware requirements of achieving similar performance with conventional spectrometer-based coherence gating. 
	
	\subsubsection{Nonlinear speckle and deep learning}\label{nonlin+deep}
	
	\begin{figure}\centering
		\includegraphics[width=\linewidth]{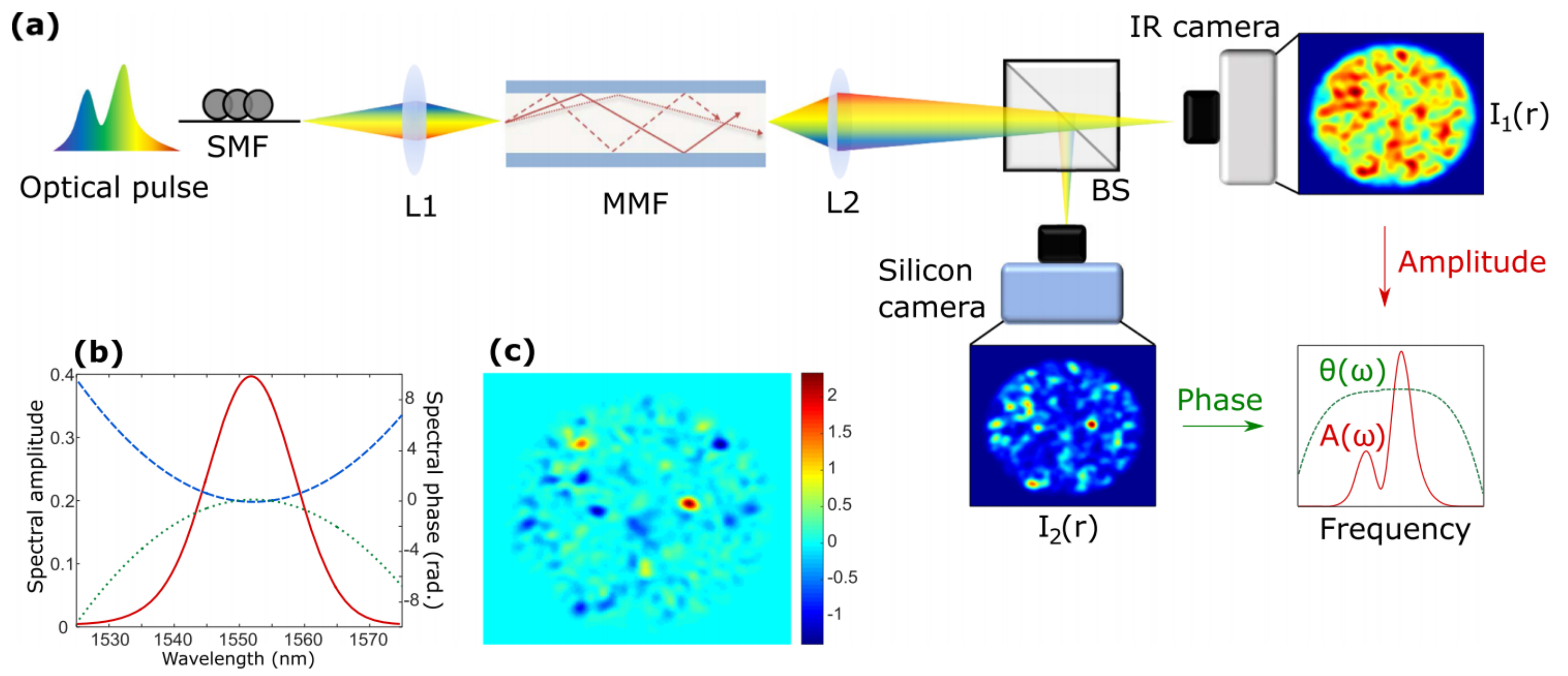}
		\caption{Full-field recovery with linear and nonlinear speckles. (a) Schematic of experimental realization. An optical pulse at $\lambda$ = 1550 nm is delivered via a single-mode fiber (SMF) to a multimode fiber (MMF). At the MMF output, a time-integrated linear speckle pattern $I_{1p}(\mathbf{r})$ is recorded by an IR camera via one-photon absorption, and a nonlinear speckle pattern $I_{2p}(\mathbf{r})$ is recorded by a silicon camera via two-photon absorption. The spectral amplitude $A(\omega)$ and phase $\theta(\omega)$ of the input pulse are recovered from $I_{1p}(\mathbf{r})$ and $I_{2p}(\mathbf{r})$, respectively. (b) An optical pulse has spectral amplitude $A(\omega)$ (red solid line) and phase $\theta(\omega)$ (blue dashed line). The green dotted line shows the phase flipped to $-\theta(\omega)$ (green dotted line), which corresponds to temporal inversion with phase conjugation. (c)  The difference in $I_{2p}(\mathbf{r})$ between the two pulses with opposite spectral phases in (b) indicates that the sign of the spectral phase can be recovered from the nonlinear speckle pattern. Figure reprinted from Ref.~\cite{xiong2020deep}, licensed under a Creative Commons Attribution (CC BY) license.}
		\label{fig:NonlinearPulse} 
	\end{figure}
	
	The method described in the last subsection relies on a known optical pulse that is mutually coherent with the unknown signal. In this subsection, a reference-free time-integrated measurement scheme will be introduced to enable stand-alone single-shot characterization of ultrafast pulses. 
	
	The key for full-field recovery is retrieving the spectral phase $\theta(\omega)$, as the spectral amplitude $A(\omega)$ is already reconstructed from the linear speckle pattern. Measuring the relative phase of different spectral components in a pulse requires these components to interfere, but distinct frequencies do not interfere in a linear, time-integrated detection. To retrieve the spectral phase without any reference, a nonlinear material is placed behind a MMF to create a speckle pattern via two-photon absorption or sum-frequency generation \cite{xiong2020deep, ziv2020deep}. 
	
	As illustrated in Fig.~\ref{fig:NonlinearPulse}(a), the propagation of a signal pulse in the MMF remains linear, so that the transmitted light is related to an incident field $E_i(t) = \mathcal{F}^{-1}[A(\omega) \, e^{i \theta(\omega)}]$ by the field transmission matrix: $E_t(\mathbf{r},t) = \mathcal{F}^{-1}[T(\mathbf{r},\omega) \, A(\omega) \, e^{i \theta(\omega)}$, where $\mathcal{F}^{-1}$ denotes the inverse Fourier transform. The time-integrated linear speckle pattern $I_{1p}(\mathbf{r}) = \int|E_t(\mathbf{r},t)|^2 \, dt 
	= \int |T(\mathbf{r},\omega)|^2 |A(\omega)|^2 \, d\omega$ encodes only the spectral amplitude $A(\omega)$ of an input signal. The time-integrated nonlinear speckle pattern $I_{2p}(\mathbf{r}) = \int |E_t(\mathbf{r},t)|^4 \, dt$ can be expressed as
	\begin{eqnarray}
		I_{2p}(\mathbf{r}) & = & 
		\iiint d\omega_1 \, d\omega_2 \, d\omega_3 \, 
		|T(\mathbf{r},\omega_1)| \, A(\omega_1) \,  |T(\mathbf{r},\omega_2)| \, A(\omega_2)  \nonumber \\
		&  & |T(\mathbf{r},\omega_3)| \, A(\omega_3) \, |T(\mathbf{r},\omega_1-\omega_2+\omega_3)| \,   \nonumber  \\
		&  & A(\omega_1-\omega_2+\omega_3)\, e^{i[\theta(\omega_1)-\theta(\omega_2) +\theta(\omega_3)-\theta(\omega_1-\omega_2+\omega_3)]}  \nonumber \\ 
		&  &  e^{i[\theta_{\rm TM}(\mathbf{r}, \omega_1)- \theta_{\rm TM}(\mathbf{r}, \omega_2) + \theta_{\rm TM}(\mathbf{r}, \omega_3)- \theta_{\rm TM}(\mathbf{r}, \omega_1 -\omega_2+\omega_3)]} \, , 
		\label{eq:I2}
	\end{eqnarray}
	where $\theta_{\rm TM}(\mathbf{r},\omega)$ is the phase of $T(\mathbf{r},\omega)$. $I_{2p}(\mathbf{r})$ encodes the spectral phase $\theta(\omega)$ of the signal, because different spectral components interfere in the nonlinear process of two-photon absorption. Moreover, the complex interference eliminates  ambiguity in the sign of the spectral phase, allowing the direction of time to be recovered [Fig.~\ref{fig:NonlinearPulse}(b,c)]. This is advantageous over other self-referenced nonlinear techniques that cannot distinguish the direction of time \cite{zahavy2018deep}. 
	
	In practice, retrieving the input spectral phase $\theta(\omega)$ from the nonlinear speckle intensity $I_{2p}(\mathbf{r})$ is highly non-trivial, because the inverse mapping from $I_{2p}(\mathbf{r})$ to $\theta(\omega)$ is complicated. Conventional phase retrieval algorithms are sensitive to noise in the nonlinear snapshot measurement, and thus cannot provide a reliable recovery. A deep neural network provides here a useful alternative to learn the nonlinear mapping from $I_{2p}(\mathbf{r})$ to $\theta(\omega)$ \cite{xiong2020deep}. Once the field transmission matrix $T(\mathbf{r},\omega)$ of the fiber is calibrated, it is straightforward to calculate the nonlinear speckle pattern for any input puls using Eq.~(\ref{eq:I2}). However, a large amount of data needs to be numerically synthesized to train a convolutional neural network. Furthermore, deep learning can be combined with compressive sensing by representing the spectral phase in a sparse basis to dramatically reduce the number of parameters that the neural network predicts. Measurement noise and fiber fluctuation are incorporated to the training data, so that the trained network is robust and outperforms classical algorithms. 
	
	Since the nonlinear speckle pattern encodes both amplitude and phase information, it may be used to recover $A(\omega)$ and $\theta(\omega)$ simultaneously, eliminating the linear speckle measurement. Numerically, a deep neural network is trained for pulse amplitude and phase recovery \cite{ziv2020deep}. This technique is shown to be robust to noise and inaccuracy in system parameters, and capable of reconstructing ultrashort pulses at low SNR. It mitigates the need for iterative optimization, which is usually slow and hampered by the presence of noise. Therefore, the deep learning method is advantageous for real time probing of ultrafast processes under noisy conditions. 
	
	The speckle-based pulse characterization scheme can be tuned to measure optical pulses of varying length. The temporal range of measurement is determined by the spectral decorrelation of the MMF, and the temporal resolution depends on the spectral bandwidth over which the field transmission matrix is calibrated. Experimentally, a 1.3-m-long step-index fiber with 105 $\mu$m core and 0.22 NA is calibrated over a wavelength range of 1525 nm - 1575 nm, and provides a temporal resolution of 160 fs over a time window of 30 ps. The resolution can be further increased by calibrating the MMF over a larger bandwidth, and the temporal range can be extended by using a longer fiber with faster spectral decorrelation. Finally, the MMF may be replaced by a random scattering medium or an on-chip multimode waveguide. 
	
\subsection{Compressive Radio-Frequency Receiver}
\label{RFreceiver}

Nyquist rate sampling of Gigahertz-band radio frequency (RF) signals rapidly generates huge amounts of data. To address this general issue, compressive sensing (CS) techniques are developed for sparse signals and images. Electronic CS systems suffer timing jitter and amplitude noise. Although microwave photonic CS systems have equivalent issues, in many cases the distortions are static or lower in frequency and amenable to calibration. Previously pseudo-random bit sequences have been created to modulate the optical carriers for compressive sensing. Recently optical speckle patterns, generated by multimode fibers, are employed for sparse RF signal recovery\cite{valley2016multimode, sefler2018demonstration}.

\begin{figure}\centering
		\includegraphics[width=0.8\textwidth]{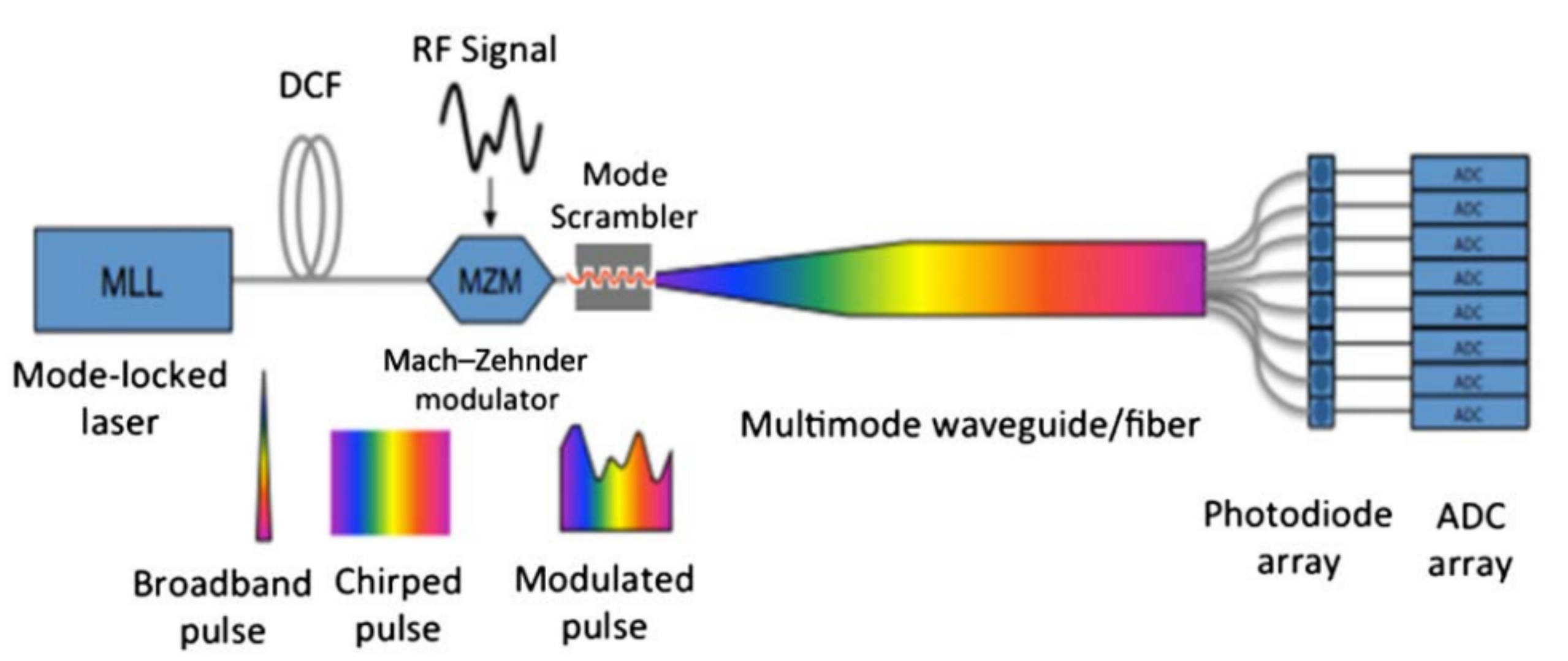}
	\caption{\label{fig:RFspeckle} Compressive sensing of RF signals. Schematic of RF signal recovery from optical speckle measurement. \textcolor{black}{Broadband} optical pulses from a mode-locked laser (MML) are stretched by a dispersion compensating fiber (DCF), and the RF signals are applied to the chirped pulse in a Mach–Zehnder modulator (MZM), then the modulated pulses are sent through a multimode fiber, and the transmitted speckle patterns are detected by a photodiode array. Figure reprinted from \cite{valley2016multimode} with permission. Copyright 2016 Optical Society of America.
	}
\end{figure}

Figure~\ref{fig:RFspeckle} is a schematic showing a multimode fiber replacing a 2D SLM in generating pseudo-random bit sequences in a modulated wideband converter \cite{valley2016multimode}. Femtosecond pulses from a mode-locked laser (MLL) are launched into a dispersion-compensating fiber (DCF), which stretches the pulse to the inter-pulse time. Then the optically chirped pulses pass through a Mach–Zehnder modulator (MZM) that imprints the RF signal on the optical intensity, and subsequently enter a multimode fiber. The transmitted signals are spatially split at the output of the guide and directed to an array of photodiodes. The integration times of the photodiodes are matched to the period of mode-locked pulses, and the electrical signals from the photodiode array are digitized by an array of analog-to-digital converters (ADCs) clocked to the pulse repetition rate. An optional fiber mode scrambler is placed near the input end of the fiber to fully excite the guided modes. The time-wavelength mapping is depicted by the rainbow-colored pulse icons. The output speckle pattern from the multimode fiber varies with wavelength, and hence with  time. The optical spectrum recovered from the speckle pattern provides temporal variation of the RF signal through the time-wavelength mapping. CS algorithms are employed to recover different types of sparse signals, e.g., sparse in time, in frequency, or after the Harr wavelet transform \cite{valley2016multimode}. 

Instead of mapping from space to wavelength and then to time, it is possible to combine them for a direct mapping from space to time or to RF frequency. In addition to the RF signal amplitude, its relative phase (with respect to the optical pulse train from the mode-locked laser) can be recovered from the speckle measurement. The mapping matrix has two columns for each RF tone, which are referred to as in-phase and quadrature components. Three methods have been explored for calibration: the first one estimates the two columns from measurements with pairs of pulses, the second method fits the data as a function of the relative RF phase, and the third one is based on singular value decomposition (SVD) of the calibration matrix \cite{sefler2018demonstration}. Finally, a penalized $l_1$ norm method recovers the amplitude, phase, and frequency of sparse RF signals. Using a 5-m-long MMF (0.22 NA step-index, 105-$\mu$m core), RF signals with one tone or two tones in the 2-19 GHz band are reconstructed with 100 MHz resolution in a single 4.5 ns pulse. 

To reduce the device footprint, the MMF is replaced by an 11-cm-long multimode waveguide (96 $\mu$m wide, and 220 nm thick), which is integrated with a multi-port splitter and a grating coupler array in a photonic circuit \cite{borlaug2021photonic}. The planar waveguide is wrapped in a spiral geometry and produces random projections for CS via optical speckle. A precise analog signal routing is supported by two integrated structures, waveguide bus trombone flare and matched 90$^{\circ}$ bus bend, in the silicon chip. 16 RF channels, each with an effective sampling rate of 35 MSps (mega samples per second), are able to recover RF signals across a 2 GHz bandwidth from 2.5 to 4.5 GHz. Compared with direct Nyquist sampling which requires 4 GSps, the CS requires $4000 / (35 \times 16) = 7$ times fewer recorded samples and a maximum sampling rate that is $4000 / 35 = 114$ times smaller \cite{borlaug2021photonic}. Sampling well below the Nyquist limit reduces the need for data storage, shortens data-payload transmission time, and reduces the overall receiver size, weight, and power.

	\subsection{Remote key establishment}
	
	
	There has been an ever growing demand for improving the security levels of optical communication networks. The challenge for establishing a secure communication channel is to distribute secret keys between remote users without exposing them to an eavesdropper. One of the most widely used solution is asymmetric keys, where the encryption is performed with a public key that is available to everyone, but the decryption requires a private key that is only available to the receiver. The security is based on the assumption that it is practically impossible to decrypt the information using only the public key (without knowing the private key). Therefore, generating and sharing secret keys are essential to secure communications. Quantum key distribution takes advantage of the ``no-cloning theorem'' and guarantees that an eavesdropper cannot reveal the key without being exposed. However, this scheme requires transmission and detection of single photons, which can be technically challenging to implement. 
	
	Single mode fibers (SMFs) have been explored for classical key generation and distribution, utilizing phase and polarization fluctuations of transmitted light caused by passive environmental changes and/or active perturbations \cite{kravtsov2013physical, hajomer2018key, zaman2018physical, zhang2019error}. Active polarization scrambling is incorporated to increase the key generation rates \cite{hajomer2019accelerated, hajomer2021284}, which has reached 2.7 Gbps (Giga-bits per second) over a 10-km SMF \cite{zhang20212}. Also lasing oscillation between the sender and receiver of a SMF link is explored for secure key distribution \cite{scheuer2006giant}. An ultra-long SMF laser, based on Raman gain, provides error-free distribution of random keys with an average rate of 100 bps (bits per second) over 500 km \cite{el2014secure}.
	
	\begin{figure}[htbp]
		\centering
		\includegraphics[width=0.7\linewidth]{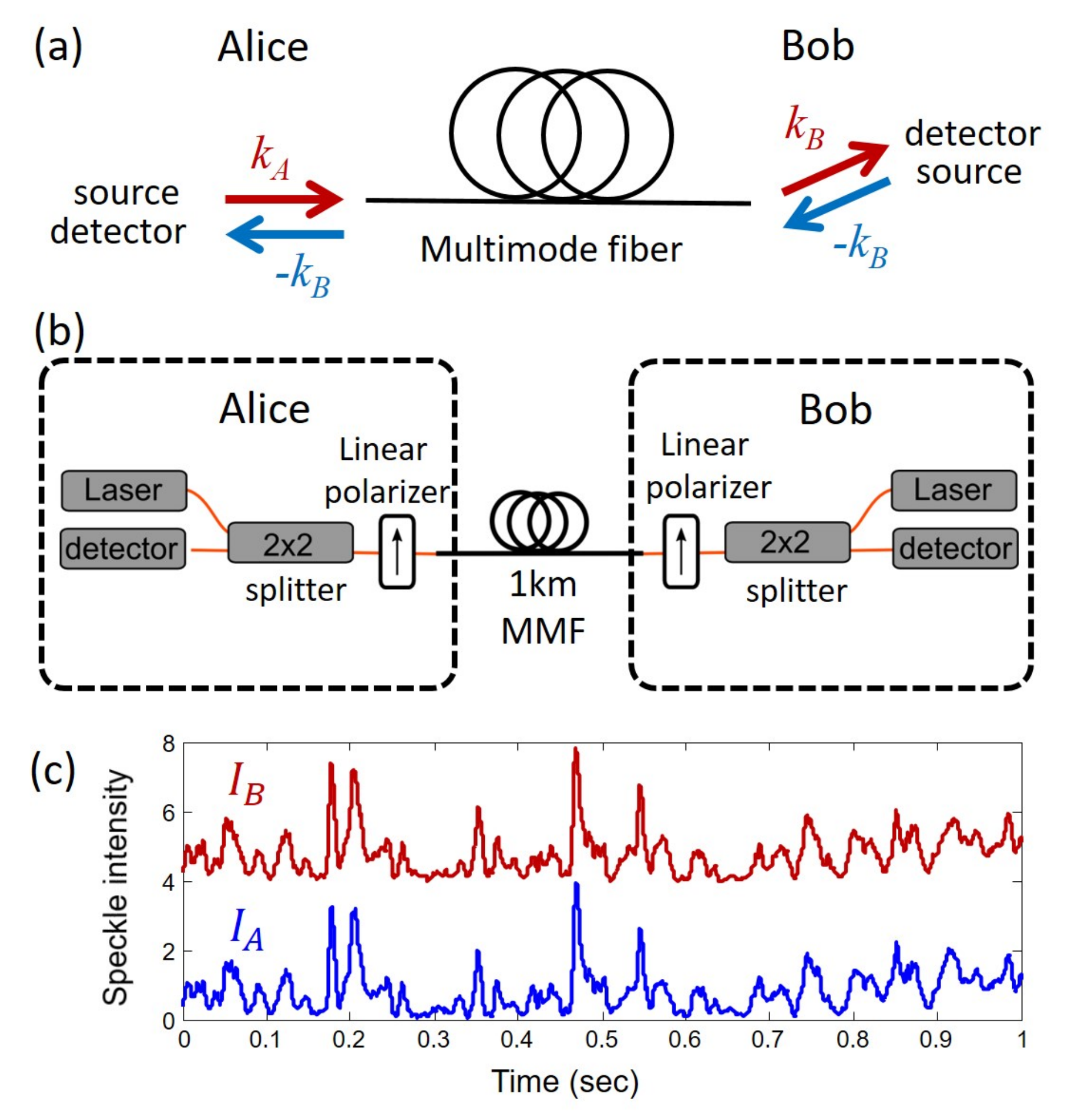}
		\caption{Classical key establishment with a multimode fiber with strong mode and polarization mixing. 
			(a) Alice (Bob) couples a laser beam into the MMF with wavevector ${\bf k}_A$ (${\bf k}_B$), and measures the transmitted intensity from Bob (Alice) $I_A$ ($I_B$) of $-{\bf k}_B$ ($-{\bf k}_A$). The optical reciprocity demands $I_A = I_B$. 
			(b) All-fiber implementation of remote key establishment between Alice and Bob through a MMF. 
			(c) Changes of environmental conditions make $I_A$ and $I_B$ fluctuate randomly in time, but their values are always correlated. $I_A$ and $I_B$ are then digitized to provide binary keys shared by Alice and Bob. Figure adpated from Ref.~\cite{bromberg2019remote} with permission. }
		\label{fig:key}
	\end{figure}
	
	Here we focus on harnessing the complexity of a MMF for secure key establishment between two remote parties - Alice and Bob \cite{bromberg2019remote}. Since ambient temperature fluctuations and mechanical strains have strong impact on random mode and polarization mixing in a MMF, the transmitted speckle pattern is extremely sensitive to environmental perturbations and constantly changes in time. Such random fluctuations are utilized for remote key establishment. By virtue of the optical reciprocity principle, Alice and Bob at the two ends of a MMF can share identical copies of the keys extracted from speckle intensity fluctuations. The keys, which are constantly updated due to intrinsic fluctuations of the fiber, can then be used to encode and decode information being sent over a standard unsecure communication channel. 
	
	Figure~\ref{fig:key} illustrates the concept of this scheme. Alice and Bob simultaneously couple a laser beam of the same frequency into the opposite ends of the MMF with wavevectors ${\bf k}_A$ and ${\bf k}_B$, respectively. Both Alice and Bob measure the speckle patterns at the far-field ($k$-plane) of the opposite fiber facets. In the presence of strong mode coupling, Alice and Bob record different speckle patterns. However, if the fiber link preserves time-reversal symmetry, then, due to reciprocity, the intensity measured by Bob at the speckle grain which corresponds to $-{\bf k}_B$, is identical to the intensity measured by Alice at the grain which corresponds to her input mode, $-{\bf k}_A$. Due to changes in the environmental conditions of the fiber, the mode mixing in the fiber constantly changes, and therefore the speckle patterns that Alice and Bob observe change as well. Nevertheless, optical reciprocity guarantees that the intensities measured at $-{\bf k}_A$ and $-{\bf k}_B$ will still be correlated, as long as the fiber is static during the time it takes for the signals from Alice and Bob (or vice versa) to propagate through the fiber. In this way Alice and Bob generate a common random analog signal, which they can further digitize for obtaining a binary key. 
	
	Alice (or Bob) does not need to know which wavevector ${\bf k}_B$ (or ${\bf k}_A$) that Bob (or Alice) chooses. Since the input and output channels are specified not only by the wavevector but also by the polarization, linear polarizers are placed at both ends of the fiber. Note that the two polarizers do not need to select the same polarization state, but can be oriented at any arbitrary angle. Since the reciprocity principle is not restricted to the wavevector space, Alice and Bob can use channels in other spaces, e.g., the position space or the guided-mode space. Moreover, either of them can choose a channel in different space without knowing which space or channel is selected by the other. 
	
	The remote key establishment relies on classical light and standard passive fibers, and can therefore be implemented in existing fiber networks \cite{bromberg2019remote}. An all-fiber setup using standard off-the-shelf components is built [Fig.~\ref{fig:key}(b)]. The nodes of Alice and Bob at each side of the multimode fiber channel consist of four elements, a continuous-wave laser at $\lambda$ = 1550 nm, a photodetector, an in-fiber 2x2 splitter and an in-fiber polarizer. The nodes are coupled to the multimode fiber channel using a SMF, which together with the polarizer assures that the detection and illumination modes at each side of the multimode link are identical. Figure~\ref{fig:key}(c) shows the time traces of intensities recorded by Alice and Bob. The intensity fluctuations are highly correlated. A key rate of 20 bps over 1 km is obtained \cite{bromberg2019remote}.
	
	Since the external perturbations distributed along the fiber act as the source of the random fluctuations in the transmittance, the key rate-distance product is limited by the speed of light. To increase the key rate and the distance of communication, a fast light modulator is added to one end of the MMF \cite{sampson2021high}. The decoupling of key rate and distance enables a higher key rate-distance product. Experimentally, the key rate-distance product is increased to 64.7 Mbps × 12 km. 
	
	The security of remote key establishment via a MMF is based on a fundamental asymmetry associated with the physical layer: the sophistication of optical tools needed by an eavesdropping adversary (Eve) to subvert the key establishment is significantly greater and more costly than the complexity needed by the legitimate parties to implement the scheme \cite{bromberg2019remote}. The signal-to-noise ratio (SNR) for Eve is significantly lower than that for Alice and Bob. Therefore, security is guaranteed as the legitimate users have access to a common source of randomness through channels that are less noisy than the channel the eavesdropper has access to \cite{cover1998comments, csiszar2011information}. 
	
	Beyond classical key distribution (CKD), quantum key distribution (QKD) is realized with transmission and detection of single photons or weak light pulses through optical fibers. QKD takes advantage of the no-cloning theorem and guarantees that an eavesdropper cannot reveal the key without being exposed. While SMFs are widely used, multi-core and multimode fibers are recently explored for high-dimensional QKD \cite{ding2017high, amitonova2020quantum}. The secure key establishment via a MMF with random mode mixing relies on secure characterization of the multimode transmission channel \cite{amitonova2020quantum}. Alice and Bob can characterize the scrambled communication channel in a calibration phase and undo the scrambling by wavefront shaping in the communication phase. Eve cannot extract enough information to identify a single transmitted symbol from a measurement. By merit of the no-cloning theorem, Eve cannot clone the transmitted symbol.
	
	Finally, analog noise has been utilized to protect optical encryption of signals sent through a SMF \cite{wu2014analog}. Without secret keys, physical layer security is introduced to enhance the information security in optical fiber networks. The information security is achieved not by exchanging a cryptographic key, but by exploiting the physical properties of the transmission channel itself. In a MMF, physical layer security is implemented by using inverse precoding of spatial wavefront with artificial noise \cite{rothe2020physical}.

	\section{Summary and Outlook}
	\label{sec:conclusion}

In the past decade, the possibility to control and exploit the complexity of light transport through MMFs has been identified simultaneously by a number of sub-communities in optics and photonics. 
While the theoretical foundation for describing the light transmission through ideal as well as imperfect MMFs is already well established, the technology needed for harnessing such complexity has only recently become broadly available with the emergence of fast and accurate wavefront-shaping technologies as well as through ever-more powerful computing algorithms and capabilities.

These advances have opened up a new route to study light transport in realistic MMFs with all sorts of imperfections, and they keep assisting greatly in broadening our understanding of various associated phenomena such as the role of disorder. Random mode coupling has been treated in statistical models, with tunable parameters to match the experimental reality. Such models are very practical and have been shown to make correct predictions, even without discriminating to what extent the mode coupling originates from the intrinsic imperfections of the MMF itself or from external perturbations including experimental limitations. Nevertheless, the amount and the origin of spatial- and polarization-mode coupling are of crucial importance for future exploitations of MMFs, particularly in scenarios where precisely structured optical signals are to be delivered through fibers under bending and twisting deformations. Isolating the disorder due to MMF imperfections from that of the experimental settings remains challenging, and significant efforts are needed to provide the necessary answers. 

In the following two sections, we will discuss newly emerging research directions with MMFs in the context of exploiting, predicting and controlling complexity. On the one had, we will present how disorder and the complex scattering it induces in MMFs is not always detrimental, but can even be beneficial to practical applications \cite{cao2022harnessing}; on the other hand, we will present recent works that show how MMFs can be used to predict and analyze other complex systems in the context of optical computing and machine learning. 

\subsection{Disordered fibers}
To achieve transverse localization of light, random fluctuations of the refractive index that remain invariant in the propagation direction have been intentionally introduced to MMFs and fiber bundles. As a result, light that is injected to one position of the input facet will be localized in the transverse cross-section while propagating longitudinally in the fiber  \cite{mafi2019disordered}. Such fibers are explored for endoscopic imaging \cite{karbasi2013image, karbasi2014image, zhao2018image, zhao2018deep, tuan2019characterization} and information transmission \cite{leonetti2016secure}. 

\begin{figure}[htbp]
	\centering
	\includegraphics[width=\linewidth]{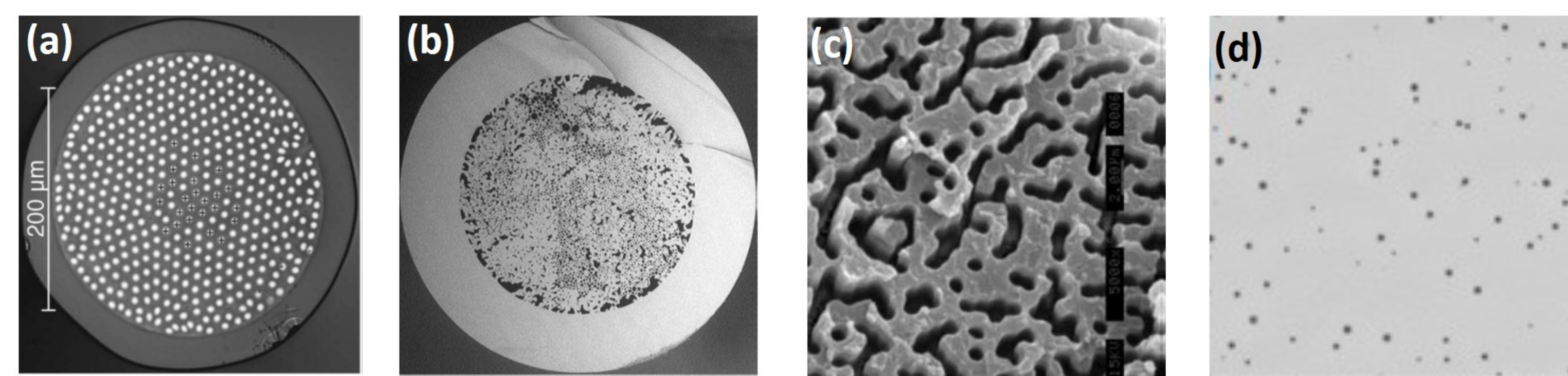}
	\caption{Various types of disordered optical fibers. Microscope images of the fiber cross-section. 
		(a) Disordered array of fiber cores with evanescent coupling.  Figure adapted from Ref.~\cite{pertsch2004nonlinearity} with permission.
		(b) Glass-air random fiber. Figure adapted from Ref.~\cite{zhao2018image}, licensed under a Creative Commons Attribution (CC BY) license. 
		(c) Polymer disordered fiber. Figure adapted from  Ref.~\cite{karbasi2012observation} with permission. Copyright 2012 Optical Society of America.
		(d) Random air line fiber.  Figure adapted from Ref.~\cite{chen2014observing} with permission. }
	\label{fig:disorderMMF}
\end{figure}

The first example is a disordered fiber bundle, i.e., a two-dimensional (2D) array of coupled optical fibers with slightly different and randomly distributed physical parameters, e.g. different radii, random locations \cite{abdullaev1980propagation, pertsch2004nonlinearity}. Because the individual fibers are evanescently coupled, light can tunnel from one fiber to another and spread over the bundle [Fig.~\ref{fig:disorderMMF}(a)]. However, the disorder can stop the lateral spread, namely, when the input light is coupled into a single fiber, the amplitude of the field, on average, decays exponentially away from it in the transverse dimension at the output facet. With sufficient disorder, the transverse decay length is shorter than the transverse dimension of the system, leading to transverse localization of light in the fiber bundle. 

The transverse localization is also achieved with random fluctuations of the refractive index in a single broad-area fiber over its cross-section, where a transverse index profile stays invariant in the longitudinal direction \cite{de1989transverse}. This was first realized \cite{karbasi2012observation} by the stack-and-draw method from two polymers with different refractive indices [Fig.~\ref{fig:disorderMMF}(c)]. Later, disordered glass-air fibers with larger refractive index contrast [Fig.~\ref{fig:disorderMMF}(b,d)] were fabricated \cite{karbasi2012transverse, chen2014observing, zhao2018image}. Transverse scattering and interference of light results in transverse localization, similar to Anderson localization in a 2D disordered structure. It has been shown that such disordered fibers can transport transversely localized beams with low cross-talk \cite{karbasi2013multiple}, facilitating image transfer \cite{karbasi2013image, karbasi2014image, zhao2018image, zhao2018deep, tuan2019characterization} and secure information transmission \cite{leonetti2016secure}.

Beyond conventional optical fibers, novel fiber structures have been designed and fabricated for target applications in recent years \cite{caravaca2021optical, singh2021tunable}. Moreover, novel applications of MMFs have been explored, e.g., optical computing and artificial neural networks, which will be described below.

\subsection{Computation and machine-learning}
		
As discussed in section 4.E.2, the intricate ways in which an input light field gets scrambled when propagating through a MMF can be ``learned'' computationally by the use of tools from artificial intelligence (AI) – enabling various applications in imaging, projection and image transmission. MMFs are, however, not just a problem that AI can be applied to, but MMFs also offer a promising solution to pressing problems in AI (see \cite{RahmaniOguzTeginHsiehPsaltisMoser+2022+1071+1082} for a recent review covering both of these aspects). Consider for this purpose that artificial neural networks, that most implementations of AI are based on, are highly-dimensional systems containing many layers with multiple neurons. The value assigned to an output neuron is typically evaluated as a weighted sum of all input neurons – an operation that constitutes a multiplication between a matrix and a vector in the language of linear algebra. As an enormous number of such operations and related ones (such as singular-value decomposition, Fourier transforms etc.) need to be executed in the training phase of a neural network, it would be highly desirable to implement them in massively parallel systems that operate in a fast and energy-efficient way – especially when compared to implementations involving electronic hardware as in the computers we use on a daily basis. 
	
As it turns out, photonic circuits offer many of these features, making them very attractive as hardware platforms for AI \cite{Wetzstein2020}. Specifically, already the propagation through simple optical elements like (i) a lens, (ii) a thin scattering layer, or (iii) a volumetric scattering medium and a disordered MMF provide very basic operations on the input wave field like (i) a Fourier transform, (ii) a multiplication with a random diagonal matrix or (iii) a multiplication with the pseudo-random transmission matrix. One may wonder at this point, whether the intrinsic randomness of the transmission matrix and the limited control one has over it, would not prevent any useful applications in AI; it turns out, however, that such ``random projections’’ are, indeed, very well suited for an implementation of ``compressive sensing’’ \cite{4472240}, in which a signal can be acquired and reconstructed based on a much reduced number of measurements as would be required based on the seminal Nyquist-Shannon sampling rate. First implemented with disordered media  \cite{Liutkus2014,Ando:15,7472872,9527089}, various tasks such as image classification and object recognition have been successfully demonstrated. 

\begin{figure}[htbp]
		\centering
		\includegraphics[width=0.8\linewidth]{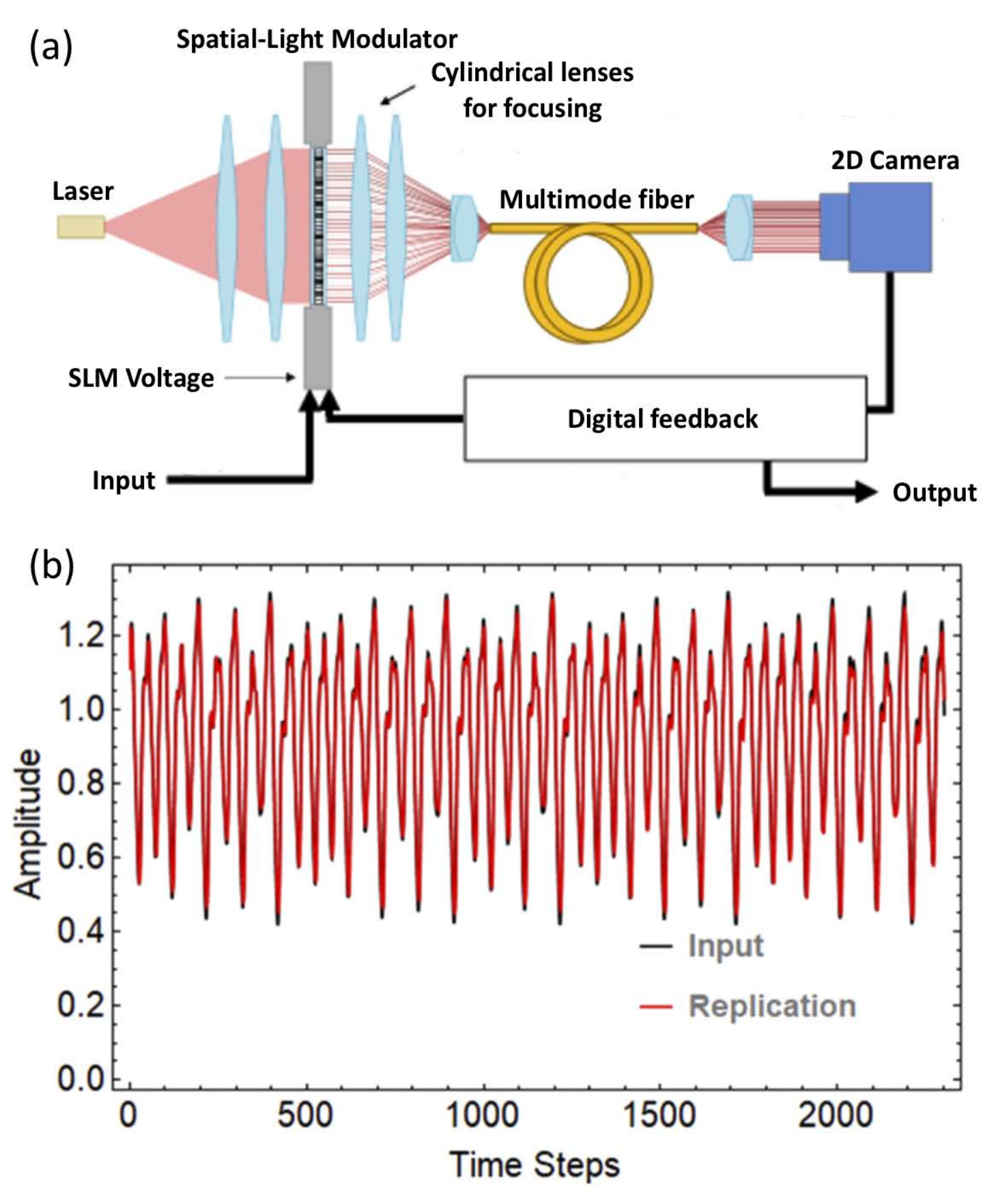}
		\caption{(a) Sketch of a reservoir computer using a MMF as a central element. In addition to the optical part involving the modulated laser field propagating through the MMF, the readout on the camera and the feedback on the SLM is provided electronically. (b) Time-evolution of a nonlinear time-delay differential equation (black) and the replication resulting from the trained weights in the output signal from a MMF (red). Figure adapted from Ref.~\cite{Paudel:20},  licensed under a Creative Commons Attribution (CC BY) license. }
		\label{fig:reservoir}
	\end{figure}

On the next level of complexity, such random projections have also been employed in the context of ``reservoir computing’’ \cite{TANAKA2019100}, in which input states are mapped into a high-dimensional recurrent neural network called a ``reservoir’’ that can capture the complex manifestations even of non-linear dynamical systems. In contrast to traditional machine-learning approaches, the weights in this network are fixed and training is only necessary at the readout stage, where an algorithm is trained to map the state of the reservoir onto a desired output. To evaluate the recursive computation of the reservoir states, disordered media \cite{8807158} and MMFs \cite{Paudel:20} have been successfully employed. In the latter case, an SLM-modulated input beam is propagated through a MMF with the output being recorded on a camera (see Fig.~\ref{fig:reservoir}a). The required non-linearity of this network is implemented through the intensity measurement of the camera, which detects the modulus square of the optical electric field and also saturates above a certain field intensity (as determined by the laser power). The intensity pattern on the camera, together with the previous input, then determines the new input to the reservoir in this iterative process. The output of the ``reservoir computer’’ is the camera image multiplied with a matrix of weights that are trained with a training data-set, like from a non-linear time-series, whose future temporal evolution one aims to predict. A first part of the data produced by the time-series is used for training the weights in the output matrix in such a way that the input wavefront is reproduced; a second part of the data is then used to check if the reservoir computer correctly predicts the input data at future time steps. As shown in Fig.~\ref{fig:reservoir}b, even the near-chaotic dynamics of a a non-linear dynamical system can be well predicted with this strategy.

In the specific MMF-based realization of a reservoir computer discussed above \cite{Paudel:20}, the non-linearity stems from the intensity-measurement on the camera and digital processing was involved to determine a new input based on the camera image. More recent studies have focused on implementing the non-linearity that is required to build a reservoir computer directly in a MMF all-optically, by using the non-linearities present in the fiber material \cite{Tegin2021}. These fiber-based implementations of photonic neural networks have shown impressive performance for classifying X-ray lung images of COVID patients, for speech recognition and for predicting the age of a person just from the image of its face. The accuracy achieved in these tasks is of comparable accuracy as with a purely digital implementation of a computer.

\subsection{Concluding remarks} 

The rapid advances of the past decade have produced a multitude of new methods and, above all, applications for MMFs. Soon we will witness the first ones to reach the level of readiness for technology transfer. This will likely include the exploitation of MMFs in imaging, particularly in the domain of \emph{in-vivo} neuroscience. In many aspects this new approach has already exceeded the current state-of-the-art, moreover with further advance of wavefront-shaping technology its performance and applicability is expected to grow rapidly. This field would benefit  immensely from the availability of phase modulators, which would reach or even exceed the speed and the pixel resolution of the current amplitude-only DMD modulators even in the binary regime (switching between phases of $0$ and $\pi$). Such a tool would  enable the use of broadband, especially pulsed light in MMF-based imaging devices so that multiphoton and non-linear imaging modalities achieve the performance required for practical applications.

With numerous degrees of freedom, a single MMF can be employed as a multi-functional photonic device. In combination with a spatial light modulator (SLM), it can function simultaneously as a temporal pulse shaper, a reconfigurable waveplate, a tunable spectral filter and a programmable polarizer.  
Moreover, the information about spectrum, temporal profile, and polarization state of an optical signal can be recovered from the speckle pattern it generates through a multimode fiber. Thus a single MMF can function as a spectrometer, a polarimeter, a hyper-spectral imager, a profiler of optical pulses and of radio-frequency signals. One issue for the MMF device is fiber stability under external perturbations, such as temperature drift and mechanical vibrations. Several methods have been developed to overcome this problem \cite{cao2017perspective}. Placing a MMF in a temperature-stabilized, low-vacuum chamber has greatly improved the long-term stability \cite{coluccelli2016optical}. Also software corrections are employed to correct the environmental effects \cite{redding2014high}. Finally, the fiber can be replaced by a multimode waveguide fabricated on a chip \cite{redding2016evanescently}, for which a robust package will eliminate most environmental perturbations. 

Finally, the combination of MMFs and light control techniques enables many applications in imaging, sensing, and spectroscopy, as well as communication,  optical computing and cryptography.  With the growing importance of these new domains, we expect the community to become increasingly inter-connected, more aware of emerging trends and equipped to benchmark emerging techniques and achievements against one another. In any case, as new methods are currently under extensive development, we expect the potential for applicability to grow at a rapid pace.  

	\begin{backmatter}
		\bmsection{Funding} HC acknowledges support from the US National Science Foundation (NSF) under Grant No.~DMR-1905465. 
  TT and T\v{C} acknowledge support from the Ministry of Education, Youth and Sports of the Czech Republic (CZ.02.1.01/0.0/0.0/15\_003/0000476) and the European Union's H2020-RIA (101016787).
ST and T\v{C} acknowledge support from the European Research Council (724530, 101069245 and 101082088) and German Federal Ministry for Economic Affairs and Climate Action (03EFRTH030 EXIST DeepEn). SR acknowledges support from the Austrian Science Fund (FWF) under Project No.~P32300 (WAVELAND).
  
		
		\bmsection{Acknowledgments} We wish to thank all members of the community who kindly provided the rights to reprint their figures and illustrations. All current and former collaborators of the authors are expressly acknowledged for their contributions to the work covered in this review. 

\bmsection{Disclosures} 
		  ST and T\v{C} received support from the German Federal Ministry of Economic Affairs and Climate Action (BMWK) to further develop research results on exploiting multimode fibres in imaging applications towards implementation in technical products and preparing startup founding.
  HC, TT and SR declare no conflicts of interest.
		\bmsection{Data Availability Statement} Data underlying the results presented in this paper are not publicly available at this time but may be obtained from the authors upon reasonable request.

	\end{backmatter}
	
	\section{References}

	
	\bibliography{AOPref}

		\section*{Author Biographies}
		\begingroup
		\setlength\intextsep{0pt}
		\begin{minipage}[t][10.3cm][t]{1.0\textwidth} 
			\begin{wrapfigure}{L}{0.25\textwidth}
				\includegraphics[width=0.25\textwidth]{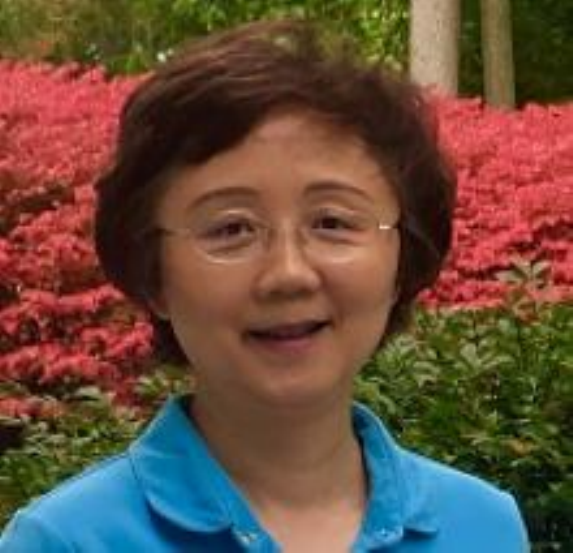}
			\end{wrapfigure}
			\noindent
			{\bfseries Hui Cao} received her Ph.D. degree in Applied Physics from Stanford University in 1997. She is the John C. Malone Professor of Applied Physics, Professor of Physics, and Professor of Electrical Engineering at Yale University. Prior to joining the Yale faculty in 2008, she was on the faculty of Northwestern University from 1997 to 2007. Her primary research interests are mesoscopic physics, complex lasers, multimode fiber optics, nanophotonics, wave chaos, and biophotonics. Cao is a Fellow of the APS, Optica (formerly OSA), IEEE and AAAS. She is an elected member of the National Academy of Sciences, and the American Academy of Arts and Sciences.\\
		\end{minipage}
  
		\begin{minipage}{1.0\textwidth}
			\begin{wrapfigure}{L}{0.25\textwidth}
				\includegraphics[width=0.25\textwidth]{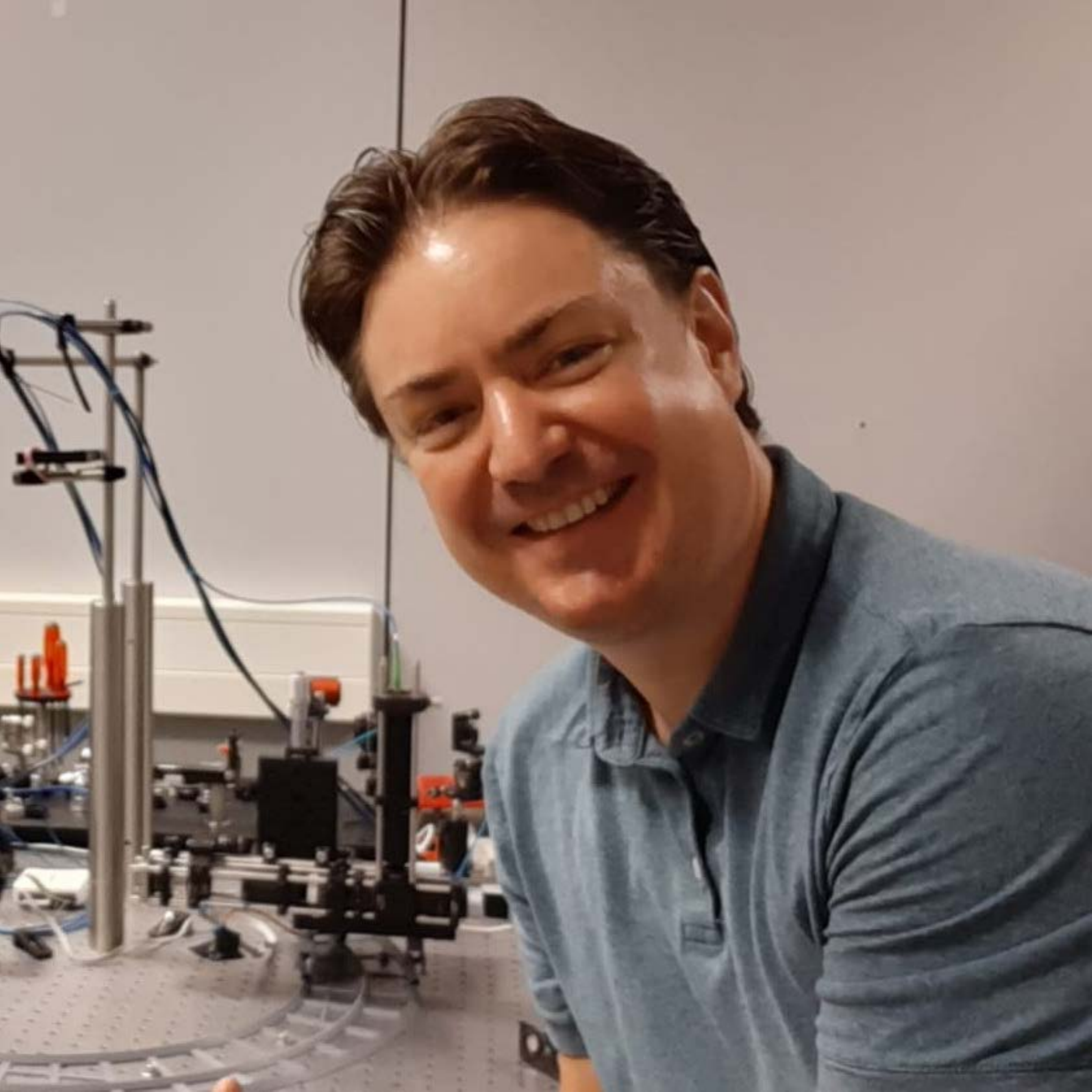}
			\end{wrapfigure}
			\noindent
			{\bfseries Tom\'{a}\v{s} \v{C}i\v{z}m\'{a}r} is a professor of Wave Optics at the Friedrich-Schiller University of Jena and the head of the Fibre Research and Technology department of the Leibniz Institute of Photonic Technology (IPHT) in Jena. He leads the Holographic Endoscopy group at IPHT and the group of Complex Photonics at the institute of Scientific Instruments in Brno. Although his scientific background is Physics, throughout his scientific career he took part in a variety of inter-disciplinary projects in Bio-Medical Photonics, mostly related to optical manipulation, digital holography, and microscopy. His recent research activities are focused on Photonics in optically random environments (particularly multimode fibres) and deep-tissue \emph{in-vivo} imaging.\\
		\end{minipage}

  \begin{minipage}{1.0\textwidth}
			\begin{wrapfigure}{L}{0.25\textwidth}
				\includegraphics[width=0.25\textwidth]{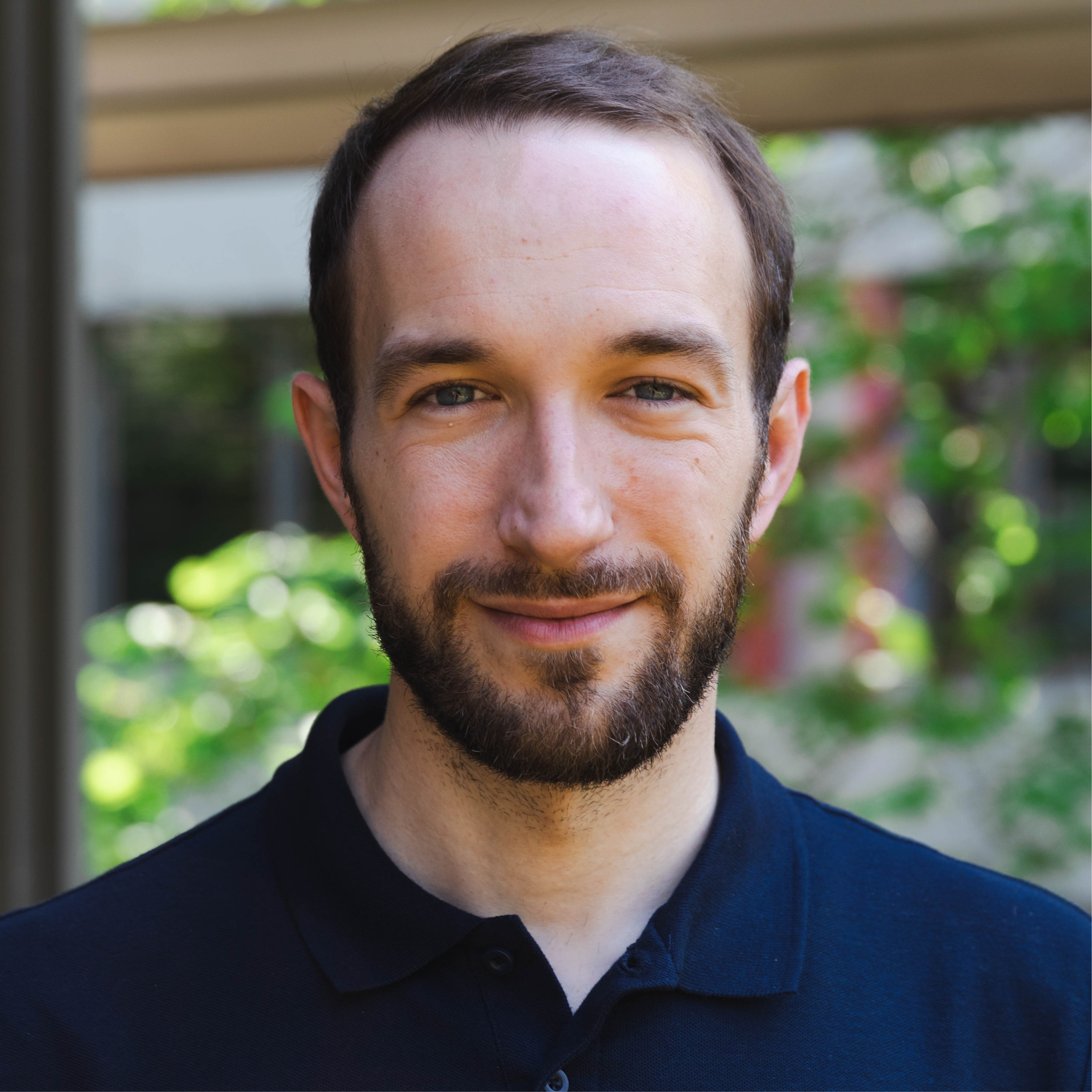}
			\end{wrapfigure}
			\noindent
			{\bfseries Sergey Turtaev} received his PhD from the University of Dundee in 2018 as a Marie-Curie PhD fellow, funded by the European Commission. Now Sergey is a senior postdoc at the Leibniz Institute of Photonic Technology (Germany), leading a startup endeavour on multimode-fibre-based endoscopes for deep-tissue imaging. His research interests include fibre optics, laser physics and wavefront shaping in application to biophotonics. \\~~\\~~\\~~\\~~\\~~
		\end{minipage}
  \begin{minipage}{1.0\textwidth}
			\begin{wrapfigure}{L}{0.25\textwidth}
				\includegraphics[width=0.25\textwidth]{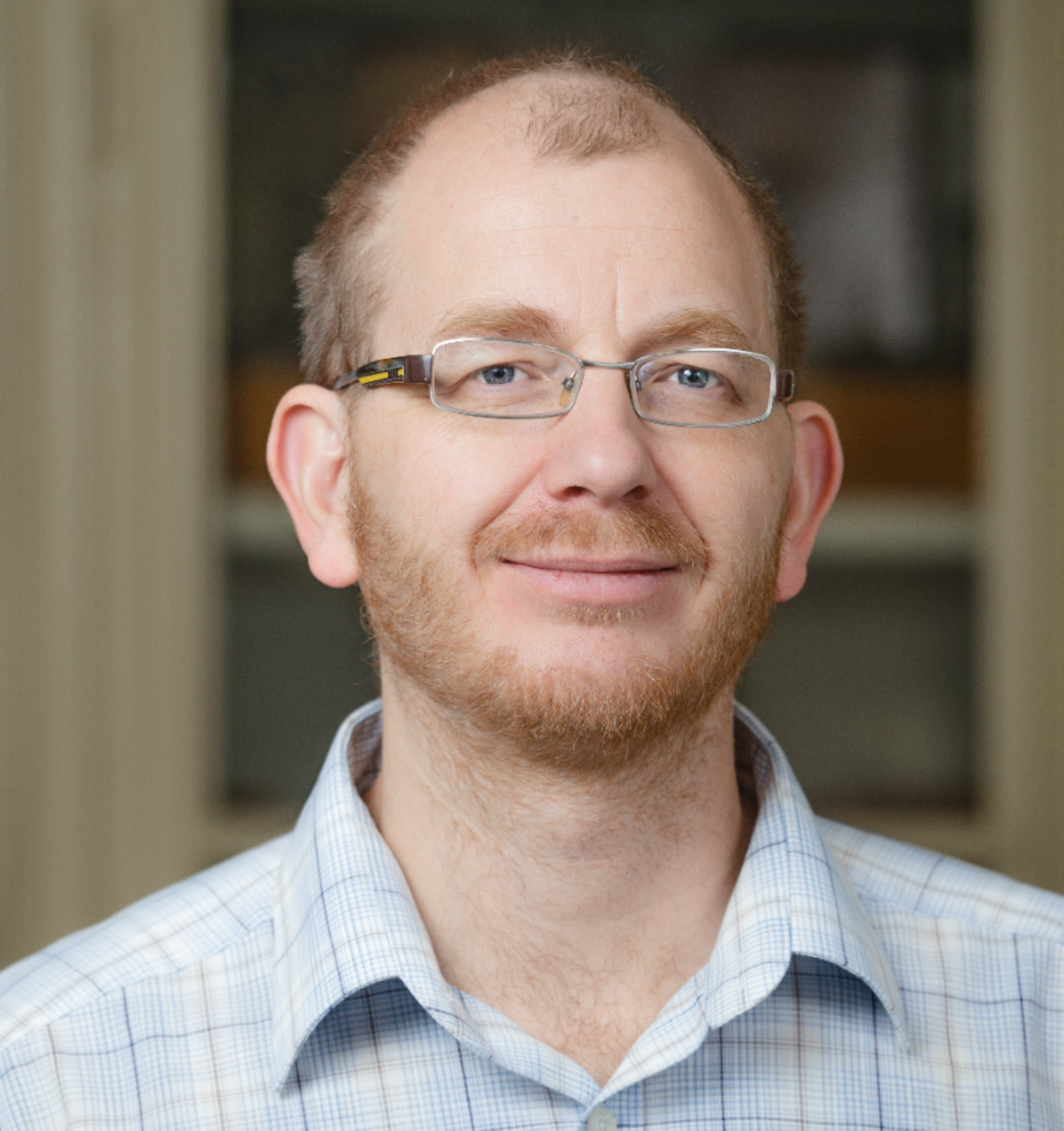}
			\end{wrapfigure}
			\noindent
			{\bfseries Tom\'{a}\v{s} Tyc} is a professor at the Department of Theoretical Physics and Astrophysics at Masaryk University, Brno, Czech Republic. After finishing his Ph.D. in 1999, he worked as a research fellow or visiting professor at universities in Vienna, Sydney, Calgary, St~Andrews and Dundee. Apart from fiber optics, his research interests include theory of geodesic lenses, invisible cloaks and absolute optical instruments.\\~~\\~~\\~~
		\end{minipage}
  
        \begin{minipage}{1.0\textwidth}
			\begin{wrapfigure}{L}{0.25\textwidth}
				\includegraphics[width=0.25\textwidth]{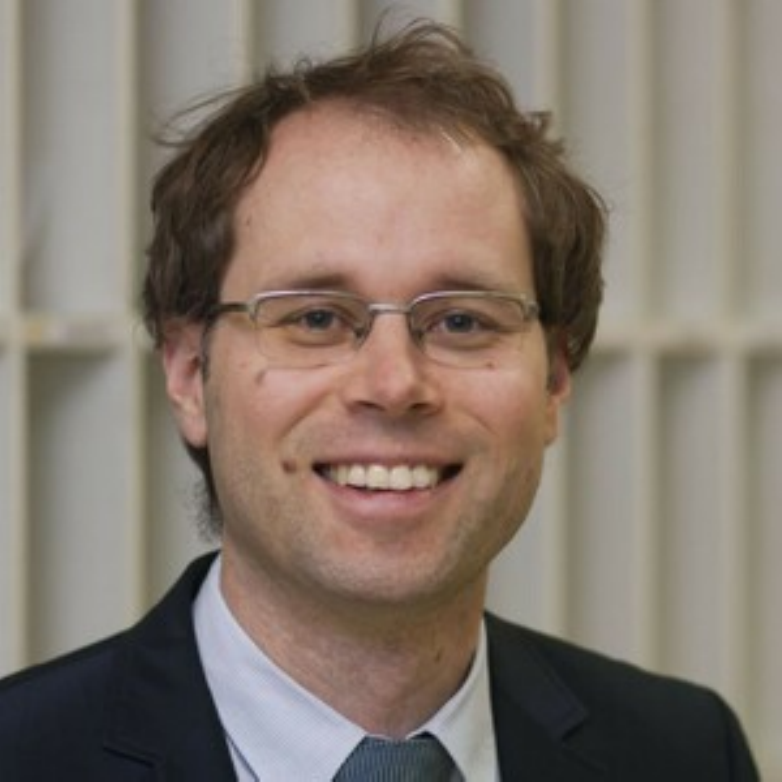}
			\end{wrapfigure}
			\noindent
			{\bfseries Stefan Rotter} is professor at the Institute for Theoretical Physics, Vienna University of Technology (TU Wien). After studies in Vienna and Lausanne, he obtained his Ph.D. in 2004, followed by a postdoctoral position at Yale University. His group was established in 2011 and focuses on non-Hermitian physics, theoretical quantum optics and on the propagation of classical or quantum waves through complex media. Rotter has been a guest professor in Paris (Joliot Chair), he has been selected as an Outstanding Referee by the American Physical Society, and two of his recent works have been included in Physics World's list of Top 10 Breakthroughs in Physics (2022).\\
		\end{minipage}
		\endgroup

\end{document}